\documentclass[longauth]{aa} 

\usepackage{threeparttable}
\usepackage{adjustbox}
\usepackage{rotating}
\usepackage{amsmath}
\usepackage{graphicx}
\usepackage{txfonts}
\usepackage[normalem]{ulem}

\usepackage{bm}

\usepackage[breaklinks=true,colorlinks=true,allcolors=blue,pagebackref=true]{hyperref}

\newcommand{\mypm}{\mathbin{\smash{%
\raisebox{0.65ex}{%
            $\underset{\raisebox{0.05ex}{$\smash -$}}{\smash+}$%
            }%
        }%
    }%
}

\newcommand{\goodchi}{\protect\raisebox{2pt}{$\chi$}}

\begin{document}

   \title{Infrared observations of the flaring maser source G358.93$-$0.03\thanks{Based on observations collected at the European Organization for Astronomical Research in the Southern Hemisphere under ESO program 0103.C-9033(A). Based on observations made with the NASA/DLR Stratospheric Observatory for Infrared Astronomy (SOFIA) under Proposal IDs 75\_0037 and 08\_0163.}}
   \subtitle{SOFIA confirms an accretion burst from a massive young stellar object
   }

   \author{B. Stecklum\inst{1}
          \and
          V. Wolf\inst{1}
          \and
          H. Linz\inst{2}
          \and
          A. Caratti o Garatti\inst{3}
          \and
          S. Schmidl\inst{1}
          \and
          S. Klose\inst{1}
          \and
          J. Eislöffel\inst{1}
          \and 
          Ch. Fischer\inst{4}
          \and
          C. Brogan\inst{5}
          \and
          R. Burns\inst{6}
          \and
          O. Bayandina\inst{7} 
          \and
          C. Cyganowski\inst{9}
          \and
          M. Gurwell\inst{10}
          \and
          T. Hunter\inst{4}
          \and
          N. Hirano\inst{11}
          \and
          K.-T. Kim\inst{12}
          \and
          G. MacLeod\inst{13}
          \and
          K. M. Menten\inst{14}
          \and
          M. Olech\inst{15}
          \and
          G. Orosz\inst{16}
          \and
          A. Sobolev\inst{17}
          \and
          T. K. Sridharan\inst{10}
          \and
          G. Surcis\inst{18}
          \and
          K. Sugiyama\inst{6}
          \and
          J. van der Walt\inst{19}
          \and
          A. Volvach\inst{20}
          \and
          Y. Yonekura\inst{21}
          }

   \institute{
            Thüringer Landessternwarte Tautenburg, Sternwarte 5, 07778 Tautenburg, Germany,
             \email{stecklum@tls-tautenburg.de}
        \and
            Max Planck Institute for Astronomy, Königstuhl 17, 69117 Heidelberg, Germany
        \and
            Dublin Institute for Advanced Studies, 31 Fitzwilliam Place, D02 XF86, Dublin, Ireland
        \and
            Deutsches SOFIA Institut, University of Stuttgart, 70569 Stuttgart, Germany
        \and
            National Radio Astronomy Observatory, 520 Edgemont Road, Charlottesville, VA 22903, USA
        \and
            Mizusawa VLBI Observatory, National Astronomical Observatory of Japan, Osawa 2-21-1, Mitaka, Tokyo 181-8588, Japan
        \and
            Joint Institute for VLBI ERIC, Oude Hoogeveensedijk 4, 7991 PD Dwingeloo, The Netherlands
         \and
            Astro Space Center, P.N. Lebedev Physical Institute of RAS, 84/32 Profsoyuznaya st., Moscow, 117997, Russia
        \and
            SUPA, School of Physics and Astronomy, University of St. Andrews, North Haugh, St. Andrews KY16 9SS, UK
        \and
            Center for Astrophysics | Harvard Smithsonian, Cambridge, MA 02138, USA
        \and
            Academia Sinica Inst of Astronomy \& Astrophysics ASIAA PO Box 23-141. Taipei 106. Taiwan
        \and
            Korea Astronomy and Space Science Institute, 776 Daedeokdae-ro, Yuseong-gu, Daejeon, 34055, Republic of Korea
        \and
            Hartebeesthoek Radio Astronomy Observatory, P.O. Box 443, Krugersdorp 1740, South Africa ; The University of Western Ontario, 1151 Richmond Street, London, ON N6A 3K7, Canada
        \and
            Max-Plank-Institut für Radioastronomie, Auf dem Hügel 69, 53121 Bonn, Germany
        \and
            Institute of Astronomy, Faculty of Physics, Astronomy and Informatics, Nicolaus Copernicus University, Grudziadzka 5, 87-100 Torun, Poland
        \and
            Xinjiang Astronomical Observatory, Chinese Academy of Sciences, Urumqi, Xinjiang, China
        \and
            Astronomical Observatory, Institute for Natural Sciences and Mathematics, Ural Federal University, 19 Mira street, Ekaterinburg 620002, Russia)
        \and
            INAF-Osservatorio Astronomico di Cagliari, Via della Scienza 5, 09047, Selargius, CA, Italy
        \and
           Space Research Unit, Physics Department, North West University, Potchefstroom 2520, South Africa 
        \and
            Radio Astronomy Laboratory of Crimean Astrophysical Observatory RAS, Katsively, RT-22 Crimea
        \and
            Center for Astronomy, Ibaraki University, 2-1-1 Bunkyo, Mito, Ibaraki 310-8512, Japan
        }

   \date{Received October 11, 2020; accepted December 21, 2020}

 
  \abstract
    {
   Class~II methanol masers are signposts of massive young stellar objects (MYSOs). Recent evidence shows that flares of these masers are driven by MYSO accretion bursts. Thus, maser monitoring can be used to identify such bursts which are hard to discover otherwise. Infrared observations reveal burst-induced changes in the spectral energy distribution - first and foremost a luminosity increase - which provide valuable information on a very intense phase of high-mass star formation.
   }
   {
   In mid-January 2019, flaring of the 6.7\,GHz CH$_3$OH maser (hereafter maser) of the MYSO G358.93-0.03 (hereafter G358) was reported. The international maser community (M2O) initiated an extensive observational campaign  which revealed extraordinary maser activity and yielded the detection of numerous new masering transitions. 
   Interferometric imaging with the Atacama Large Millimeter/submillimeter Array (ALMA) and the Submillimeter Array (SMA) resolved the maser emitting core of the  star forming region and proved the association of the masers with the brightest continuum source (MM1), which hosts a hot molecular core. These observations, however, failed to detect a significant rise in the (sub)millimeter dust continuum emission. Therefore, we performed near-infrared (NIR) and far-infrared (FIR) observations to prove or disprove whether the CH$_3$OH flare was driven by an accretion burst. 
   }
   {
    NIR imaging with the Gamma-Ray Burst Optical/Near-infrared Detector (GROND) has been acquired and integral-field spectroscopy with the Field-Imaging  Far-Infrared  Line  Spectrometer (FIFI-LS) aboard the Stratospheric Observatory for Infrared Astronomy (SOFIA) was carried out on two occasions to detect possible counterparts to the (sub)millimeter sources and compare their photometry to archival measurements. The comparison of pre-burst and burst spectral energy distributions is of crucial importance to judge whether a substantial luminosity increase, caused by an accretion burst, is present and
    if it triggered the maser flare. Radiative transfer modeling of the 
    spectral energy distribution (SED) of the dust continuum emission at multiple epochs provides valuable information on the bursting MYSO.
   }
   {
   The FIR fluxes of MM1 measured with FIFI-LS exceed those from 
   {\em Herschel} significantly, which clearly confirms the presence of an accretion burst. The second epoch data, taken 
   about 16 months later, still show increased fluxes. Our radiative transfer modeling yielded major burst parameters and suggests that the MYSO
   features a circumstellar disk which might be transient.
   From the pre-burst, burst, and post-burst SEDs, conclusions on heating and cooling time-scales could be drawn. Circumstances of the burst-induced maser relocation have been explored.
   }
   {
   The verification of the accretion burst from G358 is another confirmation that Class~II methanol maser flares represent an alert for such events. Thus, monitoring of these masers greatly enhances the chances of identifying MYSOs during periods of intense growth. 
   The few events known to date already indicate that there is a broad range in burst strength and duration as well as environmental characteristics. The G358 event is the shortest and least luminous  accretion burst known to date. According to models, bursts of this kind occur most often.
   }


   \maketitle
%

\section{Introduction}\label{intro}
The collapse of dense molecular cloud cores gives rise to the birth of stars, a process which was thought to proceed in a smooth and continuous fashion. However, the first piece of evidence for the unsteady growth of forming stars emerged by recognizing that the outburst of FU Orionis, which was thought to be a rare phase of early stellar evolution \citep{1966VA......8..109H}, was instead an episode of enhanced disk accretion \citealp{1985ApJ...299..462H}. Since then, it has been realized that episodic accretion is an intrinsic feature of forming young stars \citep{1996ARA&A..34..207H, 2014prpl.conf..387A}. While this knowledge had been established exclusively from observations of low-mass stars, which become optically visible while still accreting, it was unknown until recently whether high-mass stars ($M_{\star}{\gtrsim}8\,{\rm M}_\odot$) show the same behavior during their formation. 
Their scarcity and fast formation timescales 
imply that they are still deeply embedded in their parental core while reaching the main sequence. This suggests that similar outbursts during high-mass star formation, if present at all, might be difficult to detect. 

A candidate young massive eruptive variable, V723 Car, was identified by \citet{2015MNRAS.446.4088T}. The object brightened in the $K$ band by 3.7\,mag between 1993 and 2003. Since the outburst was found a posteriori, no information on the possible accretion luminosity is available. The post-burst luminosities range from $2.5\,{\times}\,10^3\,{\rm L}_\odot$ \citep{2011ApJS..194...14P} to ${\sim}\,4\,{\times}\,10^3\,{\rm L}_\odot$ \citep{2015MNRAS.446.4088T} which correspond to a mass of 8 -- 9\,${\rm M}_\odot$ \citep{1996MNRAS.281..257T} for a zero-age main sequence (ZAMS) object. 
In the context of the forthcoming discussion, it has to be emphasized that this source is not associated with masers.
So, while the V723 Car event might be considered to be an accretion burst from a massive young stellar object (MYSO), the lack of observational coverage before and at the time of its incidence precludes major conclusions with regard to high-mass star formation.

Recently the situation concerning MYSO accretion bursts abruptly changed following the discoveries of the almost coincident events from the MYSOs S255IR-NIRS3 \citep{2016ATel.8732....1S, 2017NatPh..13..276C, 2018ApJ...863L..12L} and NGC6334I-MM1 \citep{2017ApJ...837L..29H, 2018ApJ...854..170H}. The luminosity increase, seen at infrared (IR) and (sub)mm wavelengths for S255IR-NIRS3 and
in the (sub)mm regime for NGC6334I-MM1, provided direct evidence for enhanced accretion rates. Most notably, these outbursts were accompanied by flares of Class~II methanol masers (hereafter methanol masers; \citealp{2015ATel.8286....1F,2018A&A...617A..80S, 2018MNRAS.478.1077M}). This confirmed a radiative pumping mechanism of this kind of methanol masers \citep{1991ASPC...16..119M, 1997A&A...324..211S, 2005MNRAS.360..533C}, which is consistent with variability studies of the maser emission \citep{2018MNRAS.474..219S, 2019MNRAS.485..777D}. Since methanol masers trace the very early stages of massive star formation (for example \citealp{2013MNRAS.435..524B}), maser flares might be taken as a proxy for accretion variability of the protostellar host. Keeping this in mind, the international maser community established the Maser Monitoring Organization (M2O)\footnote{See M2O website at \url{http://MaserMonitoring.org}} to coordinate single-dish monitoring of masers and interferometric follow-up measurements.

G358.93-0.03 (hereafter G358, RA: $\rm 17^h 43^m 10.\!^s02{}$, $\rm {DEC\!:~} {-}29\degr 51'' 45\farcs8$, J2000) is a hitherto little explored massive star forming site as evident from just eight SIMBAD \citep{2000A&AS..143....9W} entries until 2018.
The kinematic distance amounts to 6.75$\,{\mypm}\,^{\,0.37}_{\,0.68}$\,kpc \citep{2019ApJ...881L..39B} which
implies a galactocentric one of 1.6\,kpc.
It is consistent with {\it Gaia} distances of visible stars in the G358 region of $\lesssim$\,5\,kpc
which impose a lower limit to the distance of the molecular cloud hosting G358
\citep{2020NatAs...4..506B}.
In mid-January 2019, flaring of the 6.7\,GHz CH$_3$OH maser line \citep{1991ApJ...380L..75M} in G358 was announced \citep{2019ATel12446....1S}. Thus, for the first time, M2O orchestrated an extensive observing campaign which became extremely successful. 

Thanks to the immediate response, an unprecedented wealth of masering lines, including numerous new transitions, could be observed \citep{2019ApJ...876L..25B, 2019ApJ...881L..39B, 2019MNRAS.489.3981M} and new maser species were discovered \citep{2020ApJ...890L..22C, 2020NatAs.tmp..144C}. Interferometric imaging in the (sub)millimeter spectral range using the Atacama Large Millimeter/submillimeter Array (ALMA) and the Submillimeter Array (SMA) dissected the star forming region and pinpointed the MYSO which hosts the flaring masers \citep{2019ApJ...881L..39B}. In all likelihood, the brightest continuum source MM1, which turned out to be a hot molecular core, experienced an accretion burst. For the first time, a spectacular confirmation of the event was achieved by high-resolution, multi-epoch observations of the 6.7-GHz methanol maser emission which revealed outward maser spot propagation, tracing the spread of the thermal radiation emanating from the burst \citep{2020NatAs...4..506B}.
However, without evidence for a significant rise in (sub)millimeter dust continuum emission from
MM1 \citep{2019ApJ...881L..39B}, the energetics of the burst remained an open issue. Therefore, we aimed for observations to identify the IR counterpart of MM1 and to verify its luminosity increase, thus independently confirming the third MYSO accretion burst witnessed so far. 

At present, the Stratospheric Observatory for Infrared Astronomy (SOFIA, \citealp{1993AdSpR..13..549E, 2012ApJ...749L..17Y}) is the only facility which offers the capability to trace the far-infrared (FIR) flux increase caused by such an event. Consequently, an attempt was made for observing G358 with SOFIA which turned out to be successful. The observational results in context with supplementary data, the analysis and interpretation are the subjects of the present paper and will be outlined in the following.

The paper is organized as follows. At the beginning the observational foundation will be explained. The next part deals with deriving constraints on the spectral energy distribution (SED) of the bursting source. The estimation of the luminosity increase, the central quantity for assessing the accretion burst, is performed in two steps. First, a simplified treatment using graybody functions is applied. Then, a more thorough analysis is performed, utilizing dust continuum radiative transfer. The discussion section concludes the paper in which results of the present investigation are put in context with regard to previous observational and theoretical findings.

We note that the term ``luminosity'', if not declared otherwise, refers to the bolometric luminosity throughout the paper. Similarly, the term or value ``error'' always implies the 1$\sigma$ error or standard deviation unless noted otherwise. Magnitudes are based on the Vega system.


\section{Observations}\label{obs}
\subsection{Archival data}\label{arch}
Archival data is essential for establishing the presence of an accretion burst since it constrains the source luminosity during the pre-burst state. Because G358 is located in the Galactic center region it has been covered by a wealth of surveys. For the present study, fluxes in the near-IR (NIR), mid-IR (MIR), and far-IR (FIR) as well as positions and images have been retrieved from the 
surveys listed in Tab.\ref{adat}

\begin{table*}
\caption[]{Archival resources}
\label{adat}
\begin{tabular}{p{0.1\linewidth}|p{0.11\linewidth}|p{0.45\linewidth}|p{0.225\linewidth}}
\hline
\noalign{\smallskip}
Survey      &  Spectral range & Facility&  Reference \\
\noalign{\smallskip}
\hline
\noalign{\smallskip}
2MASS & NIR & 1.3-m telescopes at Fred Lawrence Whipple Observatory and Cerro Tololo Inter-American Observatory &\cite{2003yCat.2246....0C}\\
VVV & NIR & 4-m VISTA telescope at ESO Cerro Paranal & \vspace{-0.6em}\cite{2010NewA...15..433M}\\
ISOGAL & MIR & 0.6-m ISO space telescope, ISOCAM camera  &\vspace{-0.6em}\cite{2003AA...403..975O}\\
GLIMPSE & MIR & 0.85-m {\em Spitzer} space telescope, IRAC camera  &\vspace{-0.6em}\cite{2009yCat.2293....0S}\\
(NEO)WISE& MIR & 0.4-m WISE space telescope &\vspace{-0.6em}\cite{2014ApJ...792...30M} \\
MIPSGAL& MIR and FIR & 0.85-m {\em Spitzer} space telescope, MIPS camera &\vspace{-0.6em}\cite{2015AJ....149...64G}  \\
FIS& FIR &0.67-m AKARI space telescope, FIS scanning photometer  &\vspace{-0.6em}\cite{2015PASJ...67...50D}  \\
HI-GAL& FIR & 3.5-m {\em Herschel} space telescope, PACS and SPIRE cameras  &\vspace{-0.6em}\cite{2016yCat..35910149M}  \\
ATLASGAL&(sub)mm  & 12-m APEX telescope at Mount Chajnantor &\vspace{-0.6em}\cite{2009AA...504..415S}  \\
\noalign{\smallskip}
\hline
\end{tabular}
\end{table*}


The comparison of the HI-GAL photometry \citep{2016yCat..35910149M} with that performed by the ATLASGAL team \citep{2013A&A...549A..45C} revealed
inconsistencies
at longer wavelengths and with regard to the error estimates. Therefore, we
performed photometry on our own on the corresponding PACS and SPIRE images  (epoch 2010 September 07), retrieved from the NASA/IPAC Infrared Science Archive (IRSA)\footnote{\url{https://irsa.ipac.caltech.edu}}, using the MPFit2DPeak function from the IDL Astronomy library \citep{1995ASPC...77..437L}. The point spread function was chosen to be a 2D Gaussian which yielded the best fit over other representations (Moffat, Lorentzian profiles). The background was estimated on a fixed annulus for all wavelengths to ensure consistency of sky estimates.

The target is not included in the AKARI Bright Source Catalog \citep{2010yCat.2298....0Y} but can be identified in Far-Infrared Surveyor images (FIS, \citealp{2015PASJ...67...50D}) taken with the N60  ($\lambda_{\rm central}\,{=65}\,\mu$m) and WIDE-S ($\lambda_{\rm central}\,{=90}\,\mu$m) filters (epoch
2007 January). Correspondingly, photometry was performed on those frames in the same way as described above. Since the WIDE-S image suffers from severe striping, the resulting flux has a substantial uncertainty and will therefore be
neglected.


\subsection{Near-infrared imaging}\label{nir}
Optical and near-infrared (NIR) imaging of G358 was performed using the seven-channel Gamma-Ray Burst Optical/Near-infrared Detector GROND \citep{2008PASP..120..405G}, using director's discretionary time (DDT) at the
MPG/ESO 2.2-m telescope at La Silla (Chile) on 2019 February 8. GROND obtains images in seven bands (optical: Sloan {\it g'\,r'\,i'\,z'}, NIR: $J H K_{\rm s}$) simultaneously. The total integration time amounts to 38 minutes. Data processing was performed by means of the GROND pipeline \citep{2008ApJ...685..376K}.

\subsection{Far-infrared integral-field spectroscopy}\label{fifi}
The Field-Imaging Far-Infrared Line Spectrometer (FIFI-LS, \citealp{2000SPIE.4014...14L, 2018JAI.....740003F, 2018JAI.....740004C}) is a far-infrared integral field spectrograph aboard SOFIA. FIFI-LS features a blue and red channel in parallel. Each channel has a field of view
(FOV) consisting of five by five spatial pixels with a
plate scale of 6\arcsec/pixel in the blue channel and
12\arcsec/pixel in the red channel. They provide an overall wavelength coverage from 51\,$\mu$m to 203\,$\mu$m. This matches very well the range of the SED, where MYSOs emit the bulk of their energy by dust continuum radiation, and where the relative flux increase due to an accretion burst is highest \citep{2019MNRAS.487.4465M}.
Thus, instruments like FIFI-LS provide the best prospects to detect the luminosity increase due to accretion bursts.

A DDT proposal to perform spectro-photometry of the FIR dust emission of G358 using FIFI-LS was submitted in mid-February 2019.
One hour of observing time was awarded 
by the Director of SOFIA Science Mission Operations. The measurements were performed on 2019 May 1, when the maser flare was still strong but already decaying (MacLeod et al., in prep.; Yonekura et al., in prep.). Several spectral bands (see Table \,\ref{fifiphot}) were chosen to sample the full spectral range of FIFI-LS and, at the same time, cover high rotational transitions of the CO molecule. The spectral scan length per sub-band ranged from 0.3 to 1.0\,$\mu$m.
A regular follow-up proposal for SOFIA Cycle 8 was submitted and accepted as well. 
Due to the flight suspension induced by the COVID-19 pandemic, the observations were delayed and eventually 
performed on 2020 August 28. The same settings and amount of observing time as for the first epoch measurements were used. However, because of a technical issue, no data could be obtained in the blue channel. Yet, 
drawing conclusions on the flux evolution from the two epochs became possible.

\section{Data analysis and results}\label{res}

As evidenced by the ALMA/SMA observations, G358 is a star forming complex harboring several massive protostars \citep{2019ApJ...881L..39B}. Because of its considerable distance and the large beam sizes of observing facilities at MIR and FIR wavelengths, the photometry obtained by the latter represents the total flux. The following 
considerations aim at identifying and characterizing the IR counterpart of MM1. This includes an approach to account for flux contributions of all other components. Eventually, conclusions on the nature of MM1 will be drawn based on radiative transfer (RT) modeling of its pre-burst and burst SEDs.

\subsection{NIR imaging and source identification}\label{rnir}
\begin{figure}   
\centering
	\includegraphics[width=\columnwidth]{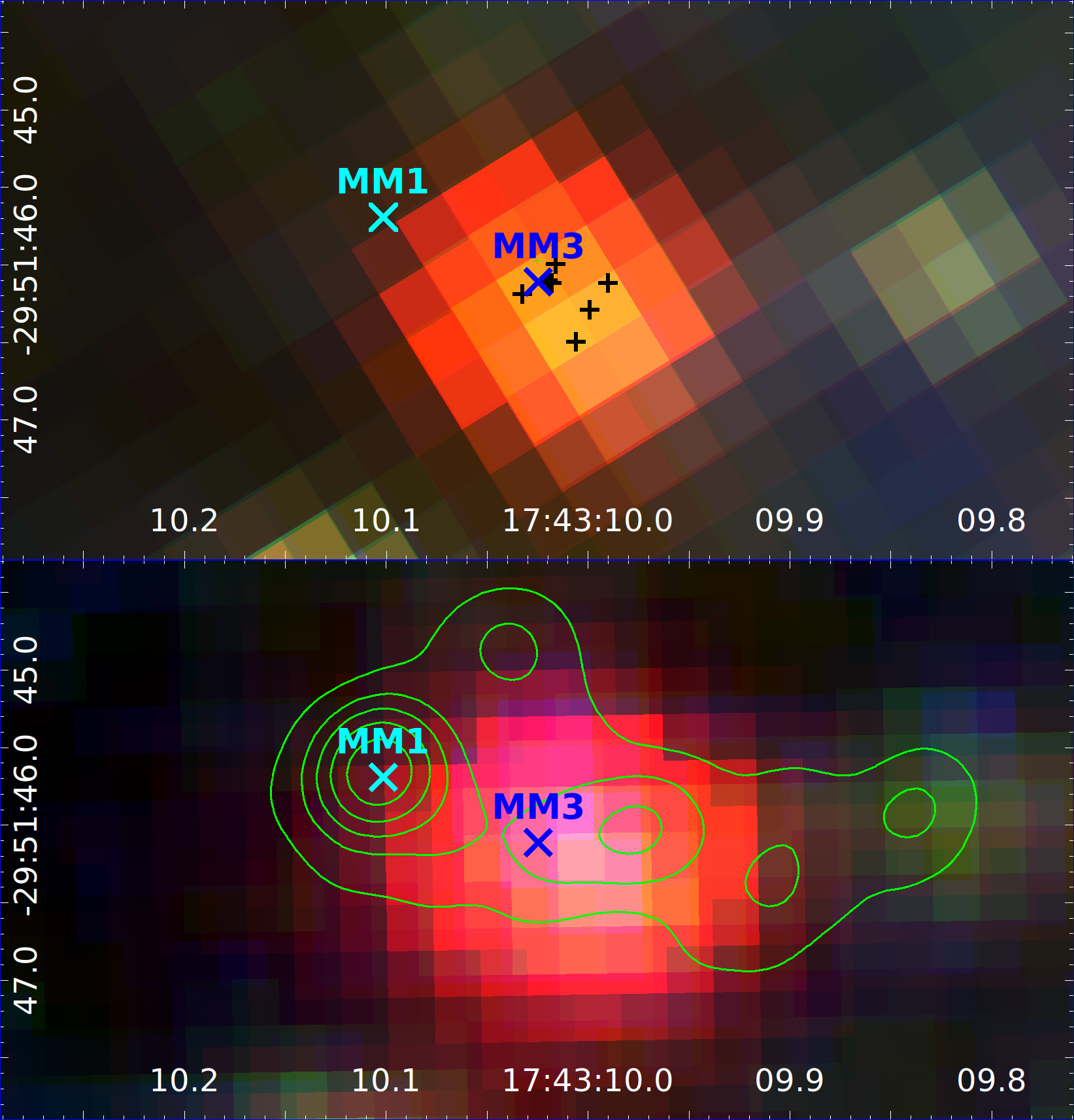}
	\caption{\textbf{top:} VVV $JHK_s$ color composite of the G358 region (epoch 2010 August 15). The positions of MM1 and MM3 from \citet{2019ApJ...881L..39B} are marked. Black plus signs denote positions from IR observations at wavelengths up to 24\,$\mu$m. \textbf{bottom:} GROND $JHK_s$ color composite (epoch 2019 February 8) with contours of the ALMA 0.89\,mm continuum map \citep{2019ApJ...881L..39B}.
	}
 \label{fig:GROND}
\end{figure}

As shown in Fig.\,\ref{fig:GROND},
a NIR source is situated close to the position of G358. It is listed as 2MASS~J17431001$-$2951460 in the 2MASS All-Sky Catalog of Point Sources \citep{2003yCat.2246....0C} but was detected only in the $K_{\rm s}$ band at 11.5$\,{\pm}\,$0.07\,mag. The deeper {\bf V}ISTA {\bf V}ariables in {\bf V}ia Lactea Survey (VVV), \citealp{2017yCat.2348....0M}) which imaged the Galactic bulge and the adjacent southern plane from 2010 to 2016
yielded an H-band detection as well. The VVV $K_{\rm s}$ and $H$ catalog magnitudes of $11.87\,{\pm}\,0.01$ and $15.54\,{\pm}\,0.06$\,mag imply a color index of $H-K_{\rm s}\,{=}\,3.67\,{\pm}\,0.06$\,mag which indicates a very red object.  In the VVV  catalog, the object is flagged as variable as it has shown rapid brightness changes within a 3$\sigma$ range of 0.4\,mag, and a peak-to-peak variation of 0.79\,mag, during five years of VVV $K_{\rm s}$-band monitoring. Its mean brightness during the VVV monitoring amounts to $K_{\rm s}\,{=}\,$12.23\,mag.

Our GROND imaging detected it in the $J, H$ and $K_{\rm s}$ bands, but not at shorter wavelengths.
The GROND Ks magnitude of $11.9\,{\pm}\,0.02$ indicates a brightness, increased by 0.34\,mag with respect to the VVV mean, which is consistent with the source variability. Its position, within 0\farcs2, is consistent with the secondary hot molecular core and dust continuum source MM3 detected by ALMA, which is located 1\farcs09 to the southwest of the main hot molecular core MM1 \citep[cf.][]{2019ApJ...881L..39B}. Therefore, the NIR source 
represents the IR counterpart of MM3 and is, thus, unrelated to the outburst. 

The VVV and GROND NIR color composite images are shown for comparison in Fig.\,\ref{fig:GROND}.
The $K_{\rm s}$ image of the difference GROND$-$VVV, obtained after flux scaling and convolving to the same spatial resolution, does not provide evidence for the presence of a light echo from an accretion outburst, unlike for the case of S255IR-NIRS3 \citep{2017NatPh..13..276C}. This may be another sign for the high extinction toward the bursting source.

\subsection{WISE/(NEO)WISE photometry}\label{rneo}

\begin{figure*}
	\sidecaption
	\includegraphics[width=12cm]{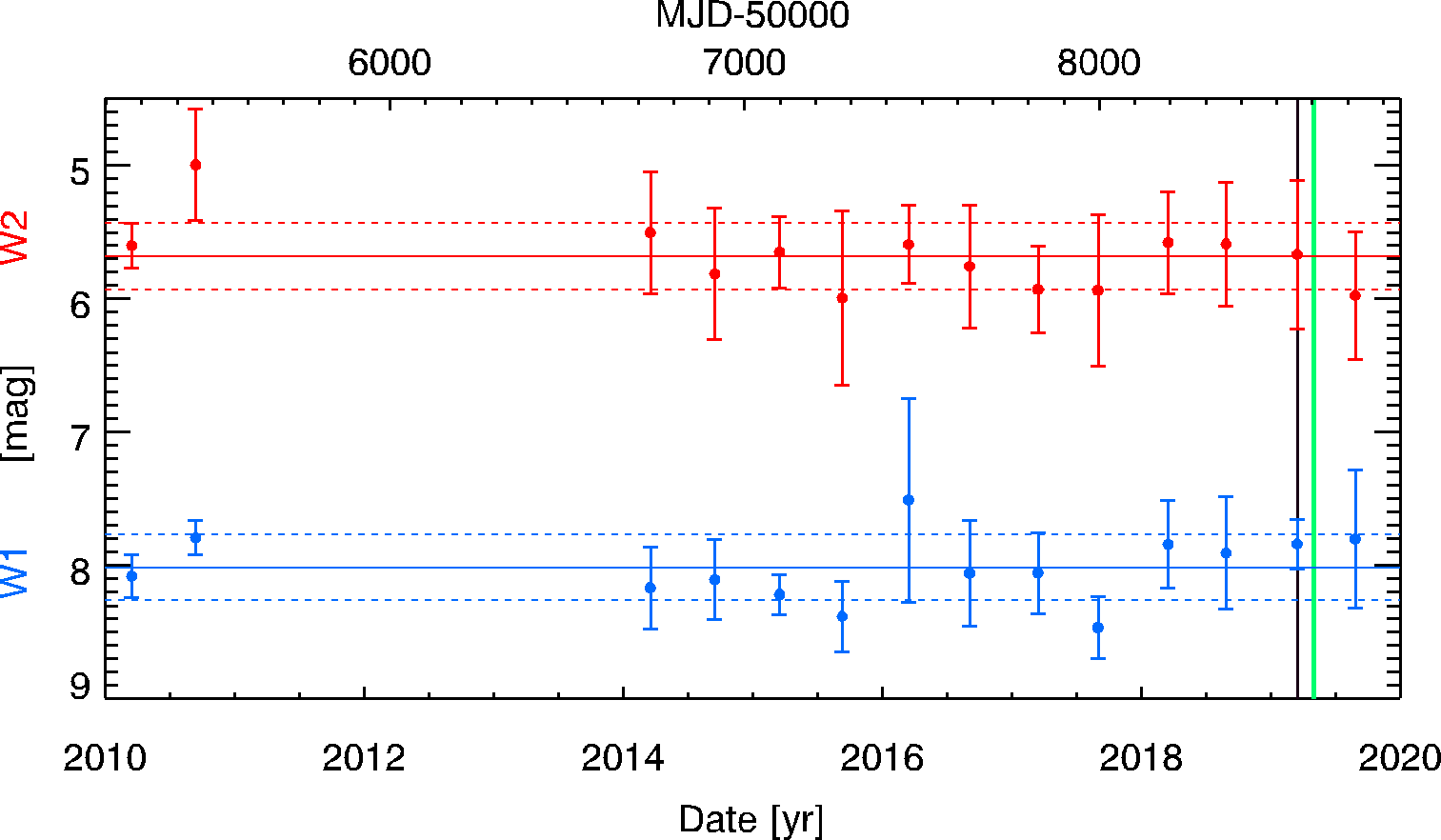}
	\caption{(NEO)WISE W1 (blue) and W2 (red) light curves based on mean magnitudes and respective errors for each visit. The first two epochs are from the WISE mission. Vertical lines mark the dates of the flare peak (black) and the first FIFI-LS epoch (green). Horizontal lines indicate mean magnitudes (solid) and their errors (dashed).
	}
 \label{fig:NW_lc}
\end{figure*}

Due to its orbit, the WISE IR space telescope \citep{2010AJ....140.1868W} visits a region in the sky twice a year. For G358 the 2019 visits occurred on March 17 and August 27. The first one almost coincided with the peak of the maser flare (MacLeod et al., in prep.; Yonekura et al., in prep.) which maximized chances for a possible MIR detection.
Photometry for G358 from the WISE and subsequent (NEO)WISE \citep{2014ApJ...792...30M} missions were retrieved from IRSA, covering observations until end of 2019. A saturation correction has been applied\footnote{See\,\,\url{http://wise2.ipac.caltech.edu/docs/release/neowise/expsup/sec2\_1civa.html}} to account for a photometric bias due to warm-up of the detector which amounts to $+0.02$ (W1) and $+0.33$ (W2) magnitudes, respectively.

The (NEO)WISE W1 (3.4\,$\mu$m) and W2 (4.6\,$\mu$m) light curves are shown in Fig.\,\ref{fig:NW_lc}. 
Because of its brightness in both filters bands, G358 led to small (W1) or mild (W2) detector saturation. In this case, the derivation of the brightness from the wings of the point spread function (c.f. section IV.4.c.iii of the WISE All-Sky Explanatory Supplement, \citealp{2012wise.rept....1C}) leads to enhanced scatter. 
Nevertheless, the discovery of an 
obvious brightening which usually accompanies enhanced accretion should have been possible. However, there is no clear-cut evidence in both light curves for a flux increase at the burst epoch or later on.
From the scatter of the photometric values, a possible increase due to the burst can be constrained. Assuming a 2$\sigma$ detection limit for W1 and W2 of $\sim$\,0.5\,mag, that is a joint 2.8$\sigma$ limit, upper bounds for a possible flux contribution caused by the burst can be derived as 0.10\,Jy and 0.45\,Jy, respectively. 
The failure of the burst detection suggests that, at any time, MM3 provided by far most, if not all, emission seen in the (NEO)WISE bands. Further details on the (NEO)WISE astrometry are outlined in Sect.\,\ref{nw}.

\subsection{Infrared photo-astrometry}\label{iras}
In case of imaging an unresolved crowded region a shift of the emission centroid may be caused by differing object SEDs which would introduce a wavelength dependence and/or by variability leading to a temporal displacement. By placing upper limits on a possible centroid shift, constraints on the contribution of a single source with known position, MM1 in the present case, to the overall emission can be established. In the following this will be done using 
the present infrared data.

\subsubsection{VVV and GROND}\label{vg}
An upper limit to the MM1 pre-burst 2.18\,$\mu$m flux can be derived from the stacked VVV $K_{\rm s}$ image. It is based on 131 exposures of 4\,s each, that is a total integration time of 8.7\,min. Confusion noise due to the high surface density of objects in the Galactic center neighborhood limits the detection of a possible faint NIR counterpart of MM1. It has to be brighter than $K_{\rm s}\,{\sim}\,$15.5\,mag to be recognized next to 2MASS~J17431001$-$2951460. This corresponds to a flux density of 0.42\,mJy \citep{2003AJ....126.1090C}. Similarly, the burst $K_{\rm s}$ magnitude can be constrained by the corresponding GROND image. Given the aperture sizes of the 2.2-m and VISTA telescopes as well as the total integration times, the GROND image should almost reach the sensitivity of the stacked VVV frame. However, inferior seeing and slightly elliptical images reduce its depth by about half a magnitude, resulting in a burst 2.15\,$\mu$m upper limit of 0.66\,mJy.

\subsubsection{(NEO)WISE}\label{nw}

\begin{figure}
    \centering
	\includegraphics[width=\columnwidth]{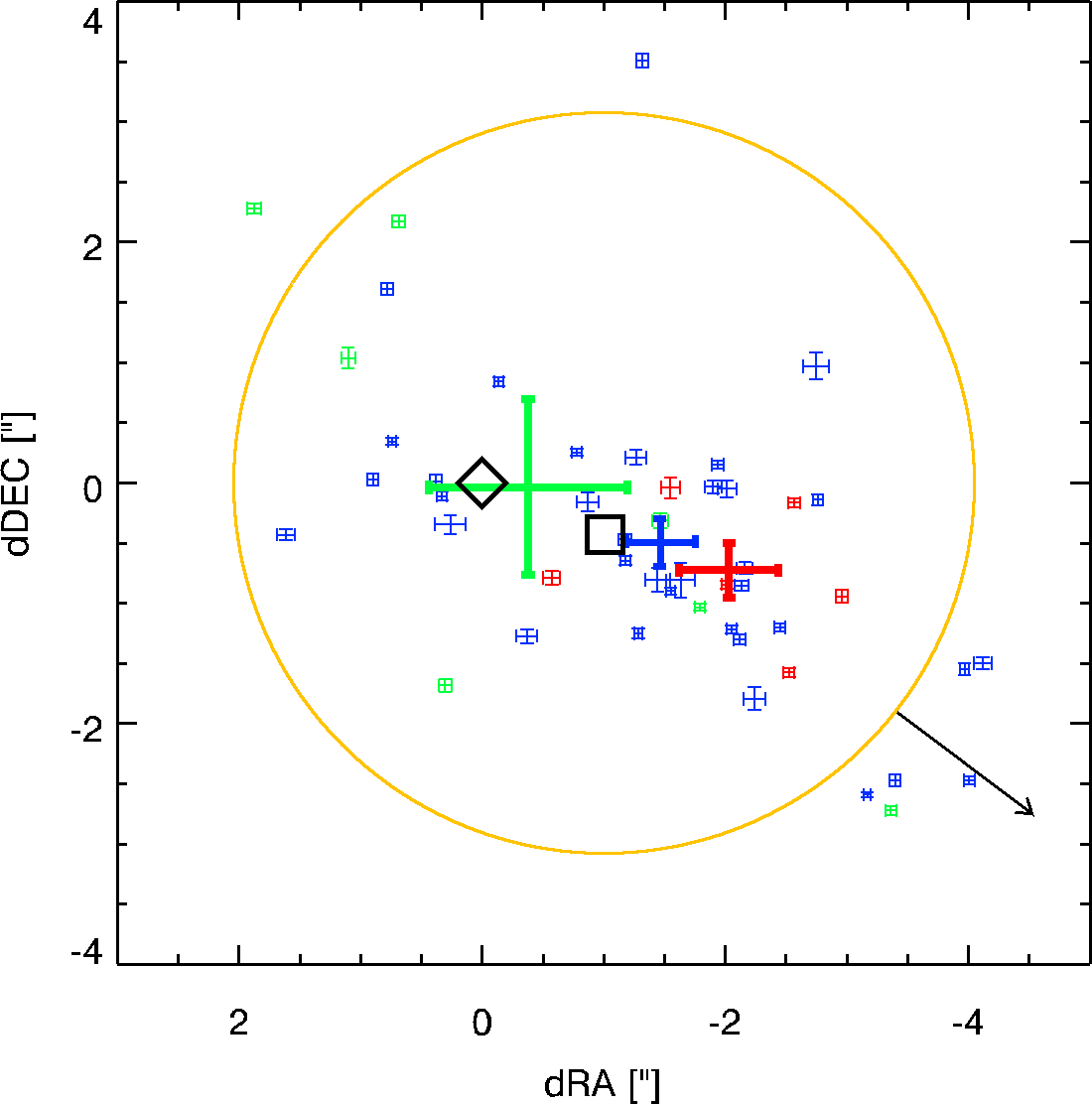}
	\caption{(NEO)WISE coordinate offsets relative to an origin at MM1 (from \cite{2019ApJ...881L..39B}). Pre-burst, burst, and post-burst data are shown in blue, red and green. Thick error bars represent mean positions while individual appear thin. Positions of MM1 and MM3 (measured with ALMA) are marked by the diamond and the square, respectively. The arrow points toward the bright 2MASS~J17431001$-$2951460, which is at a distance of 9\farcs9 from the (0,0) position. The yellow circle indicates the W1 image FWHM of 6\farcs1.
	}
 \label{fig:NW_pos}
\end{figure}

For the following analysis, (NEO)WISE positions were retrieved within 5\arcsec{} of the MM1 location  from IRSA, based on frames with the photometric quality flag A in both bands, a signal-to-noise ratio ${\ge}$\,20 and a frame quality rating of 10. The measurement quality and source reliability information flags of all frames, however, indicate contamination by the nearby 2MASS~J17430939$-$2951517 which is situated 9\farcs9 southwest of the MM3 and brighter in both (NEO)WISE bands.
This results from the image full width at half maximum (FWHM) of 6\farcs1 in the W1 band\footnote{See\,\,\url{https://wise2.ipac.caltech.edu/docs/release/allwise/expsup/sec1_2.html}}.

The offsets of the individual positions of the IR source with regard to MM1 are shown in Fig.\,\ref{fig:NW_pos} where the data was divided into three groups. The pre-burst one includes all positions until the last visit before the burst (2018 August, blue), the burst group comprises those of the 2019 March visit (red) and the post-burst group represents the 2019 August visit (green). The quantitative analysis confirms the visual impression for the mean positions that both, burst and post-burst, are consistent with the pre-burst location, situated close to MM3. So most of the emission seen by (NEO)WISE arises from the MM3 NIR counterpart 2MASS~J17431001$-$2951460, with some contamination from the nearby IR source to the southwest which causes the elongated distribution of positions.
While the mean (NEO)WISE post-burst position (green) is in between MM1 and MM3, its large position error precludes drawing any conclusions from this fact on whether or not this is a late sign of the burst. These findings are consistent with the non-detection of the burst in the (NEO)WISE light curves (Sec.\,\ref{rneo}).

\subsubsection{{\em Spitzer} MIPS}\label{mips}
As indicated by Fig.\,\ref{fig:GROND} (top) the positions of the IR counterparts at wavelengths up to 24\,$\mu$m are almost coincident with ALMA source MM3. Supposing about equal flux densities of MM1 and MM3 at this wavelength, the centroid of the MIPSGAL image of G358 should be located halfway between both sources. This is clearly not the case. Thus we conclude that the pre-burst MIPS 24\,$\mu$m flux of MM1 was considerably smaller than the joint flux of MM1 and MM3 (4.58$\,{\pm}\,$0.02\,Jy, \citealp{2009ApJS..181..227H}). An upper flux limit for MM1 can be derived by assuming that it leads to a detectable centroid shift of three times the positional error of 0\farcs02 \citep{2009ApJS..181..227H}. 
Since such a shift has not been detected, the contribution of MM1 to the total flux must be less than 0.42\,Jy. The upper limit is valuable to constrain the pre-burst luminosity.

\subsection{SOFIA FIFI-LS spectro-photometry}\label{rfifi}
The FIFI-LS data was processed by the ``FIFI\_LS\_REDUX'' pipeline  (version 1.7.0, \citealp{2015ASPC..495..355C}) and downloaded from the SOFIA Data Cycle System. The observations yielded a clear detection of continuum emission from the target in all bands. G358 is the only object in the FOV. In several bands CO line emission has been detected as well which will be discussed elsewhere. 

The continuum fluxes were derived from an image consisting of the pixel-wise median of a spectral data cube along the wavelength dimension, thus free from emission-line flux. For both epochs, central wavelengths and derived flux densities together with an error measure are given in Table \,\ref{fifiphot}. We note that our observations revealed the need for an empirical correction of the 118.9\,$\mu$m flux. It was derived from SED fits of other MYSOs of our FIFI-LS data set which show a similar flux deficit as well in this band. This was also
confirmed by looking at data from the flux calibrators Mars and Uranus.

The measurement errors derived from the error images provided by the FIFI\_LS\_REDUX pipeline do not include the calibration uncertainty. Therefore, we adopted a conservative approach by using an uncertainty of 10\%, cf. \citet{2019AAS...23420805F}. Since a narrow dust emission feature at 69\,$\mu$m \citep{2013A&A...553A...5S} - not covered by our FIFI-LS bands - is the only one in this wavelength region, a low-order polynomial fit seems to be representative for the actual SED at the time of the observing epoch. The residuals from these fits listed in Table \,\ref{fifiphot}  correspond to a mean relative error which indeed equals the above uncertainty. 

With the SOFIA flux densities at hand, the change of the SED due to the burst and its temporal evolution can be evaluated from
Fig.\,\ref{fig:SEDS} which shows, among others, the pre-burst SED (blue symbols), based on MIPS, AKARI, and
{\em Herschel}
data as well as the emission-line corrected ATLASGAL 870\,$\mu$m flux from \citet{2019ApJ...881L..39B}. The SED based on our first-epoch FIFI-LS observations (epoch 2019 May 1) and the ALMA 870\,$\mu$m integrated G358 flux from \citet{2019ApJ...881L..39B} (epoch 2019 April 12) is shown in red. The burst SED features flux densities larger than a factor of {$\gtrsim$}\,2 when compared to the pre-burst ones and a possible change of the SED shape. Thus, a luminosity increase - the prime signature of an accretion burst - has been witnessed with SOFIA. It represents the second confirmation of such an event
after the S255IR-NIRS3 burst \citep{2017NatPh..13..276C}
using this unique facility. Given the non-detection in the NIR and MIR as well as the non-significant flux increase in the (sub)mm \citep{2019ApJ...881L..39B} this promotes the G358 event to be the first NIR-, MIR- and (sub)mm-dark but FIR-loud MYSO accretion burst.

While the second epoch SOFIA observations suffer from the lack of the blue FIFI-LS bands, which sampled the flux density at wavelengths short-ward of the SED peak, the red channel data (green symbols in Fig.\,\ref{fig:SEDS}) nevertheless unambiguously indicate elevated flux levels during the post-burst stage compared to pre-burst and depressed compared to the bursting stages.


\begin{table}
\begin{threeparttable}
\caption[]{FIFI-LS photometry: Left first epoch, right second epoch}
         \label{fifiphot}
         \begin{tabular}{ccccc}
            \hline
            \noalign{\smallskip}
            Central      &  Flux & |Residual|&  Flux & |Residual| \\
            wavelength & density & & density & \\
            $[\mu m]$\ & [Jy] & [Jy]& [Jy] & [Jy] \\
            \noalign{\smallskip}
            \hline
            \noalign{\smallskip}
             ~~52.0 &  ~~95.6 &  ~~7.7 & &\\
             ~~54.8 & 126.8 & 15.0 & &\\
             ~~60.7 & 125.6 &  ~~5.7 & &\\
             ~~87.2 & 217.6 &  ~~4.4 & &\\
            118.6 & 255.1\tnote{*} & 37.0 & 216.5\tnote{*} & 42.6 \\
            124.2 & 371.5 & 72.8 & 304.5 & 52.4 \\
            142.2 & 315.4 &  ~~8.1 & 224.3 &  ~~2.2 \\
            153.3 & 270.4 & 34.5 & 221.4 & 12.8 \\
            162.8 & 297.1 &  ~~1.9 & 162.3 & 29.7 \\
            186.0 & 284.0 &  ~~9.4 & 155.9 &  ~~9.3 \\
            \noalign{\smallskip}
            \hline
         \end{tabular}
         \begin{tablenotes}\footnotesize
\item[*] recalibrated value
\end{tablenotes}
 \end{threeparttable}        
\end{table}

\subsection{FIFI-LS image analysis}

The angular distance of 1\farcs09 between MM3 and MM1 at a position angle (PA) of 247\degr{} warrants to investigate whether both sources could be marginally resolved at the shortest FIFI-LS wavelengths, provided they have similar flux contributions. The
diffraction limit for the 2.5-m mirror of SOFIA at 52.2\,$\mu$m amounts to 5\farcs3. However, due to various reasons \citep{2017SPIE10401E..12G}, the FWHM of the actual point spread function (PSF) exceeds that value. 

In order to address this point, we cannot rely on the absolute pointing accuracy of SOFIA but have to analyze the image morphology.
So the median continuum images of the blue channel were fit by a bivariate (2D) Gaussian. If MM3 
contributes a substantial flux contribution to the image, the major axis of the fit should clearly exceed the smaller one and be aligned to that direction. 
The corresponding quantities,  
axis ratio, position angle and respective errors, are given in Table \,\ref{fifibigauss}.
\begin{table}
      \caption[]{FIFI-LS bivariate Gaussfit results}
         \label{fifibigauss}
         \begin{tabular}{ccc}
            \hline
            \noalign{\smallskip}
            Effective wavelength      &  Axis ratio & Major axis PA \\
            $[\mu m]$ &  & [\degr] \\
            \noalign{\smallskip}
            \hline
            \noalign{\smallskip}
             52.0 & $1.12\pm0.03$ &  $310.1\pm6.1$\\
             54.8 & $1.22\pm0.02$ &  $347.4\pm1.9$\\
             60.7 & $1.11\pm0.02$ &  $336.4\pm3.6$\\
             87.2 & $1.10\pm0.01$ &  $330.6\pm1.1$\\
            \noalign{\smallskip}
            \hline
         \end{tabular}
\end{table}
When judging these results, it has to be kept in mind that the FIR observations were performed in chop-nod mode. Thus, deviations from a perfect image superposition may lead to an elongated image as well. 
The position angle (PA) of the major axis for the 52.2\,$\mu$m band is closer to that of MM3 compared to the other ones. However, the PAs of all 
bands 
are more aligned with the chopping angle of 293\degr{}.
Given this fact, it remains open whether the result for the shortest wavelength of the blue channel might be interpreted in favor of a noticeable contribution from MM3.

\subsection{PACS image analysis}
We checked the {\em Herschel}-PACS images at 70\,$\mu$m for potential signs of source multiplicity or source elongation. Unfortunately, the only existing data toward this region \citep[from Hi-GAL, see][]{2010PASP..122..314M} was obtained in the so-called fast-scan parallel-mode which comes with a quite peculiar PSF that is elongated along the two almost orthogonal scan directions. Its equivalent FWHM is typically 8\farcs8\,${\times}$\,9\farcs6 and is thus larger than the SOFIA PSF at the same wavelength. This impedes a study of small elongations due to source multiplicity. We nevertheless attempted a PSF photometry using the Starfinder IDL tool \citep{2000SPIE.4007..879D}, Version 1.8.2a, by using the adequate PACS PSFs (version 2.2) for the fast-scan parallel mode, provided by the instrument team\footnote{\url{https://www.cosmos.esa.int/documents/12133/996891/PACS+photometer+point+spread+function}}. The algorithm can in principle find multiple sources blended within one FWHM. While we varied several parameters for background subtraction and treatment of sub-pixel offsets, the program consistently identified just one strong source, showing a very good correlation coefficient of 0.977 with regard to the input PSF. This source is formally 1\farcs1 away from the ALMA position of component MM1, and 0\farcs7 away from ALMA component MM3. The absolute astrometry of the {\em Herschel} Spacecraft at the end of the mission came with a 3$\sigma$ uncertainty of 2\farcs7 (cf. \citealp{2014ExA....37..453S}). Therefore, based on the {\em Herschel} astrometry alone, it is not possible to decide whether the revealed {\em Herschel} compact source is coincident with the location of MM1 or MM3, or somewhere in between. Based on the SED modeling, we think that the {\em Herschel} 70\,$\mu$m emission is dominated by MM1. Given our PSF fitting experiments we conclude that a potential minor contribution at 70\,$\mu$m from MM3 must be more than 70 times weaker than the one from MM1.

\begin{figure}   
	\includegraphics[width=\columnwidth]{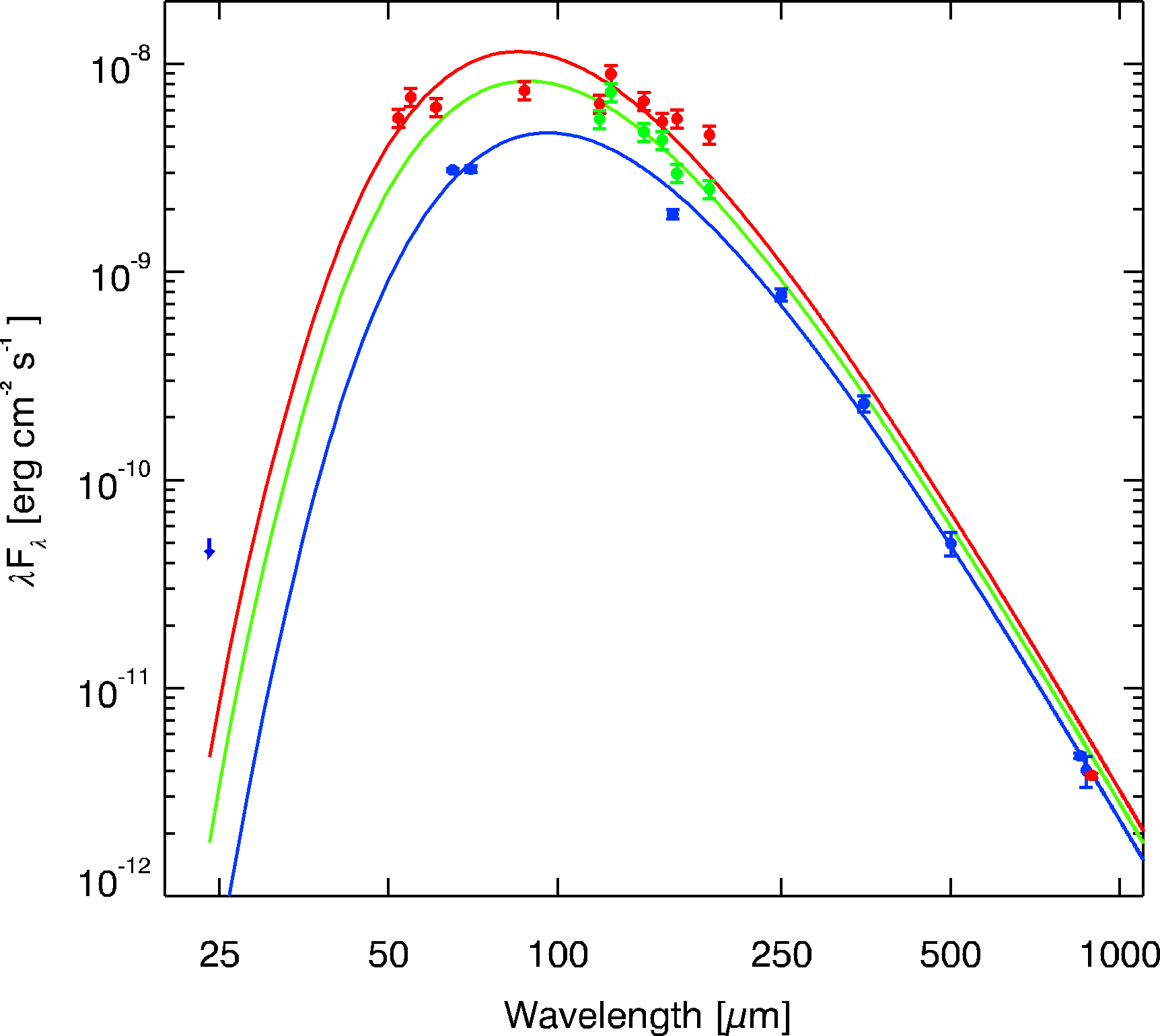}
	\caption{Observed FIR/(sub)mm SEDs showing pre-burst {\em Herschel}/AKARI data (blue) and FIFI-LS burst values (red) together with the corresponding (sub)mm fluxes from \citet{2019ApJ...881L..39B}.
	The upper 24\,$\mu$m MIPS limit
	is indicated as well. The solid lines represent the reddened graybody fits (cf. \ref{gb}). The second epoch FIFI-LS data and the corresponding fit appear {bf in} green.
	}
 \label{fig:SEDS}
\end{figure}

\section{Graybody fits and burst parameters}\label{gbp}


\subsection{Graybody fits to SEDs based on FIR and (sub)mm data}\label{gb}
Since the observed SEDs of G358 based on FIR and (sub)mm data do not strongly differ from a Planck function, the most simple approach to approximate them is using a modified blackbody, in other words a graybody (for instance \citealp{2016MNRAS.461.1328E}). 
It comprises three parameters: dust temperature, emissivity index, and solid angle of the emission.
Before performing the fits the fluxes were dereddened to account for interstellar extinction. A value of $A_{ V}\,{=}\,60$\,mag was assumed which seems to be justified (see Sec.\,\ref{drt1}). For 
flux dereddening, 
the $R_{ V}\,{=}\,5.5$ dust model of \citet{2003ARA&A..41..241D} has been used.
The fits, reddened to match the observations, are shown as solid lines in Fig.\,\ref{fig:SEDS}.

The corresponding fit to the pre-burst SED yielded  a dust temperature of $T_{\rm d}^{\rm pre}{=}\,26.4\,{\pm}\,0.5$\,K which
is slightly lower 
than the pre-burst value of $28.5\,{\pm}\,1.5$\,K by \citet{2019ApJ...881L..39B}. The dust emissivity index amounts to $\beta\,{=}\,1.83\,{\pm}\,0.06$
which is consistent with the emissivity index
close to the center of spiral galaxies in the Local Group \citep{2014MNRAS.444..797M, 2014A&A...561A..95T}
and appropriate for temperature reasons \citep{2005ApJ...633..272B}.

Since it seems plausible that the dust properties remain unchanged during the burst, except for the innermost region where modifications may have occurred (for example \citealp{2009Natur.459..224A}), 
the pre-burst emissivity index was used for the burst- and post-burst fits.

As expected, the dust temperature for the burst SED is marginally higher $T_{\rm d}^{\rm burst}{=}\,30.1\,{\pm}\,0.5$\,K as well as the 
solid angle of the emission which increased by a factor of 1.20$\,{\pm}\,0.14$.
However, the $\goodchi^2$ of the burst fit is 2.5 times as large as that of the pre-burst one. This is caused by the
lack of a pronounced peak in the burst SED.
Since our assessment of the FIFI-LS data did not reveal any issue other than that for 118.9\,$\mu$m flux, which has been corrected by our recalibration, the flat FIR burst spectrum is probably real. Future time-dependent RT modeling may show whether such a feature can be produced by a burst of short duration.


An attempt was made to come up with a graybody fit for the second FIFI-LS epoch data as well. Because of the sparse data, the pre-burst emissivity and the mean solid angle from the pre- and burst fits were adopted and only the temperature was varied. The resulting temperature amounts to $T_{\rm d}^{\rm post}{=}\,28.9\,{\pm}\,0.6$\,K.

\begin{figure*}
    \sidecaption
	\includegraphics[width=12cm]{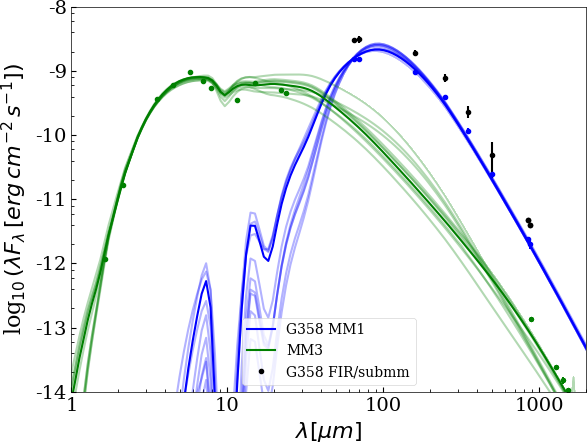}
	\caption{Pre-burst SEDs for the following
	sources: Black symbols -- total FIR/(sub)mm emission of G358, green -- 
	MM3, blue -- 
	MM1, derived from the total FIR/(sub)mm emission by removing contributions from all other sources (including MM3).
	For MM1 and MM3, the observed values are shown together with the ten best RT fits. 
	At wavelengths beyond $40\,\mu$m MM1 dominates the total flux density. The seemingly gap at $10\,\mu$m in the MM1 SED is due to strong silicate absorption.} 
 \label{fig:sed g358 pre}
\end{figure*}

\subsection{Empirical burst parameters}\label{ebp}
The FIR/(sub)mm luminosity increase due to the MM1 accretion burst can be determined by integrating the above graybody SED fits and taking source distance given in Sec.\,\ref{intro} as well as interstellar extinction into account. Here we assume that all other sources which contribute to the total G358 flux stayed constant during the pre\nobreakdash-, burst- and post-burst epochs.
The weak variability of MM3 in the NIR and MIR is of no concern here since its FIR emission is far below that of MM1 (see Sec.\,\ref{fig:sed g358 pre}).
Since the bulk of the energy is emitted in the FIR, the FIR/(sub)mm luminosity estimates only weakly depend on $ A_{\rm V}$. 
The impact of the distance ambiguity is considered in the following analysis.

Integration of the graybody fits yields the following FIR/(sub)mm luminosities: 
Pre-burst $L^{\rm pre}_{\rm FIR}\,{=}\,7600\,{\mypm}\,^{~800}_{1400}\,{\rm L}_\odot$, first FIFI-LS epoch $L^{\rm burst}_{\rm FIR}\,{=}\,19\,300\,{\mypm}\,^{2200}_{3700}\,{\rm L}_\odot$, and second epoch $L^{\rm post}_{\rm FIR}\,{=}\,12\,700\,{\mypm}\,^{1500}_{2600}\,{\rm L}_\odot$, respectively. 
As emphasized, without interstellar extinction, these values shrink but only by 9\%.
The luminosity increase due to the accretion burst at the dates of the FIFI-LS observations amounts to $\Delta L^{{\rm burst}}_{\rm FIR}\,{=}\,11\,700\,\mypm^{\,2300}_{\,3900}\,{\rm L}_\odot$ and $\Delta L^{{\rm post}}_{\rm FIR}\,{=}\,5100\,{\mypm}\,^{\,1700}_{\,2900}\,{\rm L}_\odot$. The presence of a substantial luminosity increase during the second FIFI-LS epoch, meaning about 18 months after the peak of the maser flare, is remarkable.

For deriving major parameters of the burst, we follow the approach of \citet{2017NatPh..13..276C}. In addition, for what concerns the estimate of the burst energy, the two epochs of FIFI-LS observations provide the opportunity to account for the temporal evolution of the luminosity. The simplest, yet plausible, approach is to assume a linear decrease which holds from the flare peak date to the date when the pre-burst level will be reached again. The linear flux decay approximation yields a duration $\Delta t$ of $869\,{\pm}\,303$ days, with a pre-burst-level return date of 2021 July 31.

The FIR/(sub)mm burst energy is $E^{\rm acc}_{\rm FIR}\,{=}\,{<}\Delta L^{\rm acc}_{\rm FIR}{>}\,{\times}\,\Delta t$, where ${<}\Delta L^{\rm acc}_{\rm FIR}{>}$ is the average luminosity increase which equals to half of the peak increase for a linear drop to zero.  It amounts to $E^{\rm acc}_{\rm FIR}\,{=}\,1.9\,{\times}\,10^{38}$\,J. Its upper and lower bounds from the uncertainties in both duration and luminosity are $1.0\,{\times}\,10^{38}$\,J and $2.6\,{\times}\,10^{38}$\,J, respectively. 


We emphasize that these values represent lower limits to the  luminosity increase and the corresponding energy release since they are based on the FIR/(sub)mm emission only. The ultimate quantities will be derived from the results of the RT analysis below.
Accretion related quantities require the knowledge of protostellar mass and radius and will be considered in Sec.\,\ref{sbp}.

\section{Analysis of SEDs}\label{ased}


\subsection{SED decomposition}\label{sedc}

In order to derive representative parameters of the bursting MYSO MM1 from RT modeling, the underlying SED should be as free as possible from contributions from other objects. Thanks to the availability of NIR, MIR, and (sub)mm photometry for MM3, the contamination from this source to the overall flux can be removed by using predicted flux densities from its best RT model (see Sec.\,\ref{drt3}), assuming that its SED has not changed. 
This approach has been proven successful in a similar, yet less sophisticated fashion, for a study of FIR emission from W3(OH) and the neighboring W3(H$_2$O) (\citealp{2002A&A...392.1025S}). 

Here we extend it by also taking into account the presumed contributions from the remaining sources detected 
at (sub)mm wavelengths.
Since none of these has an IR counterpart we assume that they are in an early evolutionary stage similar to MM1, with SEDs that resemble the overall pre-burst 
SED. So for MM1 and each of those (MM2, MM4-8), the 
(sub)mm fluxes from \citet{2019ApJ...881L..39B} were fit by a graybody, using the temperature and emissivity index derived from the pre-burst SED, to obtain the individual solid angles of the emission. 

For the pre-burst epoch, the following steps were performed to obtain the MM1 fluxes. First, the predicted MM3 fluxes were subtracted from the G358 total fluxes. Second, the ratio between the MM1 solid angle and the sum of all solid angles was calculated which amounts to $0.50\,{\pm}\,0.05$. It corresponds to the relative contribution of MM1 to the MM3-subtracted fluxes. So, by multiplying the latter with this ratio the pre-burst fluxes of MM1 were obtained.

Similarly, for the wavelengths of the burst as well as post-burst observations, the pre-burst fluxes were predicted from the MM3 model as well as the pre-burst graybody fit. Then, the MM3 contribution was removed. Finally, the contribution of MM2+MM4-8 to the MM3-subtracted pre-burst fluxes had to be taken out from the observed burst as well as post-burst fluxes to yield those for MM1.
These are listed in Table \ref{tab: preSED}, \ref{tab: burstSED} and \ref{tab: postSED} (for MM1 pre-burst, burst, and post-burst respectively).
They
represent the best possible approximation of the intrinsic ones of MM1.
We emphasize that the procedure to derive the MM1 fluxes is tailored in order to reproduce the total flux. Therefore, the MM1 (sub)mm fluxes exceed those given in Tab.\,2 of \cite{2019ApJ...881L..39B} by a factor of about two.

\subsection{Radiative transfer analysis}\label{rta}
Characteristic properties of YSOs, namely their luminosities as well as mass, geometry and extent of the surrounding dust, can be derived by modeling their dust continuum radiation to match the observed SEDs (for instance \citealp{2012ascl.soft04005W, 2013ApJS..207...30W}). For G358 this has been done by \citet{2019ApJ...881L..39B} to infer the  pre-burst luminosity
$L^{\rm pre}$,
using the YSO model grid of \citet{2017A&A...600A..11R}. 

Utilizing the same model pool, we performed an RT analysis of the SEDs of MM1 and MM3 using the Python implementation ``sedfitter''\footnote{\url{https://zenodo.org/record/235786}} of the SED fitter \citep{2007ApJS..169..328R} which is described below. We did not fit the G358 total flux (black symbols in Fig. \ref{fig:sed g358 pre}), since 
models with multiple sources, which would be appropriate for massive star forming regions and clearly for G358, are not included in the \citet{2017A&A...600A..11R} model grid.
Instead, we extracted the fluxes of MM1 from the total ones as described above.
From the ten best fits (per epoch/source), mean values and uncertainties for all ``free'' parameters of the models 
are derived (see Appendix \ref{app}).
By using a weighting with the respective ${\goodchi}^2$ values, we ensure that the best fits determine the corresponding mean values. For the parameters which are log-spaced in the Robitaille model pool ($r_{\rm min},\, r_{\rm max}$,\, $M_{\rm disk}^{\rm dust},\,  L$\, via $R_\star,\, T_\star$), we use the geometric mean, while for all other parameters, we use the arithmetic mean.
Since each model is comprised in the pool with nine different inclinations from 0\degr{}  to 90\degr{} to account for the inclination dependence of the SED, the ten best fits are not necessarily composed by ten different models. Instead, some models might be included more than once, but with different inclinations.

We note that these models incorporate passive disks. While this may seem inappropriate in the context of accretion bursts where active disks are often invoked, it is justified by the fact that we are primarily interested in reproducing the FIR and (sub)mm emission. Due to the strong radial dependence of their viscosity,
for instance  \citet{1981ARA&A..19..137P}, active disks differ from passive ones primarily in the very innermost region where the bulk of the dissipated energy is being released. The details of this process are not relevant for the highly reprocessed emission in the FIR and (sub)mm range, 
which is dominated by dust radiation from the outer regions.

Before describing the actual modeling, we emphasize that
additional data and results are provided in the Appendix \ref{app}. These comprise the flux tables which were used to establish the SEDs, a summary of the RT models which have been used in the analysis, tables listing the parameters of the ten best RT models for each of considered cases, and a table providing optical depths for the ten best pre-burst RT models of MM1.


\subsubsection{Modeling of the dust continuum radiation of MM3}\label{drt3}
 
For the fitting of the MM3 SED, the combined ``spubhmi'' + ``spubsmi'' data sets from \citet{2017A&A...600A..11R} have been used. They comprise 120\,000 YSO models with nine inclinations for each model 
(leading to a set of 1\,080\,000 SEDs in total).
All models employ Milky Way dust with $ R_{\rm V}\,{=}\,5.5$ and a grain size distribution from \citet{2001ApJ...548..296W}.

Their designations are based on the respective model components.
Each model (in both data sets) comprises the following components: {\bf s}tar, 
{\bf p}assive circumstellar disk,
{\bf b}ipolar cavity, {\bf U}lrich-type envelope \citep{1976ApJ...210..377U}, and ambient {\bf m}edium. 
The ``spubhmi''-setting features an inner {\bf h}ole in between star and disk, whereas for ``spubsmi'' the inner radius is governed by the dust {\bf s}ublimation radius (assuming $T_{\rm sub}\,{=}\,1600\,{\rm K}$, in accordance with \citealp{2006ApJS..167..256R}). We note that the Robitaille model pool includes other data sets, that are composed of less (or different) components and are thus less suited to represent the structure of a YSO at a very early evolutionary stage. For further details, we refer the reader to \citet{2017A&A...600A..11R}. 
A synthetic aperture size of $3\arcsec{}$ has been used for fitting the SED. The fluxes of the MM3 SED are listed in Tab.\,\ref{tab: MM3 SED}. 

We adopt the same distance range as for MM1 but allow for a smaller interstellar extinction.
Figure \ref{fig:sed g358 pre} shows the SED (green) together with the total 
pre-burst SED (black symbols) and the pre-burst flux attributed to MM1 (blue), where the contribution of the other sources has been removed as described in Sec. \ref{sedc}. The ten best fits for each SED are shown with solid lines. The best fit is dark, whereas the other are slightly transparent.
Our models suggest that the contribution of MM3 to the total flux at wavelengths beyond $\lambda\,{\ge}\,40\,\mu$m is marginal. Nevertheless we take the best MM3 fit into consideration when refining the MM1 fluxes in the following analysis. 

As indicated by its moderate obscuration in the NIR
already: MM3 is likely the most evolved object of the G358 complex. 
It features a 
disk with a dust mass of $M_{\rm disk}^{\rm dust}\,{=}\,0.068\,{\mypm}\,_{0.043}^{0.11} \,{\rm M}_\odot$. This may seem heavy keeping the canonical gas-to-dust mass ratio ($\gamma$) of 100 in mind. However, the latter is not applicable since $\gamma$ depends on the galactocentric distance  $R_{\rm GC}$ \citep[Eq. 2,]{2017A&A...606L..12G} because of the Galactic metallicity gradient. For G358, $R_{\rm GC}{=}$1.6\,kpc implies a value of $\gamma\,{=}\,38\,{\pm}\,4$ which yields a total disk mass of $2.6\,{\mypm}\,_{1.0}^{6.8}\,{\rm M}_\odot$.
The fit delivers $A_V\,{=}\,20\pm 5\,$mag and an inclination of $i\,{=}\,51\,{\pm}\,9$\degr. The inner radius of the disk seems to be close to the star, 
$r_{\rm inner}\,{<}\,3.5\,R_{\rm sub}$ holds for each of the ten best models
. The mean value of the disk outer radius amounts to $510\,{\mypm}\,_{240}^{450}\,$au, while the maximum is as high as  $1600\,$au.
The mean luminosity of $7540\,{\mypm}\,_{3130}^{5340}\,{\rm L}_\odot$ corresponds to a ZAMS star of $11\,{\rm M}_\odot$ and $4.2\,{\rm R}_\odot$ \citep{1996MNRAS.281..257T}. While most of the models have luminosities below $10\,000\,{\rm L}_\odot$, one has a luminosity of $43\,000\,{\rm L}_\odot$. Such a high effective temperature would imply the presence of a compact H{\sc ii} region. However, MM3 escaped the detection in the radio continuum at a sensitivity level of ${\sim}\,50\,\mu$Jy\,beam$^{-1}$
in the survey of
\citet{2016ApJ...833...18H} which probably rules out this model. Recently, Bayandina et al. (in prep.) succeeded to detect faint radio continuum emission from MM3.
The whole parameter set for the ten best models (including their respective $\goodchi^2$ values) can be found (together with the weighted means and standard deviation $\sigma$) in Table \ref{tab MM3 fit}.

As described in Sec.\,\ref{sedc}  the MM3 SED fit is used for the SED decomposition. In order to check the robustness of the FIR part of the best models, two more fits were performed with reduced data sets. At first, the ALMA fluxes were omitted which yielded slightly lower FIR fluxes compared to the nominal fit described above. Then, also the WISE W4 and MIPS 24\,$\mu$m fluxes were dropped. The best model for this SED shows fluxes exceeding those of the nominal fit (cf. Fig.\,\ref{fig:MM3 contri}). Thus, we conclude that the FIR part of the nominal MM3 SED fit is well constrained by the incorporated fluxes longward of 20\,$\mu$m.

\subsubsection{Modeling of the dust continuum radiation of MM1}\label{drt1}

\begin{figure*}   
    \sidecaption
	\includegraphics[width=12cm]{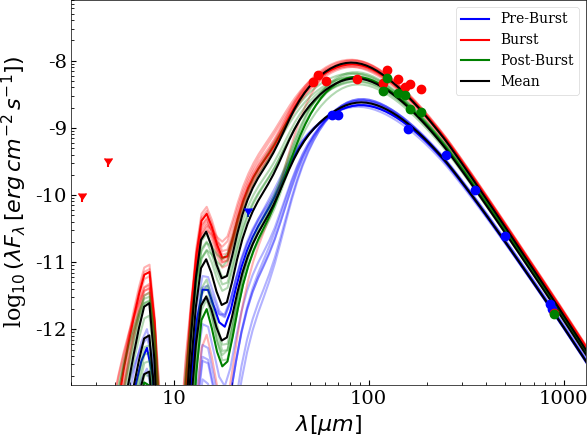}
	\caption{Modeled pre-burst MM1 SEDs (blue), together with its burst (red) and post-burst SED (green). Triangles mark upper limits. The weighted mean-model (see text) is shown in black. It is shown three times but with different source luminosities for the respective observing epochs. 
	Since the $870$ (pre-) and $889\,\mu$m (post- and burst) fluxes are very similar, their symbols cannot easily be distinguished.
	}
 \label{fig:sed g358}
\end{figure*}

With the refined MM1 SEDs at hand, a comparison of the results from RT modeling for the pre-burst, post- and burst-states becomes possible.
Before describing this analysis, a remark must be made concerning the interstellar extinction $A_V$ which is a free parameter of the SED fitter. Because of the wavelength dependence of the dust optical properties, extinction is most pronounced at short wavelengths. Thus, for SEDs like that of MM3 it can be well constrained by the best models. However, for the SED of MM1 which is almost exclusively defined by FIR and (sub)mm measurements, interstellar extinction is less influential and, therefore, harder to derive. Since higher $A_V$ implies larger source luminosities to reproduce the observed fluxes, an upper limit needs to be established to constrain $L$. The recalibration of a Galactic dust-reddening model based on IRAS and COBE/DIRBE results by \citet{2011ApJ...737..103S} suggests a value of $A_V\,{=}\,115\,{\pm}\,4.2$\,mag along the sight line toward G358. Since this holds for the whole path across the Galaxy, while G358 is in front of the Galactic center region, the actual $A_V$ will in fact be lower. Therefore, we assumed an interstellar extinction range of $A_V\,{=}\,30\,{-}\,70$\,mag.

To begin with, we fit the pre-burst MM1 SED which represents the stationary state using the distance and interstellar extinction ranges given above. This established the ten best pre-burst 
fits. 
For this purpose,
the ``spubsmi'' data set from \citet{2017A&A...600A..11R} has been used which includes 
40\,000 models 
at nine inclinations (360\,000 SEDs in total).
For these models, the inner radius of the dust disk corresponds to the sublimation radius $R_{\rm sub}$. This constraint seems to be justified given the high accretion rates at which massive stars form (for example \citealp{2007ARA&A..45..481Z, 2016ApJ...832...40K}), and in particular before and during an accretion burst. It requires a few remarks concerning the understanding of ``heating'' and ``cooling'' in this case, since the dust disk cannot get any hotter than the sublimation temperature. What happens due to a luminosity increase is that enhanced dust sublimation at the inner rim $R_{\rm sub}$ pushes the latter outward. This is accompanied by an adjustment of the temperature profile $T(r)$ via absorption and re-emission such that, for a given radius (beyond the actual $R_{\rm sub}$), the temperature exceeds the previous value. Conversely, for a given temperature, the growth in radius implies a larger radiating area and, therefore, leads to a flux increase. 
The reverse process happens once the burst ceased. $R_{\rm sub}$ will shift back inward, allowing the dust replenishment by accretion and/or dust reformation. This will lead to a flux decrease and might be considered as ``cooling'' but, yet, the inner rim is still at the dust sublimation temperature.

The next step was to establish a new model pool from the ten best pre-burst-SEDs which serves to fit the post-/burst SEDs. This pool contains 100 SEDs in total, where we reuse the best 10 models, with a source luminosity increasing in nine linearly spaced steps from $2$ to $6\,L^{\rm pre}$, respectively. 
Inclination, interstellar extinction and the distance
were set to the values of the underlying model from the pre-burst-fit
since none of these parameters is expected to change during the burst.
Additionally, the original pre-burst-SEDs are included. 
Since the inner disk radius is governed by the dust sublimation temperature, assumed to be $1600\,$K, it shifts outward for those models with increasing luminosity. Otherwise, the system geometry is kept the same.
While this ignores possible changes of the disk due to the burst, for instance in the density structure, it is the simplest approach to model the dust continuum emission due to the accretion burst, and feasible to be treated using common RT codes which are generally static. 
While a proper burst modeling requires time dependent RT, possibly coupled to hydrodynamics for utmost consistency, our simplified treatment nevertheless allows us to draw major conclusions.

For computing the burst models, the Hyperion code 
\citep{Robitaille2011} has been used. 
The so obtained database is the foundation to fit the SEDs of burst and post-burst epochs. 
We note that by constraining the model pool, we ensure consistency of the results, which means the best fits for  pre-burst, burst, and post-burst SEDs are based on the same models. 
A sketch of all SEDs that are included in burst model database, can be found in the Appendix (Fig. \ref{fig:sed g358 modelpool}.) The range of the observed fluxes from the burst- and post-burst-epoch is well covered by the models. Only the burst observations at $\lambda\,{=}\,163$ and $186\,\mu$m are outside of the flux range covered by the models.
Only one model out of the ten best burst fits has the maximal luminosity increase included in the pool. Therefore, it is not necessary to include models with higher source luminosities.

Before presenting the results, a few remarks have to be made concerning the fitting. We exclude the sub-mm observations at $\lambda\,{>}\,890\,\mu$m from the SED-fit of the burst since their deviation from the stationary models is biggest at those wavelengths (see Sec. \ref{misf}). The post-burst is observed only at wavelengths greater than $118\,\mu$m, meaning that there are no constraints in the MIR. Since the maser flux did not fall below its pre-burst level until now (MacLeod et al., in prep.; Yonekura et al., in prep.), we assume that the post-burst flux in the MIR is also not below the pre-burst level. Therefore, models from the post-burst fits that have MIR fluxes below those of the mean pre-burst model were excluded.
An aperture size of $3\arcsec{}$ has been applied in for the fits. With this choice the $\goodchi^2$-value of the pre-burst fit becomes smallest. All obtained parameters are stable against a variation of the aperture size (they agree within their respective errors).



The results of the RT modeling of the MM1 SEDs are visualized in
Fig.\,\ref{fig:sed g358}, which shows the ten best fits to the pre-burst (blue), burst (red) and post-burst SED (green).
The best model is shown with the darkest line, while the nine other ones are slightly transparent.
In addition to the ten best fits, the so called ``mean''-model is shown in black. This model is computed using the weighted mean parameters from the combination of pre-burst, post-burst and burst fits (see below). The stellar parameters, which are obviously changing during the burst, are defined by the pre-burst models alone. The mean model is not only shown for the pre-burst, but also for source luminosities
corresponding to the
mean values during burst and post-burst. As expected, the mean model lies within the range, spanned by the ten best models at each epoch. The results are summarized in Table \ref{tab MM1 fit}.


The RT modeling indicates that during the burst the luminosity increases from pre-burst level of $L^{\rm pre}\,{=}\,5000\,{\mypm}\, _{\,900}^{1100} \,{\rm L}_\odot$ to $L^{\rm burst}\,{=}\,23\,400\,{\mypm}\,_{3700}^{4400}\,{\rm L}_\odot$.
At the post-burst epoch, the luminosity is still elevated at a level of $L^{\rm post}\,{=}\,12\,400\,{\mypm}\,_{1700}^{2000}\,{\rm L}_\odot$.

This corresponds to a relative luminosity gain by a factor of 
$4.7\,{\mypm}\, _{1.5}^{2.1}$ during the burst and by
$2.5\,{\mypm}\, _{0.8}^{1.1}$ for the post-burst epoch.
The increase in luminosity during the burst amounts to  $\Delta L^{\rm burst}\,{=}\,18\,000\,{\mypm}\,_{4900}^{6100}\,{\rm L}_\odot$ which implies a total burst energy of 
$E_{\rm acc}\,{=}\,2.9\mypm_{0.8}^{1.1}\,{\times}\,10^{38}$\,J, where the calculation has been done similar to Sec. \ref{ebp}. 
The derived linear decay time of $\Delta t\,{=}\,907\,{\mypm}\,_{236}^{318}$\,d predicts an expected return to the pre-burst luminosity in 2021 September, which agrees with the previous estimate based on the gray-body fits within the errors (see Sec.\,\ref{ebp}).

A comparison of the RT results to those obtained from graybody fits 
in Sec. \ref{gb} shows that the luminosity increase and energy release from the RT models are indeed higher, although the luminosities at burst and post-burst epochs agree within their respective errors. 
The main reasons
are the inclusion of more energetic emission from hotter regions in the models which lacks in the graybody fits and the correction of the flux contribution from the other G358 sources as outlined in Sec.\,\ref{sedc}.

The estimated parameters from the RT modeling are given below.
The interstellar extinction indicated by the fits is $A_V\,{=}\,60\,{\pm}\, 10$\,mag. The system has a low inclination of $22\,{\pm}\,11\degr$, which means it is seen close to pole on. This viewing geometry agrees with that of the spiral-arm accretion flow of \citet{2020NatAs.tmp..144C} with $i\,{=}\,25\,{\pm}\,10$\degr{}.
The stellar radius amounts to $8.4\,{\mypm}\,_{5.5}^{15.7}{\, \rm R}_\odot$. It has been obtained from the pre-burst-models alone. The burst might cause an increase in the stellar radius (bloating) although it is unclear, whether the protostar will respond on time scales of a few months.
The models show a considerable scatter in the derived disk properties. The derived outer radii are between $140$ and 3800\,au, with a mean value of $950\,{\mypm}\,_{580}^{1500}$\,au. 
The favored dust mass of the circumstellar disk is as low as $8.4\,{\times}\, 10^{-5}\,{\rm M}_\odot$, where the $1\sigma$-confidence interval extends from $1.1\,{\times}\, 10^{-6}$ to $6.1\,{\times}\, 10^{-3}\,{\rm M}_\odot$. 
Using the appropriate gas-to-dust mass ratio (see Sec.\,\ref{drt3}) this corresponds to a most probable total disk mass of 
$3.2\,{\times}\, 10^{-3}\,{\rm M}_\odot$ within an uncertainty range of $4.4\,{\times}\, 10^{-5}$ to $0.24\,{\rm M}_\odot$.
Remarkably, the corresponding total disk masses are lower, and mostly much smaller, than those derived for MYSOs from SED fitting (for example \citealp{2010Natur.466..339K, 2015ApJ...813L..19J}).  
Among the best models there is only one (90Yt0exl\_03) with a total disk mass of $1.4\,{\rm M}_\odot$ which is not that much below present estimates. In contrast to MM3, the much higher scatter of the disk mass of MM1 compared to the other log-spaced parameters indicates that it cannot be reliably estimated by the fitter. In other words, a substantial flux contribution from a disk is not necessarily required to reproduce the SEDs. Thus, while we cannot reliably estimate its mass, we may conclude that the disk is quite likely lightweight.

We note that during the burst the disk dust mass might decrease due to dust sublimation. The total mass will be unaffected, assuming that all sublimated dust just increases the gas mass. 
The whole parameter set for the ten best models for each of the three epochs (including their respective $\goodchi^2$ values) can be found (together with the weighted means and standard deviation $\sigma$) in Table \ref{tab MM1 fit}.

The above given properties have been derived from the results of all MM1 SED-fits, pre-burst, post-burst and burst. The respective $\goodchi^2$-values were used to weight the obtained parameters, similar to what has been done for MM3. We normalized the sum of the respective weighting factors to unity, 
split up to 0.5 for the pre-burst and 0.25 for the other two epochs.
With this we ensure an equal contribution of the fit to the ``stationary''- (pre-burst) and ``non-stationary''-system (burst and post-burst), where the ``non stationary''-contribution is obtained taking into account burst and post-burst epoch equally.
The $\goodchi^2$-values of the burst models are about 5 times higher than for the pre-burst. This might be related to the scatter in the FIFI-LS data, to a coarser sampling of the model-parameters (because of the fact that we only consider models that fit the pre-burst SED pretty well), and to differences in comparison to the static case (whereas the pre-burst-system most probably is stationary, this does not hold for the post-/bursting stage).
We note, that in the burst case the relative weights of the models are on the same order, whereas in the pre-burst case they differ by a factor of 1.5 at most (for the post-burst it is 1.3).

\section{Stellar-dependent burst parameters}\label{sbp}
The RT modeling of the MM1 SEDs aimed at determining the luminosities of the MYSO for the various epochs to infer the total energy released by the accretion burst.
For deriving the accreted mass and the mass accretion rate, protostellar mass and radius have to be known. For YSOs approaching the ZAMS, the corresponding values which match the  luminosity are a good approximation. At first, we use this approach to obtain mass and radius estimates
which reproduce the MM1 pre-burst luminosity, assuming solar metallicity. For
$L^{\rm pre}\,{=}\,5000\,{\mypm}\, _{\,900}^{1100} \,{\rm L}_\odot$,
these correspond to 
$9.7\,{\mypm}\,_{0.6}^{0.3}\,{\rm M}_\odot$ and $3.9\,{\mypm}\,_{0.2}^{0.1}\,{\rm R}_\odot$
\citep{1996MNRAS.281..257T}, respectively.
The accreted mass is inferred from $E_{\rm acc}\,{=}\,GM_*M_{\rm acc}/R_*$, where G is the gravitational constant, $M_*$ is 
the stellar mass, $M_{\rm acc}$ is the accreted mass, and $E_{\rm acc}$ the released energy derived 
in the previous section. Here it is implicitly assumed that the potential energy is released as well when the matter eventually reaches the protostar.

\begin{figure}
    \centering
	\resizebox{\hsize}{!}{\includegraphics{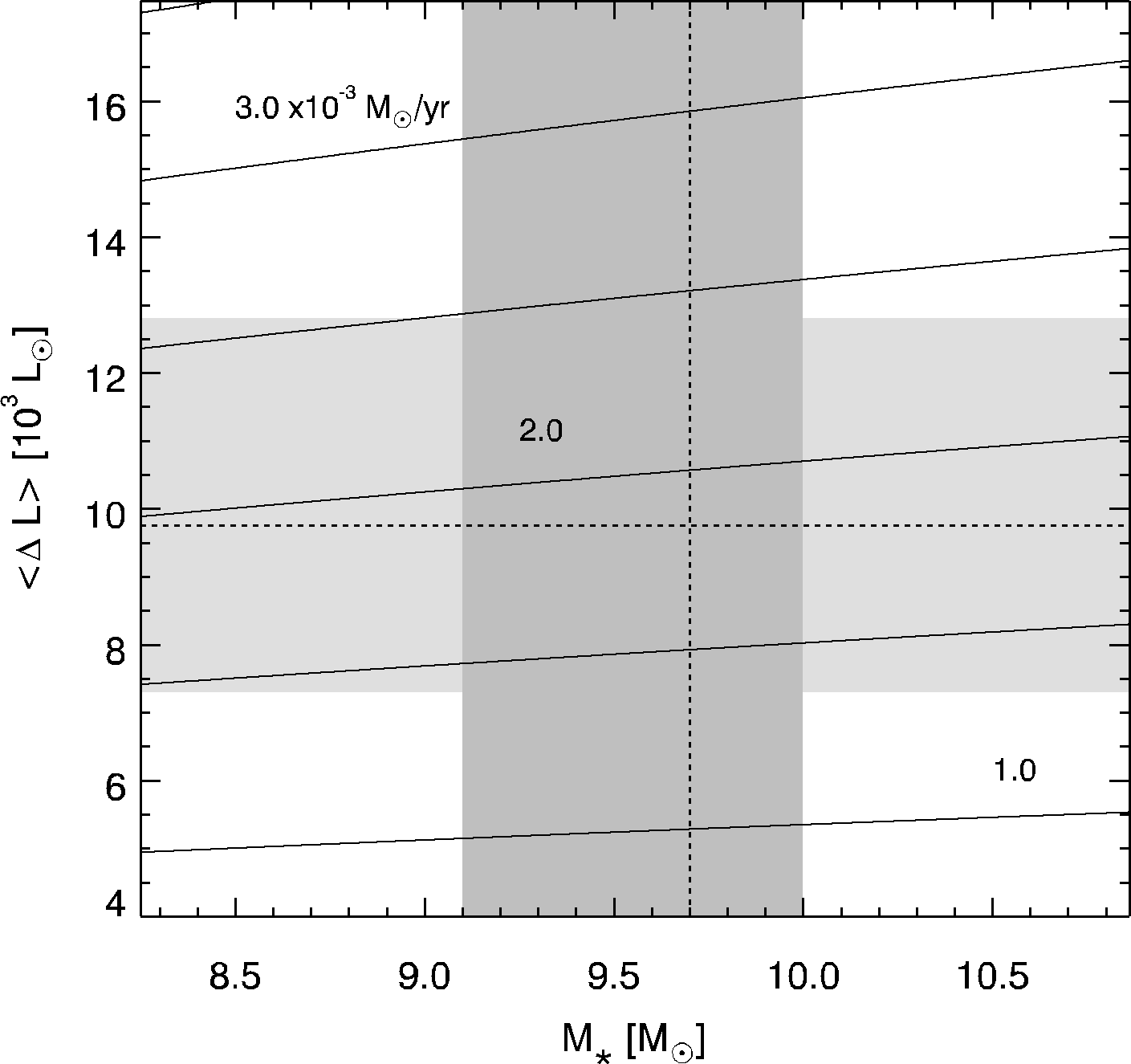}}
	\caption{Dependence of the derived accretion rate on the average luminosity increase ${<}\Delta L^{\rm acc}{>}$ and assumed ZAMS stellar mass for the parameter range bracketing MM1. The horizontal dashed line corresponds to ${<}\Delta L^{\rm acc}{>}$ and the light gray region marks its 1$\sigma$ uncertainty while the vertical dashed line and the darker gray region indicate the stellar mass and its uncertainty. The values of the accretion rate are indicated.
	}
 \label{fig:acc}
\end{figure}

Finally, the mass accretion rate will be obtained from $\dot{M}_{\rm acc}{=}M_{\rm acc}/\Delta t_{\rm acc}$. We have to emphasize here that, generally, the duration of the enhanced accretion $\Delta t_{\rm acc}$ will be shorter than that of the elevated emission $\Delta t$ which has to be used to come up with an overall burst energy estimate. Since the maser excitation is due to MIR dust emission, the total maser flux can be taken as a proxy for the accretion strength  (see Sec.\,\ref{bd}). From the rise and fall of the maser light curve (MacLeod et al., in prep.; Yonekura et al., in prep.), an effective duration $\Delta t_{\rm acc}$ of about two months can be derived with a presumed uncertainty range of ${\pm}$5\,days.
With the above quantities we obtain 
$M_{\rm acc}\,{=}\,3.1\mypm^{1.2}_{0.9}\,{\times}\,10^{-4}\,{\rm M}_{\odot}$, and $\dot{M}_{\rm acc}{=}1.8\mypm^{1.2}_{1.1}\,{\times}\,10^{-3}\,{\rm M}_{\odot}\rm yr^{-1}$,
where the errors are dominated by the error of 
$\Delta t_{\rm acc}$ and
$E_{\rm acc}$. 
Using the ZAMS mass-radius relation the range of the accretion rate for a given average luminosity increase is shown in Fig.\,\ref{fig:acc}.

However, since MM1 is likely in an earlier evolutionary stage, preceding the ZAMS, the above assumption may not hold. A different and presumably more realistic approach is possible using the
stellar radius from the RT modeling of the pre-burst SED together with the stellar mass of $12\,{\pm}3\,{\rm M}_\odot$ derived from the kinematic model of the spiral-arm accretion flows \citep{2020NatAs.tmp..144C}. This leads to
$M_{\rm acc}\,{=}\,5.3\mypm_{4.4}^{11.1}\,{\times}\,10^{-4}\,{\rm M}_{\odot}$, and $\dot{M}_{\rm acc}\,{=}\,3.2\mypm^{5.4}_{3.0}\,{\times}\,10^{-3}\,{\rm M}_{\odot}\rm yr^{-1}$. The large positive error range is mainly due to the corresponding large uncertainty of the stellar radius.
To put this into perspective, during its
short 
burst G358 MM1 consumed about 180 Earth masses. Notably, because of the small disk mass, the accreted fraction represents 16\% of the total. This raises the question whether the lightweight disk is a stable or transient feature.


\begin{figure*}
\centering
	\includegraphics[width=\textwidth]{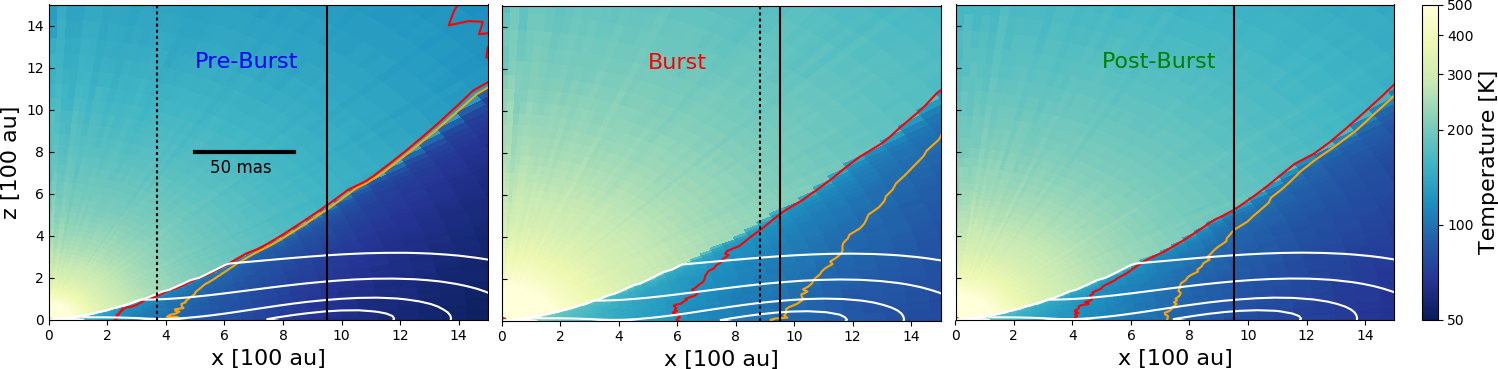}
\caption{Temperature distribution
of the mean model in the x-z plane (innermost part of first quadrant) for the pre-burst ({\bf left}), burst- ({\bf center}) and post-burst epochs ({\bf right}).  The orange and red lines enclose the temperature range from 94\,K (orange) to 120\,K (red) during the pre- and burst epochs, respectively. The white contours mark gas particle volume densities of 
$n_{\rm H_2}\,{=}\,[0.2,0.3,0.5,1]\,{\times}\,10^{8}\,\rm cm^{-3}$ which decrease with increasing z. The vertical solid black line indicates the outer radius of the disk. The dashed black lines mark the radius of the maser ring from the first and 4th epoch of the VLBI observations \citep[Burns et al., in prep.;]{2020NatAs...4..506B}. The length of the black bar corresponds to 50\,mas.
	}
 \label{fig: T, rho}
\end{figure*}

\section{Methanol maser relocation}


It has been emphasized that methanol masers play a crucial role in identifying accretion bursts from MYSOs. The maser activity of G358 during its burst was extraordinary and unique in several aspects \citep{2019ApJ...876L..25B, 2019ApJ...881L..39B, 2019MNRAS.489.3981M}.
The excitation of new maser spots at larger distance has been observed for the first time during the accretion burst of S255IR-NIRS3 \citep{2017A&A...600L...8M}. Notably, for G358 a ring-like propagation of maser spots, likely excited by the heat wave due to the burst, has been witnessed \citep{2020NatAs...4..506B}. Further evidence for spatial changes of the G358 maser distribution during and after the burst has been gained
(Bayandina et al., in prep.). 

Both methanol in the gas phase and the proper IR radiation are two of the major requirements for maser excitation. 
While cosmic ray sputtering of dust grain mantles can also release methanol to the gas phase \citep{2020A&A...634A.103D}, thermal desorption due to grain heating will be the dominant process near the MYSO. It 
requires temperatures of at least 94\,K \citep{2018MNRAS.473.1967L} while the maximum desorption rate is reached at about 120-125\,K \citep{2004MNRAS.354.1133C}. 
For such temperatures,
the peak flux of the thermal IR radiation from the warm dust is in the proper wavelength range needed to excite these masers \citep{2002IAUS..206..183O, 2005MNRAS.360..533C}. 
The RT models of MM1 during the pre-burst, burst, and post-burst epochs described in Sec.\,\ref{drt1} yielded spatial dust temperature distributions which can be used to address which regions of the circumstellar environment are potential sites of Class II 6.7\,GHz methanol masers, and how they change due to the burst.


Figure \ref{fig: T, rho} shows the central part of temperature distribution of the mean model in the first quadrant of the x-z plane for the pre-burst (left), burst (middle) and post-burst (right) epochs. The white lines mark the following gas densities $n_{\rm H_2}\,{=}\,[1, 0.5, 0.3, 0.2]\,{\times}\,10^{8}\,\rm cm^{-3}$ (density decreases with z). To transform the dust densities from the RT-simulation to the above given gas densities, we use the relation $\rho\,{=}\,{m_{\rm H_2}}{N_A^{-1}}\,{\times}\,n_{\rm H_2}\gamma^{-1}$, where $m_{\rm H_2}$ is the molar mass of H$_2$, $N_A$ is the Avogadro constant and $\gamma$ is the dust to gas ratio introduced in Sec.\,\ref{drt3}. Because of its low mass, the disk, which is embedded in the rotationally flattened envelope, cannot be  easily recognized. The vertical solid black line indicates the outer radius of the disk.

For gas densities exceeding $\rm 10^{8}\,cm^{-3}$, the maser brightness drops rapidly
due to collisional de-excitation \citep[Fig. 2]{2005MNRAS.360..533C}. Therefore, within the densest regions (innermost contour - disk mid-plane, envelope at the centrifugal radius) the excitation of Class II methanol masers is basically ruled out. 

The orange and red lines of Fig. \ref{fig: T, rho} indicate temperatures of $\rm 94\,K$ (orange, minimum temperature for thermal methanol desorption to occur) and $\rm 120\, K$ (red, optimum temperature for methanol desorption) during the pre-burst, burst, and post-burst, respectively. At each epoch the temperatures right of the solid orange line are too low for desorption of methanol, which means that it remains bound within icy dust grain mantles. Thus, these lines represent the methanol snow line beyond which masers cannot occur.
Due to the heating by the burst, 
it will be shifted outward
and move inward again once the disk started to cool after the burst ceased.

During the pre-burst stage (Fig.\,\ref{fig: T, rho} left), possible maser sites are likely confined to a region which originates in the disk around ${\sim}\,300$\,au and stretches along the surface of the cavity wall. The situation is different for the burst epoch (Fig.\,\ref{fig: T, rho} center). As expected, the methanol snow line moved outward and is located at about 1000\,au. At the same time the region below the disk/envelope surface has become warmer and presumably got enriched with gaseous methanol. These conclusions from the RT modeling can be compared to the multi-epoch VLBI observations of the expanding maser ring (\citealp{2020NatAs...4..506B}; Burns et al., in prep.).
The vertical dashed black lines mark the circumference of the maser ring from the first and 4th VLBI epoch. The first epoch observations were obtained already two weeks after the flare started. It is reasonable to assume that during the very early flare rise the excitation conditions were not too far off the pre-burst case. In fact the location of the maser circumference during the first epoch is within the region where phase transition from the solid to the gaseous methanol seems to occur in the pre-burst state. The data of the fourth VLBI epoch was taken about three weeks after the first FIFI-LS observations (''burst'' epoch). The maser circumference at that time is confined to the desorption temperature interval for regions not too far off the disk plane. The fact that the extent of the maser ring matches the expected maser positions for both epochs suggests that our RT modeling, although static, nevertheless describes major changes of the circumstellar environment due to the burst properly.

Moreover, Fig.\,\ref{fig: T, rho} seems to indicate that at the first VLBI epoch, the maser sites were likely close to the interface between the outflow lobe and the disk/envelope,
that is in a region of relatively low optical depth. The subliminal maser propagation speed reported by \citet{2020NatAs...4..506B} implies substantial optical depths. This suggests that the masers likely propagated into the disk rather than on its surface.
If so it could be expected that the ring expansion slowed down over time. This conclusion is fairly speculative and its verification requires time-dependent RT simulations. 

The right panel of Fig.\,\ref{fig: T, rho} indicates the readjustment of temperature distribution after the burst. Thus, the desorption range moved inward and is basically in between those for both pre-burst and burst.


Another important parameter for the maser excitation is the specific column density, $N_{\rm M} {\Delta V}^{-1}$, where $N_{\rm M}$ is the methanol column density along the line of sight and ${\Delta V}$ the velocity range of the molecules. 
The low inclination $i$ obtained from our RT modeling and by \citet{2020NatAs.tmp..144C} implies that $\Delta V$ is only slightly larger than thermal line-width since the radial velocity component of the orbital motion will be small. This increases the prospects for maser to occur.

When assessing these results it should be kept in mind that the circumstellar environment of G358 is almost certainly more structured than our RT model. Thus, sight lines of lower optical depth as well as density fluctuations may leave an imprint on the actual distribution of the maser spots.

\section{Wavelength and time dependence of the flux variation}\label{wtd}

\begin{figure}
	\centering
    \includegraphics[width=\hsize]{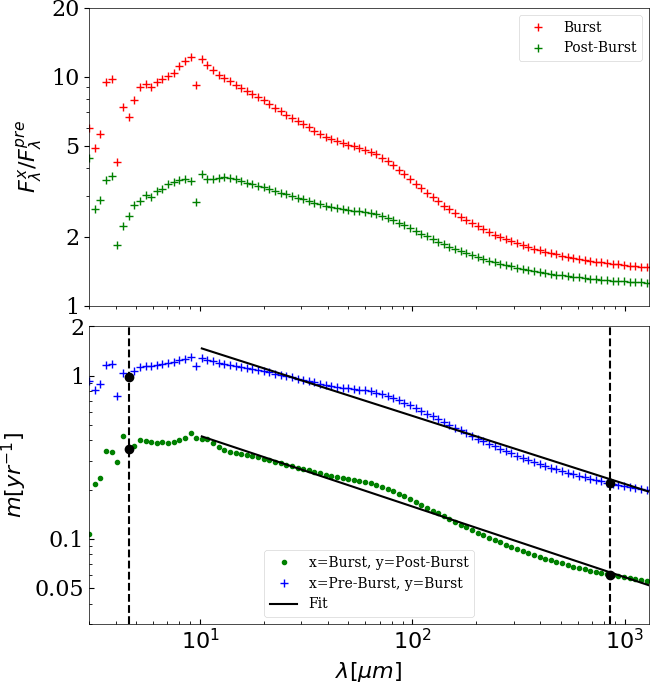}
	\caption{({\bf top}) Wavelength dependence of the relative flux increase for the burst (red) and post-burst (green) epochs for mean model. Markers denote grid points.
	({\bf bottom}) Wavelength dependence of the speed of the relative flux change $m$ for the pre-burst to burst (blue) and burst to post-burst (green) periods. The black solid lines mark linear fits to $\lambda\,{>}\,10\mu m$, the dashed vertical lines indicates the wavelengths used by \citet{2020MNRAS.495.3614C}. The black dots mark the values for $m$ used to compare our models to \citet{2020MNRAS.495.3614C}.	}
    \label{fig:wfr}
\end{figure}

The RT results for G358 can be used to address the question which wavelength range is suited best to detect the flux increase induced by a MYSO burst. For this purpose, the mean model SEDs for burst and post-burst were normalized by the pre-burst model SED. It has to be kept in mind that dividing the flux values cancels the extinction correction. The result is shown in the upper panel of  Fig.\,\ref{fig:wfr}. It can be seen that the peak values of the relative flux $m$ exceed the luminosity gains derived in the previous section. For our MYSO models, the largest relative flux increase occurs around 10\,$\mu$m. This agrees with the results of \citet{2019MNRAS.487.4465M} for similar luminosity gains. 
For their low-mass YSO models, the maximum flux ratio shifts toward the FIR for stronger accretion bursts.
The least relative flux increase occurs in the (sub)mm. This is because the emission in this part of the spectrum arises from the Rayleigh-Jeans tail of the Planck function, where the spectral radiance only linearly depends on temperature. Since the temperature increase of the outer circumstellar regions is lower than for the inner ones, their flux increase is accordingly smaller. 

The wavelength dependence of the relative flux increase agrees qualitatively with that of the outburst of OO Serpentis (OO Ser), as found by the first YSO outburst monitoring which covered the full wavelength range from NIR to FIR  \citep{2007A&A...470..211K}. These observations also revealed a wavelength-dependent time shift of the peak brightness which had not been witnessed for other eruptive YSOs before.

A recent study by \citet{2020MNRAS.495.3614C}, based on MIR (NEOWISE W2, 4.6\,$\mu$m) and (sub)mm (SCUBA-2, 850\,$\mu$m) variability of deeply embedded protostars, shed more light on the wavelength dependence of the rise/drop speed of the relative flux $m$. From our SED fits, we can assess this issue in a ``semi-empirical'' manner for the whole wavelength range of the SED. The relative rise/drop speeds are obtained for the pre-burst to the burst (Pr$\to$B) and the burst to post-burst (B$\to$Po) epochs, respectively. The results are presented in the lower panel of Fig.\,\ref{fig:wfr}. The two curves show $m(\lambda)=|\,log(F_{\rm x}(\lambda)/F_{\rm y}(\lambda))\,{\times}\,{\Delta t}^{-1}\,|$ for each wavelength of the modeled SEDs. The indices $x, y$ denote the respective epoch pairs and $\Delta t$ is the
corresponding epoch difference (in years). 
The upper curve holds for the flux rise while the lower one for the flux decrease. Since we do not know exactly when the rise started, the upper curve may be shifted while its slope is unaffected from the actual date.

With only two wavelengths at hand, \citet{2020MNRAS.495.3614C} were not able to infer the functional form of $m(\lambda)$ and had to assume a proportionality such that $m({\rm 4.6\mu m})\,{=}\,\eta\, m({\rm 850\mu m})$. From objects with the most significant variability at both wavelengths, they derived $\eta\,{=}\,5.53\,{\pm}\,0.29$. This can be compared to the $\eta$ values based on $m$ at 4.6 and 850\,$\mu$m from our results (black dots in Fig. \ref{fig:wfr}). These amount to $\eta_{\rm Pr \to B}\,{=}\,4.48$ and $\eta_{\rm B\to Po}\,{=}\,5.88$ for rise and drop, respectively. The values from our models are similar, although $\eta_{\rm Pr \to B}$ is somewhat lower.

More importantly, our approach allowed us to derive $m(\lambda)$ for the whole range of the SED. The corresponding curves show that
beyond $\lambda\,{\gtrsim}\,10\,\mu$m, $log(m)$ depends approximately linear on $log(\lambda)$. Thus, the wavelength dependence $m(\lambda)$ actually resembles a power law. The black lines show the respective fits to both epoch pairs. They yielded the following relations for $m(\lambda)$: $m_{\rm Pr \to B}\,{\propto}\, \lambda^{-0.42 \pm 0.01}$ and $m_{\rm B \to Po}\,{=}\, (1.16 \mypm^{0.04}_{0.03})\times \, \lambda^{-0.43 \pm 0.01}$, for the rise and drop, respectively. 
The spectral indices of both relations agree within the errors, suggesting that the sense of the flux variability does not affect the speed of the relative flux change.
These results indicate that a) burst induced flux variations are quicker at shorter wavelengths and b) the rise and fall of the fluxes take equal times.

Yet, a cautionary note is required here. First of all, our assessment of the correlated flux variability is based on YSO models with passive disks. The viscous energy release in an active disk leads to a temperature distribution that differs from the passive case in particular in the innermost region. This will lead to a higher value of $\eta$ as shown by \citet{2020MNRAS.495.3614C}. Moreover, we have to mention that the models used by these authors and us are based on static RT. Time-dependent RT shall be used to verify the above considerations.
Moreover, it has to be emphasized that the wavelength dependence of the fading rate as obtained by \citet{2020MNRAS.495.3614C} and in the present paper has not been found for the OO Ser event \citep{2007A&A...470..211K}. In that case, the fading rate is only time-dependent and slows down over time. The reason for the different behavior still needs to be understood.

\section{Discussion}\label{disc}
In the following, particular results are evaluated with regard to their credibility and impact concerning our understanding of MYSO accretion bursts. The G358 event is compared to the previous cases. Before we address individual topics, we note for completeness, that our investigation only covers the surplus dust continuum emission caused by the accretion burst. While the fraction of
the energy consumed by phase transitions (sublimation of grain ice mantles, dust evaporation and dissociation of molecules)
is negligible in the present context, it nevertheless seriously affects the chemistry of the YSO \citep{2017A&A...604A..15R, 2019MNRAS.485.1843W}.

\subsection{Misfit of the (sub)mm fluxes}\label{misf}
Figure \ref{fig:sed g358} shows that the best burst and post-burst fits predict (sub)mm fluxes that exceed the measured values. 
The observed values, indicated by the red and green symbols at $\lambda\,{=}\,889\,\mu$m amount to 60-70\% of the respective mean burst and post-burst model.
Since attempts to solve this discrepancy failed or required non-plausible assumptions, for instance on grain properties or geometry, we conclude that this mismatch indicates a deviation of the temperature distribution of the circumstellar environment from the static case. The accretion burst causes an additional inside-out heating of the circumstellar environment where inner regions try to reach a new thermal equilibrium while outer ones are still in the previous steady-state.  Time-dependent RT simulations suggest that the heating timescales 
strongly increase with wavelength
\citep{2011MNRAS.416.1500H, 2013ApJ...765..133J}.  
Since the MM1 SEDs are dominated by the increased FIR fluxes, their fits will, therefore, yield over-predicted (sub)mm fluxes.

For deeply embedded YSOs, the contribution of the envelope to the thermal FIR/(sub)mm emission may be significant (or even dominant), for example \cite{2020MNRAS.495.3614C}. For about half of the best MM1 pre-burst models, the overall optical depth of the envelope exceeds by far that of the disk (see Table \ref{tab: tau_V}). In these cases the envelope produces the overwhelming fraction of the (sub)mm emission. 
Furthermore, the heating of parts of the envelope, which are in the shadow of the disk, will be suppressed or (strongly) delayed. This effect will be especially important for systems with strongly flared disks. 

Not only the heating time scales have an influence on the burst-SED, but also the characteristics of the burst itself. Short/weak bursts might be not strong or long enough to heat the 
whole circumstellar environment which is probably the case for G358. 
This will lead to an overestimation of the sub-mm fluxes when using stationary models as well. Because of that,
fluxes beyond $889\,\mu $m were not included to fit the MM1 burst SED in order to avoid their systematic overestimation induced by our method.


\subsection{Uncertainty of the accretion parameters}

It was mentioned above that the ZAMS values for $R_*$ might not apply since the high accretion rates during the growth of a massive protostar will temporarily bloat its radius which could then be many times larger than that of a main sequence star (\citealp{2009ApJ...691..823H, 2010ApJ...721..478H}).
Recent simulations suggest that bloating is even intensified by accretion bursts \citep{2019MNRAS.484.2482M}. 
As a consequence of bloating,
more mass needs to be accreted to produce the observed luminosity increase. 
In the cold disk accretion model of \citet{2010ApJ...721..478H}, the protostellar swelling occurs in the mass range between 5-9\,M$_\odot$. Since the mass of MM1 is likely of that order, the supposition of being bloated is perhaps valid. If so, the accretion rate may be elevated as it is proportional to the stellar radius. Unfortunately, due to the high extinction of deeply embedded protostars, it is hard to derive their surface gravity from photospheric absorption lines because of strong veiling. 
Only recently this has been achieved for the first time. The MYSO G015.1288--00.6717 shows an A-type spectrum with a relatively low surface gravity, pointing to a bloated protostar \citep{2017MNRAS.472.3624P}.

For that reason,
caution is required for what concerns the comparison of stellar-related burst properties. For a MYSO in a relatively evolved stage, like S255IR-NIRS3 which shows a flat pre-burst SED, the derivation of an accretion rate using the ZAMS approach seems to be justified. However, its application to MYSOs in earlier stages, like G358 or NGC6334I-MM1, is questionable.
In addition, the ZAMS approach has its own challenges because, unlike low-mass stars, massive stars arrive on the ZAMS while accreting. Therefore, their pre-burst luminosity includes a likely small but unknown fraction of accretion luminosity. For low- and intermediate mass-stars, accretion rates can be derived from emission-line tracers (for example \citealp{2011A&A...535A..99M}) which, however, might be only exceptionally seen in scattered light from MYSOs \citep{1995ApJ...448..832N}. 

In any case the burst energy $E_{\rm acc}$ provides a good proxy on the energetics of the event. In this regard, the values for G358-MM1, S255IR-NIRS3, and NGC6334I-MM1 of 
$2.9\mypm_{0.8}^{1.1}\,{\times}\,10^{38}$\,J
(see Sec.\,\ref{drt1}), $1.2\pm{0.4}\,{\times}\,10^{39}$\,J \citep{2017NatPh..13..276C}, and $8\,{\times}\,10^{38}$\,J \citep{2017ApJ...837L..29H} already indicate a range of about an order of magnitude. Since the luminosities of S255IR-NIRS3 and NGC6334I-MM1 were still elevated when the aforementioned papers appeared, their final numbers will be even higher.

\subsection{Burst strength bias}
It is well established that young massive stars have a high stellar multiplicity \citep{2007ARA&A..45..481Z, 2019MNRAS.484..226P}. Given their distances, high spatial resolution obtained by interferometric observations is required to resolve the individual components of MYSOs like G358 or to prove their single nature. For less embedded objects, this can be achieved with the VLTI in the thermal IR, while for deeply embedded sources (sub)mm interferometers like NOEMA or ALMA have to be used. 
FIR interferometry, which requires groups of satellites in space, does not yet exist, although it has been envisaged \citep{2020AdSpR..65..831L}. Keeping in mind that MYSOs emit the bulk of their energy in the FIR, the coarse spatial resolution in the FIR implies that in most if not all cases, their luminosities consist of individual contributions. Thus, relating the luminosity increase caused by an accretion burst of single source to the total pre-burst luminosity only yields a lower limit to the burst strength. In case of G358 the luminosity gain derived from the graybody fits of pre- and burst total fluxes amounts to 2.5 while the RT results based on the SED decomposition (see Sec.\,\ref{sedc}) yielded a value of 4.7 for MM1.


\subsection{Burst duration and temporal evolution of the IR emission}\label{bd}
Class II methanol masers are excited by MIR thermal emission from warm dust grains with temperatures exceeding 100\,K
\citep{2002IAUS..206..183O, 2005MNRAS.360..533C}. Regardless of the details of the enhanced accretion, the increase of released energy will in all likelihood cause the evaporation of dust grains at the inner boundary of the disk and thus shift its rim outward. Since the radiative transfer will adjust the temperature profile accordingly, the MIR-emitting region moves toward a larger radius.
Due to its larger surface area
the emission increases and that of the masers as well. Thus, the rise and fall of the maser emission is coupled to the accretion variability through the dust MIR continuum emission. First direct evidence for this relation has been found by us for the S255IR-NIRS3 burst \citep{hansfest}, see \cite{2019MNRAS.485..777D} and \cite{2020A&A...634A..41O} for other cases. 

Since the maser flare of G358 lasted only for a few months 
(MacLeod et al., in prep.; Yonekura et al., in prep.), it can be concluded that the increased accretion rate dropped shortly before the end of the flare. So the question arises how does it come that elevated FIR emission is still present after 18 months past the flare peak? The current notion is that dust cooling should be quick because of the low optical depth at those wavelengths. The bulk of the FIR/(sub)mm emission comes from outer regions of the circumstellar environment. Thus, their optical depth toward the observer is low indeed. However, this is not necessarily the case for the optical depth toward the protostar where the energy is being released. An effective optical depth of a few will slow down the heating of the outer YSO environment and ``trap'' burst energy in the dusty medium. The latter will be radiated away and cause elevated FIR emission for quite some time after the burst terminated. 
Evidence of post-burst elevated FIR emission has been found for the OO Ser event \citep{2007A&A...470..211K} and by us for S255IR-NIRS3 (Stecklum et al., in prep.).
G358 is another example which shows this behavior as well. 

The projected linear separations between nearest neighbors of the G358 complex of up to 10$^4$\,au \citep{2019ApJ...881L..39B} imply individual object sizes with corresponding light travel times of up to two months. These can easily stretch to one year or more for moderate efficient optical depths. Furthermore, it has to be kept in mind that there is a variety of sight lines with different optical depths along which outer regions will be heated. The different propagation speeds along those lines incur a broadening of the heating duration. A subluminal speed  (0.04$\dots0.08\,c$) of the heat wave during the burst has been witnessed in G358 by tracing, for the first time, the outward motion of maser spots \citep{2020NatAs...4..506B}. Subluminal speed for the relocation of maser spots was found for S255IR-NIRS3 \citep{2018IAUS..336...37S} as well. This illustrates the imprint of the optical depth on the radiative transfer of the burst energy throughout the YSO.


\subsection{Accretion burst drivers}

YSO accretion bursts are thought to be caused by enhanced mass transfer through the circumstellar disk, triggered by various reasons, which may drive the inner disk to become active and self-luminous \citep{1996ARA&A..34..207H}. Among the possible causes, erratic infall from a protostellar envelope (cf. \citealp{2015ApJ...805..115V, 2017MNRAS.464L..90M})  seems to be favorable for a MYSO in an evolutionary stage as early as G358. Interestingly, recent high-resolution radio imaging of MM1 suggests the presence of two spiral-arm accretion flows, winding toward the protostar as traced by methanol masers \citep{2019ApJ...881L..39B, 2020NatAs.tmp..144C}. This is in line with very high-resolution ALMA imaging of high-mass protostars that provided evidence for filamentary streamers pointing onto the central sources \citep{2018arXiv180505364G, 2020A&A...634L..11J}. These might represent multi-directional accretion channels which possibly inhibit the formation of a large, steady disk at the very early stages of massive star formation. Thus, the small disk mass suggested by our best RT SED fits may be seen as a hint for the extreme youth of MM1. Possibly, its environment is not relaxed enough to allow the formation of a larger sustainable circumstellar disk. Such a disk might be prone to gravitational instability, considered to be another cause for episodic accretion \citep{2015ApJ...805..115V}. 
Faint radio continuum emission from MM1 has been detected during the burst as well (Bayandina et al., in prep.). Whether it is variable and perhaps related to an active disk needs to be explored.

\subsection{G358 in the accretion burst context}
The G358 burst is the second one from a MYSO for which the accretion luminosity could be measured from the SED changes. This allows us to compare the derived quantities with the results obtained for S255IR-NIRS3 \citep{2017NatPh..13..276C}, which is the first step toward a statistics of MYSO burst properties. 
Assuming a ZAMS star, an accreted mass of ${\sim}\, 3.4\,{\times}\,$10$^{-3}$\,M$_\odot$ has been estimated for S255IR-NIRS3. The corresponding accretion rate amounts to
5$\,{\pm}\,$2\,${\times}\,$10$^{-3}$\,M$_\odot$\,yr$^{-1}$. While these values will likely be revised once the comprehensive monitoring data set has been analyzed, we can assume that their order of magnitude is correct.
The accretion rate of the G358 burst almost corresponds to that of the S255IR-NIRS3 event. However, the accreted mass is about one order of magnitude smaller due to the shorter duration of the burst. Thus,
compared to S255IR-NIRS3, the G358 burst was a minor one.
Simulations of MYSO accretion \citep{2019MNRAS.482.5459M} suggest that minor bursts similar to that of G358 are much more frequent compared to major ones, and occur predominantly at very early stages of protostellar evolution. We emphasize that the above-mentioned disk accretion rates during the burst are at least an order of magnitude larger than those derived from SED modeling of non-bursting high-mass protostellar objects \citep{2008ApJ...688L..41F, 2009A&A...498..147G}.

While the MYSO accretion burst sample is still small, it will grow in the mid-term by following up methanol maser alerts on a regular basis. Recently, the suggestion by \citet{2019MNRAS.487.2407P} that the periodicity of methanol masers in G323.46--0.08 could be due to an accretion burst was confirmed (Wolf et al., in prep.) using (NEO)WISE and VVV data. While this is another event which was identified a posteriori, it raised the total number of known MYSO bursts to five,
including V723~Car.
So a  thorough comparison of MYSO burst properties with corresponding models (for example \citealp{2019MNRAS.482.5459M}), should become possible in the foreseeable future.



\section{Conclusions}\label{conc}

The SOFIA FIFI-LS observations of G358 during two epochs, supplemented by NIR imaging and supplementary archival data, and their RT analysis yield the following major results:
\begin{itemize}
    \item Increased FIR fluxes measured with FIFI-LS confirmed the accretion burst of G358 which was alerted by the maser flare. 
    Since no excess emission associated with the burst was detected in NIR, MIR or (sub)mm observations,
    it represents the first NIR-, FIR- and (sub)mm-dark but FIR-loud MYSO accretion burst.
    \item By means of the SED decomposition approach actual luminosity estimates of the driving source of the accretion burst could be obtained which yielded more representative burst parameters.
    \item From the change of the luminosities at the two FIFI-LS epochs the decay-time time of the FIR excess emission could be determined. This yielded more reliable estimates of the burst energy and the derived parameters. Moreover, wavelength-dependent rates for the rise and fall of the relative fluxes could be obtained.
    \item The circumstellar disk of MM1 is less massive than those found for other MYSOs and possibly transient. This may be an indication of the extreme youth of the source.
    \item A comparison with previous MYSO bursts indicates a considerable range of burst characteristics. The burst of G358 is the least energetic of the small sample.
    \item The RT modeling allowed us to draw conclusions on the possible sites of maser excitation and their relocation due to the burst which seem to be supported by VLBI radio observations.
   \item As already demonstrated for  the S255IR-NIRS3 event, the G358 results underline once more that the observational capabilities of SOFIA in the FIR are of utmost importance to study the erratic growth of stars.
 
\end{itemize}


As soon as a maser parallax for G358 becomes available, a reanalysis should be performed to narrow down the uncertainties of the derived parameters which will increase their credibility.

\begin{acknowledgements}
We thank the anonymous referee for accurate review and highly
appreciated comments and suggestions which helped
to improve the quality of this paper.
VW is supported by the by the German Aerospace Center (DLR) under grant number 50OR1718.
ACG has received funding from the European Research Council (ERC) under the European Union’s Horizon 2020 research and innovation programme (grant agreement No. 743029).
AS acknowledges support by the Ministry of Science and Higher Education of the Russian Federation under the grant 075-15-2020-780 (contract 780-10).
AV acknowledges the support from the Government of Russian Federation and Ministry of Science and Higher Education of the Russian Federation (grant N13.1902.21.0039) in the part of the data analysis.
Part of the funding for GROND (both hardware and personnel) was generously granted by the Leibniz-Prize to G. Hasinger (DFG grant HA 1850/28-1) and by the Th\"uringer Landessternwarte Tautenburg.
Based on observations made with the NASA/DLR Stratospheric Observatory for Infrared Astronomy (SOFIA). SOFIA is jointly operated by the Universities Space Research Association, Inc. (USRA), under NASA contract NNA17BF53C, and the Deutsches SOFIA Institut (DSI) under DLR contract 50 OK 0901 to the University of Stuttgart.
This research has made use of the NASA/IPAC Infrared Science Archive, which is funded by the National Aeronautics and Space Administration and operated by the California Institute of Technology.
This publication makes use of data products from the Near-Earth Object Wide-field Infrared Survey Explorer ((NEO)WISE), which is a joint project of the Jet Propulsion Laboratory/California Institute of Technology and the University of Arizona. (NEO)WISE is funded by the National Aeronautics and Space Administration.
The ATLASGAL project is a collaboration between the Max-Planck-Gesellschaft, the European Southern Observatory (ESO) and the Universidad de Chile. It includes projects E-181.C-0885,  E-78.F-9040(A), M-079.C-9501(A), M-081.C-9501(A) plus Chilean data. This research has made use of NASA''s Astrophysics Data System.
\end{acknowledgements}

\bibliographystyle{aa_url} 

\bibliography{G358-SOFIA} 

\begin{thebibliography}{118}
\expandafter\ifx\csname natexlab\endcsname\relax\def\natexlab#1{#1}\fi

\bibitem[{{{\'A}brah{\'a}m} {et~al.}(2009){{\'A}brah{\'a}m}, {Juh{\'a}sz},
  {Dullemond}, {K{\'o}sp{\'a}l}, {van Boekel}, {Bouwman}, {Henning},
  {Mo{\'o}r}, {Mosoni}, {Sicilia-Aguilar}, \& {Sipos}}]{2009Natur.459..224A}
{{\'A}brah{\'a}m}, P., {Juh{\'a}sz}, A., {Dullemond}, C.~P., {et~al.} 2009,
  \href{http://dx.doi.org/10.1038/nature08004}{\color{magenta}\nat},
  \href{https://ui.adsabs.harvard.edu/abs/2009Natur.459..224A}{459, 224}

\bibitem[{{Audard} {et~al.}(2014){Audard}, {{\'A}brah{\'a}m}, {Dunham},
  {Green}, {Grosso}, {Hamaguchi}, {Kastner}, {K{\'o}sp{\'a}l}, {Lodato},
  {Romanova}, {Skinner}, {Vorobyov}, \& {Zhu}}]{2014prpl.conf..387A}
{Audard}, M., {{\'A}brah{\'a}m}, P., {Dunham}, M.~M., {et~al.} 2014, in
  Protostars and Planets VI, ed. H.~{Beuther}, R.~S. {Klessen}, C.~P.
  {Dullemond}, \& T.~{Henning},
  \href{https://ui.adsabs.harvard.edu/abs/2014prpl.conf..387A}{387}

\bibitem[{{Boudet} {et~al.}(2005){Boudet}, {Mutschke}, {Nayral}, {J{\"a}ger},
  {Bernard}, {Henning}, \& {Meny}}]{2005ApJ...633..272B}
{Boudet}, N., {Mutschke}, H., {Nayral}, C., {et~al.} 2005,
  \href{http://dx.doi.org/10.1086/432966}{\color{magenta}\apj},
  \href{https://ui.adsabs.harvard.edu/abs/2005ApJ...633..272B}{633, 272}

\bibitem[{{Breen} {et~al.}(2013){Breen}, {Ellingsen}, {Contreras}, {Green},
  {Caswell}, {Stevens}, {Dawson}, \& {Voronkov}}]{2013MNRAS.435..524B}
{Breen}, S.~L., {Ellingsen}, S.~P., {Contreras}, Y., {et~al.} 2013,
  \href{http://dx.doi.org/10.1093/mnras/stt1315}{\color{magenta}\mnras},
  \href{https://ui.adsabs.harvard.edu/abs/2013MNRAS.435..524B}{435, 524}

\bibitem[{{Breen} {et~al.}(2019){Breen}, {Sobolev}, {Kaczmarek}, {Ellingsen},
  {McCarthy}, \& {Voronkov}}]{2019ApJ...876L..25B}
{Breen}, S.~L., {Sobolev}, A.~M., {Kaczmarek}, J.~F., {et~al.} 2019,
  \href{http://dx.doi.org/10.3847/2041-8213/ab191c}{\color{magenta}\apjl},
  \href{https://ui.adsabs.harvard.edu/abs/2019ApJ...876L..25B}{876, L25}

\bibitem[{{Brogan} {et~al.}(2019){Brogan}, {Hunter}, {Towner}, {McGuire},
  {MacLeod}, {Gurwell}, {Cyganowski}, {Brand}, {Burns}, {Caratti o Garatti},
  {Chen}, {Chibueze}, {Hirano}, {Hirota}, {Kim}, {Kramer}, {Linz}, {Menten},
  {Remijan}, {Sanna}, {Sobolev}, {Sridharan}, {Stecklum}, {Sugiyama}, {Surcis},
  {Van der Walt}, {Volvach}, \& {Volvach}}]{2019ApJ...881L..39B}
{Brogan}, C.~L., {Hunter}, T.~R., {Towner}, A.~P.~M., {et~al.} 2019,
  \href{http://dx.doi.org/10.3847/2041-8213/ab2f8a}{\color{magenta}\apjl},
  \href{https://ui.adsabs.harvard.edu/abs/2019ApJ...881L..39B}{881, L39}

\bibitem[{{Burns} {et~al.}(2020){Burns}, {Sugiyama}, {Hirota}, {Kim},
  {Sobolev}, {Stecklum}, {MacLeod}, {Yonekura}, {Olech}, {Orosz}, {Ellingsen},
  {Hyland}, {Caratti o Garatti}, {Brogan}, {Hunter}, {Phillips}, {van den
  Heever}, {Eisl{\"o}ffel}, {Linz}, {Surcis}, {Chibueze}, {Baan}, \&
  {Kramer}}]{2020NatAs...4..506B}
{Burns}, R.~A., {Sugiyama}, K., {Hirota}, T., {et~al.} 2020,
  \href{http://dx.doi.org/10.1038/s41550-019-0989-3}{\color{magenta}Nature
  Astronomy}, \href{https://ui.adsabs.harvard.edu/abs/2020NatAs...4..506B}{4,
  506}

\bibitem[{{Caratti o Garatti} {et~al.}(2017){Caratti o Garatti}, {Stecklum},
  {Garcia Lopez}, {Eisl{\"o}ffel}, {Ray}, {Sanna}, {Cesaroni}, {Walmsley},
  {Oudmaijer}, {de Wit}, {Moscadelli}, {Greiner}, {Krabbe}, {Fischer}, {Klein},
  \& {Iba{\~n}ez}}]{2017NatPh..13..276C}
{Caratti o Garatti}, A., {Stecklum}, B., {Garcia Lopez}, R., {et~al.} 2017,
  \href{http://dx.doi.org/10.1038/nphys3942}{\color{magenta}Nature Physics},
  \href{https://ui.adsabs.harvard.edu/abs/2017NatPh..13..276C}{13, 276}

\bibitem[{{Chen} {et~al.}(2020{\natexlab{a}}){Chen}, {Sobolev}, {Breen},
  {Shen}, {Ellingsen}, {MacLeod}, {Li}, {Voronkov}, {Kaczmarek}, {Zhang},
  {Ren}, {Wang}, {Linz}, {Hunter}, {Brogan}, {Sugiyama}, {Burns}, {Menten},
  {Sanna}, {Stecklum}, {Hirota}, {Kim}, {Chibueze}, \&
  {Heever}}]{2020ApJ...890L..22C}
{Chen}, X., {Sobolev}, A.~M., {Breen}, S.~L., {et~al.} 2020{\natexlab{a}},
  \href{http://dx.doi.org/10.3847/2041-8213/ab72a5}{\color{magenta}\apjl},
  \href{https://ui.adsabs.harvard.edu/abs/2020ApJ...890L..22C}{890, L22}

\bibitem[{{Chen} {et~al.}(2020{\natexlab{b}}){Chen}, {Sobolev}, {Ren},
  {Parfenov}, {Breen}, {Ellingsen}, {Shen}, {Li}, {MacLeod}, {Baan}, {Brogan},
  {Hirota}, {Hunter}, {Linz}, {Menten}, {Sugiyama}, {Stecklum}, {Gong}, \&
  {Zheng}}]{2020NatAs.tmp..144C}
{Chen}, X., {Sobolev}, A.~M., {Ren}, Z.-Y., {et~al.} 2020{\natexlab{b}},
  \href{https://ui.adsabs.harvard.edu/abs/2020NatAs.tmp..144C}{\href{http://dx.doi.org/10.1038/s41550-020-1144-x}{\color{magenta}Nature
  Astronomy}}

\bibitem[{{Clarke} {et~al.}(2015){Clarke}, {Vacca}, \&
  {Shuping}}]{2015ASPC..495..355C}
{Clarke}, M., {Vacca}, W.~D., \& {Shuping}, R.~Y. 2015, in Astronomical Society
  of the Pacific Conference Series, Vol. 495, Astronomical Data Analysis
  Software an Systems XXIV (ADASS XXIV), ed. A.~R. {Taylor} \& E.~{Rosolowsky},
  \href{https://ui.adsabs.harvard.edu/abs/2015ASPC..495..355C}{355}

\bibitem[{{Cohen} {et~al.}(2003){Cohen}, {Wheaton}, \&
  {Megeath}}]{2003AJ....126.1090C}
{Cohen}, M., {Wheaton}, W.~A., \& {Megeath}, S.~T. 2003,
  \href{http://dx.doi.org/10.1086/376474}{\color{magenta}\aj},
  \href{https://ui.adsabs.harvard.edu/abs/2003AJ....126.1090C}{126, 1090}

\bibitem[{{Colditz} {et~al.}(2018){Colditz}, {Beckmann}, {Bryant}, {Fischer},
  {Fumi}, {Geis}, {Hamidouche}, {Henning}, {H{\"o}nle}, {Iserlohe}, {Klein},
  {Krabbe}, {Looney}, {Poglitsch}, {Raab}, {Rebell}, {Rosenthal}, {Savage},
  {Schweitzer}, \& {Vacca}}]{2018JAI.....740004C}
{Colditz}, S., {Beckmann}, S., {Bryant}, A., {et~al.} 2018,
  \href{http://dx.doi.org/10.1142/S2251171718400044}{\color{magenta}Journal of
  Astronomical Instrumentation},
  \href{https://ui.adsabs.harvard.edu/abs/2018JAI.....740004C}{7, 1840004}

\bibitem[{{Collings} {et~al.}(2004){Collings}, {Anderson}, {Chen}, {Dever},
  {Viti}, {Williams}, \& {McCoustra}}]{2004MNRAS.354.1133C}
{Collings}, M.~P., {Anderson}, M.~A., {Chen}, R., {et~al.} 2004,
  \href{http://dx.doi.org/10.1111/j.1365-2966.2004.08272.x}{\color{magenta}\mnras},
  \href{https://ui.adsabs.harvard.edu/abs/2004MNRAS.354.1133C}{354, 1133}

\bibitem[{{Contreras} {et~al.}(2013){Contreras}, {Schuller}, {Urquhart},
  {Csengeri}, {Wyrowski}, {Beuther}, {Bontemps}, {Bronfman}, {Henning},
  {Menten}, {Schilke}, {Walmsley}, {Wienen}, {Tackenberg}, \&
  {Linz}}]{2013A&A...549A..45C}
{Contreras}, Y., {Schuller}, F., {Urquhart}, J.~S., {et~al.} 2013,
  \href{http://dx.doi.org/10.1051/0004-6361/201220155}{\color{magenta}\aap},
  \href{https://ui.adsabs.harvard.edu/abs/2013A&A...549A..45C}{549, A45}

\bibitem[{{Contreras Pe{\~n}a} {et~al.}(2020){Contreras Pe{\~n}a}, {Johnstone},
  {Baek}, {Herczeg}, {Mairs}, {Scholz}, {Lee}, \& {JCMT Transient
  Team}}]{2020MNRAS.495.3614C}
{Contreras Pe{\~n}a}, C., {Johnstone}, D., {Baek}, G., {et~al.} 2020,
  \href{http://dx.doi.org/10.1093/mnras/staa1254}{\color{magenta}\mnras},
  \href{https://ui.adsabs.harvard.edu/abs/2020MNRAS.495.3614C}{495, 3614}

\bibitem[{{Cragg} {et~al.}(2005){Cragg}, {Sobolev}, \&
  {Godfrey}}]{2005MNRAS.360..533C}
{Cragg}, D.~M., {Sobolev}, A.~M., \& {Godfrey}, P.~D. 2005,
  \href{http://dx.doi.org/10.1111/j.1365-2966.2005.09077.x}{\color{magenta}\mnras},
  \href{http://adsabs.harvard.edu/abs/2005MNRAS.360..533C}{360, 533}

\bibitem[{{Cutri} \& {et al.}(2014)}]{2014yCat.2328....0C}
{Cutri}, R.~M. \& {et al.} 2014,
  \href{https://ui.adsabs.harvard.edu/abs/2014yCat.2328....0C}{VizieR Online
  Data Catalog, II/328}

\bibitem[{{Cutri} {et~al.}(2003){Cutri}, {Skrutskie}, {van Dyk}, {Beichman},
  {Carpenter}, {Chester}, {Cambresy}, {Evans}, {Fowler}, {Gizis}, {Howard},
  {Huchra}, {Jarrett}, {Kopan}, {Kirkpatrick}, {Light}, {Marsh}, {McCallon},
  {Schneider}, {Stiening}, {Sykes}, {Weinberg}, {Wheaton}, {Wheelock}, \&
  {Zacarias}}]{2003yCat.2246....0C}
{Cutri}, R.~M., {Skrutskie}, M.~F., {van Dyk}, S., {et~al.} 2003,
  \href{https://ui.adsabs.harvard.edu/abs/2003yCat.2246....0C}{VizieR Online
  Data Catalog, II/246}

\bibitem[{{Cutri} {et~al.}(2012){Cutri}, {Wright}, {Conrow}, {Bauer},
  {Benford}, {Brandenburg}, {Dailey}, {Eisenhardt}, {Evans}, {Fajardo-Acosta},
  {Fowler}, {Gelino}, {Grillmair}, {Harbut}, {Hoffman}, {Jarrett},
  {Kirkpatrick}, {Leisawitz}, {Liu}, {Mainzer}, {Marsh}, {Masci}, {McCallon},
  {Padgett}, {Ressler}, {Royer}, {Skrutskie}, {Stanford}, {Wyatt}, {Tholen},
  {Tsai}, {Wachter}, {Wheelock}, {Yan}, {Alles}, {Beck}, {Grav}, {Masiero},
  {McCollum}, {McGehee}, {Papin}, \& {Wittman}}]{2012wise.rept....1C}
{Cutri}, R.~M., {Wright}, E.~L., {Conrow}, T., {et~al.} 2012,
  \href{https://wise2.ipac.caltech.edu/docs/release/allsky/expsup/sec4\_4c.html\#wpro}{Explanatory
  Supplement to the WISE All-Sky Data Release Products, 1}

\bibitem[{{Dartois} {et~al.}(2020){Dartois}, {Chabot}, {Bacmann}, {Boduch},
  {Domaracka}, \& {Rothard}}]{2020A&A...634A.103D}
{Dartois}, E., {Chabot}, M., {Bacmann}, A., {et~al.} 2020,
  \href{http://dx.doi.org/10.1051/0004-6361/201936934}{\color{magenta}\aap},
  \href{https://ui.adsabs.harvard.edu/abs/2020A&A...634A.103D}{634, A103}

\bibitem[{{Diolaiti} {et~al.}(2000){Diolaiti}, {Bendinelli}, {Bonaccini},
  {Close}, {Currie}, \& {Parmeggiani}}]{2000SPIE.4007..879D}
{Diolaiti}, E., {Bendinelli}, O., {Bonaccini}, D., {et~al.} 2000, in Society of
  Photo-Optical Instrumentation Engineers (SPIE) Conference Series, Vol. 4007,
  Adaptive Optical Systems Technology, ed. P.~L. {Wizinowich},
  \href{https://ui.adsabs.harvard.edu/abs/2000SPIE.4007..879D}{879--888}

\bibitem[{{Doi} {et~al.}(2015){Doi}, {Takita}, {Ootsubo}, {Arimatsu}, {Tanaka},
  {Kitamura}, {Kawada}, {Matsuura}, {Nakagawa}, {Morishima}, {Hattori},
  {Komugi}, {White}, {Ikeda}, {Kato}, {Chinone}, {Etxaluze}, \&
  {Cypriano}}]{2015PASJ...67...50D}
{Doi}, Y., {Takita}, S., {Ootsubo}, T., {et~al.} 2015,
  \href{http://dx.doi.org/10.1093/pasj/psv022}{\color{magenta}\pasj},
  \href{https://ui.adsabs.harvard.edu/abs/2015PASJ...67...50D}{67, 50}

\bibitem[{{Draine}(2003)}]{2003ARA&A..41..241D}
{Draine}, B.~T. 2003,
  \href{http://dx.doi.org/10.1146/annurev.astro.41.011802.094840}{\color{magenta}\araa},
  \href{https://ui.adsabs.harvard.edu/abs/2003ARA&A..41..241D}{41, 241}

\bibitem[{{Durjasz} {et~al.}(2019){Durjasz}, {Szymczak}, \&
  {Olech}}]{2019MNRAS.485..777D}
{Durjasz}, M., {Szymczak}, M., \& {Olech}, M. 2019,
  \href{http://dx.doi.org/10.1093/mnras/stz472}{\color{magenta}\mnras},
  \href{https://ui.adsabs.harvard.edu/abs/2019MNRAS.485..777D}{485, 777}

\bibitem[{{Elia} \& {Pezzuto}(2016)}]{2016MNRAS.461.1328E}
{Elia}, D. \& {Pezzuto}, S. 2016,
  \href{http://dx.doi.org/10.1093/mnras/stw1399}{\color{magenta}\mnras},
  \href{https://ui.adsabs.harvard.edu/abs/2016MNRAS.461.1328E}{461, 1328}

\bibitem[{{Erickson} \& {Davidson}(1993)}]{1993AdSpR..13..549E}
{Erickson}, E.~F. \& {Davidson}, J.~A. 1993,
  \href{http://dx.doi.org/10.1016/0273-1177(93)90164-7}{\color{magenta}Advances
  in Space Research},
  \href{https://ui.adsabs.harvard.edu/abs/1993AdSpR..13..549E}{13, 549}

\bibitem[{{Fadda} {et~al.}(2019){Fadda}, {Vacca}, {Fischer}, {Colditz},
  {Minchin}, \& {Klein}}]{2019AAS...23420805F}
{Fadda}, D., {Vacca}, W.~D., {Fischer}, C., {et~al.} 2019, in American
  Astronomical Society Meeting Abstracts, Vol.~51, American Astronomical
  Society Meeting Abstracts \#234,
  \href{https://ui.adsabs.harvard.edu/abs/2019AAS...23420805F}{208.05}

\bibitem[{{Fazal} {et~al.}(2008){Fazal}, {Sridharan}, {Qiu}, {Robitaille},
  {Whitney}, \& {Zhang}}]{2008ApJ...688L..41F}
{Fazal}, F.~M., {Sridharan}, T.~K., {Qiu}, K., {et~al.} 2008,
  \href{http://dx.doi.org/10.1086/593975}{\color{magenta}\apjl},
  \href{https://ui.adsabs.harvard.edu/abs/2008ApJ...688L..41F}{688, L41}

\bibitem[{{Fischer} {et~al.}(2018){Fischer}, {Beckmann}, {Bryant}, {Colditz},
  {Fumi}, {Geis}, {Hamidouche}, {Henning}, {H{\"o}nle}, {Iserlohe}, {Klein},
  {Krabbe}, {Looney}, {Poglitsch}, {Raab}, {Rebell}, {Rosenthal}, {Savage},
  {Schweitzer}, {Trinh}, \& {Vacca}}]{2018JAI.....740003F}
{Fischer}, C., {Beckmann}, S., {Bryant}, A., {et~al.} 2018,
  \href{http://dx.doi.org/10.1142/S2251171718400032}{\color{magenta}Journal of
  Astronomical Instrumentation},
  \href{https://ui.adsabs.harvard.edu/abs/2018JAI.....740003F}{7, 1840003}

\bibitem[{{Fujisawa} {et~al.}(2015){Fujisawa}, {Yonekura}, {Sugiyama},
  {Horiuchi}, {Hayashi}, {Hachisuka}, {Matsumoto}, \&
  {Niinuma}}]{2015ATel.8286....1F}
{Fujisawa}, K., {Yonekura}, Y., {Sugiyama}, K., {et~al.} 2015, The Astronomer's
  Telegram, \href{https://ui.adsabs.harvard.edu/abs/2015ATel.8286....1F}{8286,
  1}

\bibitem[{{Giannetti} {et~al.}(2017){Giannetti}, {Leurini}, {K{\"o}nig},
  {Urquhart}, {Pillai}, {Brand}, {Kauffmann}, {Wyrowski}, \&
  {Menten}}]{2017A&A...606L..12G}
{Giannetti}, A., {Leurini}, S., {K{\"o}nig}, C., {et~al.} 2017,
  \href{http://dx.doi.org/10.1051/0004-6361/201731728}{\color{magenta}\aap},
  \href{https://ui.adsabs.harvard.edu/abs/2017A&A...606L..12G}{606, L12}

\bibitem[{{Goddi} {et~al.}(2018){Goddi}, {Ginsburg}, {Maud}, {Zhang}, \&
  {Zapata}}]{2018arXiv180505364G}
{Goddi}, C., {Ginsburg}, A., {Maud}, L., {Zhang}, Q., \& {Zapata}, L. 2018,
  \href{https://ui.adsabs.harvard.edu/abs/2018arXiv180505364G}{arXiv e-prints,
  arXiv:1805.05364}

\bibitem[{{Graf} {et~al.}(2017){Graf}, {Reinacher}, {Spohr}, {Jakob}, \&
  {Fasoulas}}]{2017SPIE10401E..12G}
{Graf}, F., {Reinacher}, A., {Spohr}, D., {Jakob}, H., \& {Fasoulas}, S. 2017,
  in Society of Photo-Optical Instrumentation Engineers (SPIE) Conference
  Series, Vol. 10401, \procspie,
  \href{https://ui.adsabs.harvard.edu/abs/2017SPIE10401E..12G}{1040112}

\bibitem[{{Grave} \& {Kumar}(2009)}]{2009A&A...498..147G}
{Grave}, J.~M.~C. \& {Kumar}, M.~S.~N. 2009,
  \href{http://dx.doi.org/10.1051/0004-6361/200810902}{\color{magenta}\aap},
  \href{https://ui.adsabs.harvard.edu/abs/2009A&A...498..147G}{498, 147}

\bibitem[{{Greiner} {et~al.}(2008){Greiner}, {Bornemann}, {Clemens}, {Deuter},
  {Hasinger}, {Honsberg}, {Huber}, {Huber}, {Krauss}, {Kr{\"u}hler},
  {K{\"u}pc{\"u} Yolda{\c s}}, {Mayer-Hasselwander}, {Mican}, {Primak},
  {Schrey}, {Steiner}, {Szokoly}, {Th{\"o}ne}, {Yolda{\c s}}, {Klose}, {Laux},
  \& {Winkler}}]{2008PASP..120..405G}
{Greiner}, J., {Bornemann}, W., {Clemens}, C., {et~al.} 2008,
  \href{http://dx.doi.org/10.1086/587032}{\color{magenta}\pasp},
  \href{http://adsabs.harvard.edu/abs/2008PASP..120..405G}{120, 405}

\bibitem[{{Gutermuth} \& {Heyer}(2015)}]{2015AJ....149...64G}
{Gutermuth}, R.~A. \& {Heyer}, M. 2015,
  \href{http://dx.doi.org/10.1088/0004-6256/149/2/64}{\color{magenta}\aj},
  \href{https://ui.adsabs.harvard.edu/abs/2015AJ....149...64G}{149, 64}

\bibitem[{{Harries}(2011)}]{2011MNRAS.416.1500H}
{Harries}, T.~J. 2011,
  \href{http://dx.doi.org/10.1111/j.1365-2966.2011.19147.x}{\color{magenta}\mnras},
  \href{https://ui.adsabs.harvard.edu/abs/2011MNRAS.416.1500H}{416, 1500}

\bibitem[{{Hartmann} \& {Kenyon}(1985)}]{1985ApJ...299..462H}
{Hartmann}, L. \& {Kenyon}, S.~J. 1985,
  \href{http://dx.doi.org/10.1086/163713}{\color{magenta}\apj},
  \href{https://ui.adsabs.harvard.edu/abs/1985ApJ...299..462H}{299, 462}

\bibitem[{{Hartmann} \& {Kenyon}(1996)}]{1996ARA&A..34..207H}
{Hartmann}, L. \& {Kenyon}, S.~J. 1996,
  \href{http://dx.doi.org/10.1146/annurev.astro.34.1.207}{\color{magenta}\araa},
  \href{https://ui.adsabs.harvard.edu/abs/1996ARA&A..34..207H}{34, 207}

\bibitem[{{Herbig}(1966)}]{1966VA......8..109H}
{Herbig}, G.~H. 1966,
  \href{http://dx.doi.org/10.1016/0083-6656(66)90025-0}{\color{magenta}Vistas
  in Astronomy},
  \href{https://ui.adsabs.harvard.edu/abs/1966VA......8..109H}{8, 109}

\bibitem[{{Hinz} {et~al.}(2009){Hinz}, {Rieke}, {Yusef-Zadeh}, {Hewitt},
  {Balog}, \& {Block}}]{2009ApJS..181..227H}
{Hinz}, J.~L., {Rieke}, G.~H., {Yusef-Zadeh}, F., {et~al.} 2009,
  \href{http://dx.doi.org/10.1088/0067-0049/181/1/227}{\color{magenta}\apjs},
  \href{https://ui.adsabs.harvard.edu/abs/2009ApJS..181..227H}{181, 227}

\bibitem[{{Hosokawa} \& {Omukai}(2009)}]{2009ApJ...691..823H}
{Hosokawa}, T. \& {Omukai}, K. 2009,
  \href{http://dx.doi.org/10.1088/0004-637X/691/1/823}{\color{magenta}\apj},
  \href{https://ui.adsabs.harvard.edu/abs/2009ApJ...691..823H}{691, 823}

\bibitem[{{Hosokawa} {et~al.}(2010){Hosokawa}, {Yorke}, \&
  {Omukai}}]{2010ApJ...721..478H}
{Hosokawa}, T., {Yorke}, H.~W., \& {Omukai}, K. 2010,
  \href{http://dx.doi.org/10.1088/0004-637X/721/1/478}{\color{magenta}\apj},
  \href{https://ui.adsabs.harvard.edu/abs/2010ApJ...721..478H}{721, 478}

\bibitem[{{Hu} {et~al.}(2016){Hu}, {Menten}, {Wu}, {Bartkiewicz}, {Rygl},
  {Reid}, {Urquhart}, \& {Zheng}}]{2016ApJ...833...18H}
{Hu}, B., {Menten}, K.~M., {Wu}, Y., {et~al.} 2016,
  \href{http://dx.doi.org/10.3847/0004-637X/833/1/18}{\color{magenta}\apj},
  \href{https://ui.adsabs.harvard.edu/abs/2016ApJ...833...18H}{833, 18}

\bibitem[{{Hunter} {et~al.}(2017){Hunter}, {Brogan}, {MacLeod}, {Cyganowski},
  {Chandler}, {Chibueze}, {Friesen}, {Indebetouw}, {Thesner}, \&
  {Young}}]{2017ApJ...837L..29H}
{Hunter}, T.~R., {Brogan}, C.~L., {MacLeod}, G., {et~al.} 2017,
  \href{http://dx.doi.org/10.3847/2041-8213/aa5d0e}{\color{magenta}\apjl},
  \href{https://ui.adsabs.harvard.edu/abs/2017ApJ...837L..29H}{837, L29}

\bibitem[{{Hunter} {et~al.}(2018){Hunter}, {Brogan}, {MacLeod}, {Cyganowski},
  {Chibueze}, {Friesen}, {Hirota}, {Smits}, {Chandler}, \&
  {Indebetouw}}]{2018ApJ...854..170H}
{Hunter}, T.~R., {Brogan}, C.~L., {MacLeod}, G.~C., {et~al.} 2018,
  \href{http://dx.doi.org/10.3847/1538-4357/aaa962}{\color{magenta}\apj},
  \href{https://ui.adsabs.harvard.edu/abs/2018ApJ...854..170H}{854, 170}

\bibitem[{{Johnston} {et~al.}(2020){Johnston}, {Hoare}, {Beuther}, {Kuiper},
  {Kee}, {Linz}, {Boley}, {Maud}, {Ahmadi}, \&
  {Robitaille}}]{2020A&A...634L..11J}
{Johnston}, K.~G., {Hoare}, M.~G., {Beuther}, H., {et~al.} 2020,
  \href{http://dx.doi.org/10.1051/0004-6361/201937154}{\color{magenta}\aap},
  \href{https://ui.adsabs.harvard.edu/abs/2020A&A...634L..11J}{634, L11}

\bibitem[{{Johnston} {et~al.}(2015){Johnston}, {Robitaille}, {Beuther}, {Linz},
  {Boley}, {Kuiper}, {Keto}, {Hoare}, \& {van Boekel}}]{2015ApJ...813L..19J}
{Johnston}, K.~G., {Robitaille}, T.~P., {Beuther}, H., {et~al.} 2015,
  \href{http://dx.doi.org/10.1088/2041-8205/813/1/L19}{\color{magenta}\apjl},
  \href{https://ui.adsabs.harvard.edu/abs/2015ApJ...813L..19J}{813, L19}

\bibitem[{{Johnstone} {et~al.}(2013){Johnstone}, {Hendricks}, {Herczeg}, \&
  {Bruderer}}]{2013ApJ...765..133J}
{Johnstone}, D., {Hendricks}, B., {Herczeg}, G.~J., \& {Bruderer}, S. 2013,
  \href{http://dx.doi.org/10.1088/0004-637X/765/2/133}{\color{magenta}\apj},
  \href{https://ui.adsabs.harvard.edu/abs/2013ApJ...765..133J}{765, 133}

\bibitem[{{K{\'o}sp{\'a}l} {et~al.}(2007){K{\'o}sp{\'a}l}, {{\'A}brah{\'a}m},
  {Prusti}, {Acosta-Pulido}, {Hony}, {Mo{\'o}r}, \&
  {Siebenmorgen}}]{2007A&A...470..211K}
{K{\'o}sp{\'a}l}, {\'A}., {{\'A}brah{\'a}m}, P., {Prusti}, T., {et~al.} 2007,
  \href{http://dx.doi.org/10.1051/0004-6361:20066108}{\color{magenta}\aap},
  \href{https://ui.adsabs.harvard.edu/abs/2007A&A...470..211K}{470, 211}

\bibitem[{{Kraus} {et~al.}(2010){Kraus}, {Hofmann}, {Menten}, {Schertl},
  {Weigelt}, {Wyrowski}, {Meilland}, {Perraut}, {Petrov}, {Robbe-Dubois},
  {Schilke}, \& {Testi}}]{2010Natur.466..339K}
{Kraus}, S., {Hofmann}, K.-H., {Menten}, K.~M., {et~al.} 2010,
  \href{http://dx.doi.org/10.1038/nature09174}{\color{magenta}\nat},
  \href{https://ui.adsabs.harvard.edu/abs/2010Natur.466..339K}{466, 339}

\bibitem[{{Kr{\"u}hler} {et~al.}(2008){Kr{\"u}hler}, {K{\"u}pc{\"u} Yolda{\c
  s}}, {Greiner}, {Clemens}, {McBreen}, {Primak}, {Savaglio}, {Yolda{\c s}},
  {Szokoly}, \& {Klose}}]{2008ApJ...685..376K}
{Kr{\"u}hler}, T., {K{\"u}pc{\"u} Yolda{\c s}}, A., {Greiner}, J., {et~al.}
  2008, \href{http://dx.doi.org/10.1086/590240}{\color{magenta}\apj},
  \href{http://adsabs.harvard.edu/abs/2008ApJ...685..376K}{685, 376}

\bibitem[{{Kuiper} {et~al.}(2016){Kuiper}, {Turner}, \&
  {Yorke}}]{2016ApJ...832...40K}
{Kuiper}, R., {Turner}, N.~J., \& {Yorke}, H.~W. 2016,
  \href{http://dx.doi.org/10.3847/0004-637X/832/1/40}{\color{magenta}\apj},
  \href{https://ui.adsabs.harvard.edu/abs/2016ApJ...832...40K}{832, 40}

\bibitem[{{Landsman}(1995)}]{1995ASPC...77..437L}
{Landsman}, W.~B. 1995, in Astronomical Society of the Pacific Conference
  Series, Vol.~77, Astronomical Data Analysis Software and Systems IV, ed.
  R.~A. {Shaw}, H.~E. {Payne}, \& J.~J.~E. {Hayes},
  \href{https://ui.adsabs.harvard.edu/abs/1995ASPC...77..437L}{437}

\bibitem[{{Linz} {et~al.}(2020){Linz}, {Bhatia}, {Buinhas}, {Lezius}, {Ferrer},
  {F{\"o}rstner}, {Frankl}, {Philips-Blum}, {Steen}, {Bestmann}, {H{\"a}nsel},
  {Holzwarth}, {Krause}, \& {Pany}}]{2020AdSpR..65..831L}
{Linz}, H., {Bhatia}, D., {Buinhas}, L., {et~al.} 2020,
  \href{http://dx.doi.org/10.1016/j.asr.2019.06.022}{\color{magenta}Advances in
  Space Research},
  \href{https://ui.adsabs.harvard.edu/abs/2020AdSpR..65..831L}{65, 831}

\bibitem[{{Liu} {et~al.}(2018){Liu}, {Su}, {Zinchenko}, {Wang}, \&
  {Wang}}]{2018ApJ...863L..12L}
{Liu}, S.-Y., {Su}, Y.-N., {Zinchenko}, I., {Wang}, K.-S., \& {Wang}, Y. 2018,
  \href{http://dx.doi.org/10.3847/2041-8213/aad63a}{\color{magenta}\apj},
  \href{https://ui.adsabs.harvard.edu/abs/2018ApJ...863L..12L}{863, L12}

\bibitem[{{Looney} {et~al.}(2000){Looney}, {Geis}, {Genzel}, {Park},
  {Poglitsch}, {Raab}, {Rosenthal}, {Urban}, \&
  {Henning}}]{2000SPIE.4014...14L}
{Looney}, L.~W., {Geis}, N., {Genzel}, R., {et~al.} 2000, in Society of
  Photo-Optical Instrumentation Engineers (SPIE) Conference Series, Vol. 4014,
  Airborne Telescope Systems, ed. R.~K. {Melugin} \& H.-P. {R{\"o}ser},
  \href{https://ui.adsabs.harvard.edu/abs/2000SPIE.4014...14L}{14--22}

\bibitem[{{Luna} {et~al.}(2018){Luna}, {Satorre}, {Domingo}, {Mill{\'a}n},
  {Luna-Ferr{\'a}ndiz}, {Gisbert}, \& {Santonja}}]{2018MNRAS.473.1967L}
{Luna}, R., {Satorre}, M.~{\'A}., {Domingo}, M., {et~al.} 2018,
  \href{http://dx.doi.org/10.1093/mnras/stx2473}{\color{magenta}\mnras},
  \href{https://ui.adsabs.harvard.edu/abs/2018MNRAS.473.1967L}{473, 1967}

\bibitem[{{MacFarlane} {et~al.}(2019){MacFarlane}, {Stamatellos}, {Johnstone},
  {Herczeg}, {Baek}, {Chen}, {Kang}, \& {Lee}}]{2019MNRAS.487.4465M}
{MacFarlane}, B., {Stamatellos}, D., {Johnstone}, D., {et~al.} 2019,
  \href{http://dx.doi.org/10.1093/mnras/stz1570}{\color{magenta}\mnras},
  \href{https://ui.adsabs.harvard.edu/abs/2019MNRAS.487.4465M}{487, 4465}

\bibitem[{{MacLeod} {et~al.}(2018){MacLeod}, {Smits}, {Goedhart}, {Hunter},
  {Brogan}, {Chibueze}, {van den Heever}, {Thesner}, {Banda}, \&
  {Paulsen}}]{2018MNRAS.478.1077M}
{MacLeod}, G.~C., {Smits}, D.~P., {Goedhart}, S., {et~al.} 2018,
  \href{http://dx.doi.org/10.1093/mnras/sty996}{\color{magenta}\mnras},
  \href{https://ui.adsabs.harvard.edu/abs/2018MNRAS.478.1077M}{478, 1077}

\bibitem[{{MacLeod} {et~al.}(2019){MacLeod}, {Sugiyama}, {Hunter}, {Quick},
  {Baan}, {Breen}, {Brogan}, {Burns}, {Caratti o Garatti}, {Chen}, {Chibueze},
  {Houde}, {Kaczmarek}, {Linz}, {Rajabi}, {Saito}, {Schmidl}, {Sobolev},
  {Stecklum}, {van den Heever}, \& {Yonekura}}]{2019MNRAS.489.3981M}
{MacLeod}, G.~C., {Sugiyama}, K., {Hunter}, T.~R., {et~al.} 2019,
  \href{http://dx.doi.org/10.1093/mnras/stz2417}{\color{magenta}\mnras},
  \href{https://ui.adsabs.harvard.edu/abs/2019MNRAS.489.3981M}{489, 3981}

\bibitem[{{Mainzer} {et~al.}(2014){Mainzer}, {Bauer}, {Cutri}, {Grav},
  {Masiero}, {Beck}, {Clarkson}, {Conrow}, {Dailey}, {Eisenhardt}, {Fabinsky},
  {Fajardo-Acosta}, {Fowler}, {Gelino}, {Grillmair}, {Heinrichsen}, {Kendall},
  {Kirkpatrick}, {Liu}, {Masci}, {McCallon}, {Nugent}, {Papin}, {Rice},
  {Royer}, {Ryan}, {Sevilla}, {Sonnett}, {Stevenson}, {Thompson}, {Wheelock},
  {Wiemer}, {Wittman}, {Wright}, \& {Yan}}]{2014ApJ...792...30M}
{Mainzer}, A., {Bauer}, J., {Cutri}, R.~M., {et~al.} 2014,
  \href{http://dx.doi.org/10.1088/0004-637X/792/1/30}{\color{magenta}\apj},
  \href{https://ui.adsabs.harvard.edu/abs/2014ApJ...792...30M}{792, 30}

\bibitem[{{Mattsson} {et~al.}(2014){Mattsson}, {Gomez}, {Andersen}, {Smith},
  {De Looze}, {Baes}, {Viaene}, {Gentile}, {Fritz}, \&
  {Spinoglio}}]{2014MNRAS.444..797M}
{Mattsson}, L., {Gomez}, H.~L., {Andersen}, A.~C., {et~al.} 2014,
  \href{http://dx.doi.org/10.1093/mnras/stu1228}{\color{magenta}\mnras},
  \href{https://ui.adsabs.harvard.edu/abs/2014MNRAS.444..797M}{444, 797}

\bibitem[{{Mendigut{\'\i}a} {et~al.}(2011){Mendigut{\'\i}a}, {Calvet},
  {Montesinos}, {Mora}, {Muzerolle}, {Eiroa}, {Oudmaijer}, \&
  {Mer{\'\i}n}}]{2011A&A...535A..99M}
{Mendigut{\'\i}a}, I., {Calvet}, N., {Montesinos}, B., {et~al.} 2011,
  \href{http://dx.doi.org/10.1051/0004-6361/201117444}{\color{magenta}\aap},
  \href{https://ui.adsabs.harvard.edu/abs/2011A&A...535A..99M}{535, A99}

\bibitem[{{Menten}(1991{\natexlab{a}})}]{1991ASPC...16..119M}
{Menten}, K.~M. 1991{\natexlab{a}}, in Astronomical Society of the Pacific
  Conference Series, Vol.~16, Atoms, Ions and Molecules: New Results in
  Spectral Line Astrophysics, ed. A.~D. {Haschick} \& P.~T.~P. {Ho},
  \href{https://ui.adsabs.harvard.edu/abs/1991ASPC...16..119M}{119--136}

\bibitem[{{Menten}(1991{\natexlab{b}})}]{1991ApJ...380L..75M}
{Menten}, K.~M. 1991{\natexlab{b}},
  \href{http://dx.doi.org/10.1086/186177}{\color{magenta}\apjl},
  \href{https://ui.adsabs.harvard.edu/abs/1991ApJ...380L..75M}{380, L75}

\bibitem[{{Meyer} {et~al.}(2019{\natexlab{a}}){Meyer}, {Haemmerl{\'e}}, \&
  {Vorobyov}}]{2019MNRAS.484.2482M}
{Meyer}, D.~M.~A., {Haemmerl{\'e}}, L., \& {Vorobyov}, E.~I.
  2019{\natexlab{a}},
  \href{http://dx.doi.org/10.1093/mnras/sty3527}{\color{magenta}\mnras},
  \href{https://ui.adsabs.harvard.edu/abs/2019MNRAS.484.2482M}{484, 2482}

\bibitem[{{Meyer} {et~al.}(2019{\natexlab{b}}){Meyer}, {Vorobyov}, {Elbakyan},
  {Stecklum}, {Eisl{\"o}ffel}, \& {Sobolev}}]{2019MNRAS.482.5459M}
{Meyer}, D.~M.~A., {Vorobyov}, E.~I., {Elbakyan}, V.~G., {et~al.}
  2019{\natexlab{b}},
  \href{http://dx.doi.org/10.1093/mnras/sty2980}{\color{magenta}\mnras},
  \href{https://ui.adsabs.harvard.edu/abs/2019MNRAS.482.5459M}{482, 5459}

\bibitem[{{Meyer} {et~al.}(2017){Meyer}, {Vorobyov}, {Kuiper}, \&
  {Kley}}]{2017MNRAS.464L..90M}
{Meyer}, D.~M.~A., {Vorobyov}, E.~I., {Kuiper}, R., \& {Kley}, W. 2017,
  \href{http://dx.doi.org/10.1093/mnrasl/slw187}{\color{magenta}\mnras},
  \href{https://ui.adsabs.harvard.edu/abs/2017MNRAS.464L..90M}{464, L90}

\bibitem[{{Minniti} {et~al.}(2017){Minniti}, {Lucas}, \& {VVV
  Team}}]{2017yCat.2348....0M}
{Minniti}, D., {Lucas}, P., \& {VVV Team}. 2017, VizieR Online Data Catalog,
  \href{http://adsabs.harvard.edu/abs/2017yCat.2348....0M}{2348}

\bibitem[{{Minniti} {et~al.}(2010){Minniti}, {Lucas}, {Emerson}, {Saito},
  {Hempel}, {Pietrukowicz}, {Ahumada}, {Alonso}, {Alonso-Garcia}, {Arias},
  {Bandyopadhyay}, {Barb{\'a}}, {Barbuy}, {Bedin}, {Bica}, {Borissova},
  {Bronfman}, {Carraro}, {Catelan}, {Clari{\'a}}, {Cross}, {de Grijs},
  {D{\'e}k{\'a}ny}, {Drew}, {Fari{\~n}a}, {Feinstein}, {Fern{\'a}ndez
  Laj{\'u}s}, {Gamen}, {Geisler}, {Gieren}, {Goldman}, {Gonzalez}, {Gunthardt},
  {Gurovich}, {Hambly}, {Irwin}, {Ivanov}, {Jord{\'a}n}, {Kerins}, {Kinemuchi},
  {Kurtev}, {L{\'o}pez-Corredoira}, {Maccarone}, {Masetti}, {Merlo},
  {Messineo}, {Mirabel}, {Monaco}, {Morelli}, {Padilla}, {Palma}, {Parisi},
  {Pignata}, {Rejkuba}, {Roman-Lopes}, {Sale}, {Schreiber}, {Schr{\"o}der},
  {Smith}, {}, {Soto}, {Tamura}, {Tappert}, {Thompson}, {Toledo}, {Zoccali}, \&
  {Pietrzynski}}]{2010NewA...15..433M}
{Minniti}, D., {Lucas}, P.~W., {Emerson}, J.~P., {et~al.} 2010,
  \href{http://dx.doi.org/10.1016/j.newast.2009.12.002}{\color{magenta}\na},
  \href{https://ui.adsabs.harvard.edu/abs/2010NewA...15..433M}{15, 433}

\bibitem[{{Molinari} {et~al.}(2016){Molinari}, {Schisano}, {Elia},
  {Pestalozzi}, {Traficante}, {Pezzuto}, {Swinyard}, {Noriega-Crespo}, {Bally},
  {Moore}, {Plume}, {Zavagno}, {di Giorgio A.~M.}, {Liu}, {Pilbratt},
  {Mottram}, {Russeil}, {Piazzo}, {Veneziani}, {Benedettini}, {Calzoletti},
  {Faustini}, {Natoli}, {Piacentini}, {Merello}, {Palmese}, {Del Grand e},
  {Polychroni}, {Rygl}, {Polenta}, {Barlow}, {Bernard}, {Martin}, {Testi},
  {Ali}, {Andre}, {Beltran}, {Billot}, {Carey}, {Cesaroni}, {Compiegne},
  {Eden}, {Fukui}, {Garcia-Lario}, {Hoare}, {Huang}, {Joncas}, {Lim}, {Lord},
  {Martinavarro-Armengol}, {Motte}, {Paladini}, {Paradis}, {Peretto},
  {Robitaille}, {Schilke}, {Schneider}, {Schulz}, {Sibthorpe}, {Strafella},
  {Thompson}, {Umana}, {Ward-Thompson}, \& {Wyrowski}}]{2016yCat..35910149M}
{Molinari}, S., {Schisano}, E., {Elia}, D., {et~al.} 2016,
  \href{https://ui.adsabs.harvard.edu/abs/2016yCat..35910149M}{VizieR Online
  Data Catalog, J/A+A/591/A149}

\bibitem[{{Molinari} {et~al.}(2010){Molinari}, {Swinyard}, {Bally}, {Barlow},
  {Bernard}, {Martin}, {Moore}, {Noriega-Crespo}, {Plume}, {Testi}, {Zavagno},
  {Abergel}, {Ali}, {Andr{\'e}}, {Baluteau}, {Benedettini}, {Bern{\'e}},
  {Billot}, {Blommaert}, {Bontemps}, {Boulanger}, {Brand}, {Brunt}, {Burton},
  {Campeggio}, {Carey}, {Caselli}, {Cesaroni}, {Cernicharo}, {Chakrabarti},
  {Chrysostomou}, {Codella}, {Cohen}, {Compiegne}, {Davis}, {de Bernardis}, {de
  Gasperis}, {Di Francesco}, {di Giorgio}, {Elia}, {Faustini}, {Fischera},
  {Fukui}, {Fuller}, {Ganga}, {Garcia-Lario}, {Giard}, {Giardino}, {Glenn},
  {Goldsmith}, {Griffin}, {Hoare}, {Huang}, {Jiang}, {Joblin}, {Joncas},
  {Juvela}, {Kirk}, {Lagache}, {Li}, {Lim}, {Lord}, {Lucas}, {Maiolo},
  {Marengo}, {Marshall}, {Masi}, {Massi}, {Matsuura}, {Meny}, {Minier},
  {Miville-Desch{\^e}nes}, {Montier}, {Motte}, {M{\"u}ller}, {Natoli}, {Neves},
  {Olmi}, {Paladini}, {Paradis}, {Pestalozzi}, {Pezzuto}, {Piacentini},
  {Pomar{\`e}s}, {Popescu}, {Reach}, {Richer}, {Ristorcelli}, {Roy}, {Royer},
  {Russeil}, {Saraceno}, {Sauvage}, {Schilke}, {Schneider-Bontemps},
  {Schuller}, {Schultz}, {Shepherd}, {Sibthorpe}, {Smith}, {Smith},
  {Spinoglio}, {Stamatellos}, {Strafella}, {Stringfellow}, {Sturm}, {Taylor},
  {Thompson}, {Tuffs}, {Umana}, {Valenziano}, {Vavrek}, {Viti}, {Waelkens},
  {Ward-Thompson}, {White}, {Wyrowski}, {Yorke}, \&
  {Zhang}}]{2010PASP..122..314M}
{Molinari}, S., {Swinyard}, B., {Bally}, J., {et~al.} 2010,
  \href{http://dx.doi.org/10.1086/651314}{\color{magenta}\pasp},
  \href{https://ui.adsabs.harvard.edu/abs/2010PASP..122..314M}{122, 314}

\bibitem[{{Moscadelli} {et~al.}(2017){Moscadelli}, {Sanna}, {Goddi},
  {Walmsley}, {Cesaroni}, {Caratti o Garatti}, {Stecklum}, {Menten}, \&
  {Kraus}}]{2017A&A...600L...8M}
{Moscadelli}, L., {Sanna}, A., {Goddi}, C., {et~al.} 2017,
  \href{http://dx.doi.org/10.1051/0004-6361/201730659}{\color{magenta}\aap},
  \href{https://ui.adsabs.harvard.edu/abs/2017A&A...600L...8M}{600, L8}

\bibitem[{{Neckel} \& {Staude}(1995)}]{1995ApJ...448..832N}
{Neckel}, T. \& {Staude}, H.~J. 1995,
  \href{http://dx.doi.org/10.1086/176011}{\color{magenta}\apj},
  \href{https://ui.adsabs.harvard.edu/abs/1995ApJ...448..832N}{448, 832}

\bibitem[{{Olech} {et~al.}(2020){Olech}, {Szymczak}, {Wolak}, {G{\'e}rard}, \&
  {Bartkiewicz}}]{2020A&A...634A..41O}
{Olech}, M., {Szymczak}, M., {Wolak}, P., {G{\'e}rard}, E., \& {Bartkiewicz},
  A. 2020,
  \href{http://dx.doi.org/10.1051/0004-6361/201936943}{\color{magenta}\aap},
  \href{https://ui.adsabs.harvard.edu/abs/2020A&A...634A..41O}{634, A41}

\bibitem[{{Omont} {et~al.}(2003){Omont}, {Gilmore}, {Alard}, {Aracil},
  {August}, {Baliyan}, {Beaulieu}, {B{\'e}gon}, {Bertou}, {Blommaert},
  {Borsenberger}, {Burgdorf}, {Caillaud}, {Cesarsky}, {Chitre}, {Copet}, {de
  Batz}, {Egan}, {Egret}, {Epchtein}, {Felli}, {Fouqu{\'e}}, {Ganesh},
  {Genzel}, {Glass}, {Gredel}, {Groenewegen}, {Guglielmo}, {Habing},
  {Hennebelle}, {Jiang}, {Joshi}, {Kimeswenger}, {Messineo},
  {Miville-Desch{\^e}nes}, {Moneti}, {Morris}, {Ojha}, {Ortiz}, {Ott},
  {Parthasarathy}, {P{\'e}rault}, {Price}, {Robin}, {Schultheis}, {Schuller},
  {Simon}, {Soive}, {Testi}, {Teyssier}, {Tiph{\`e}ne}, {Unavane}, {van Loon},
  \& {Wyse}}]{2003AA...403..975O}
{Omont}, A., {Gilmore}, G.~F., {Alard}, C., {et~al.} 2003,
  \href{http://dx.doi.org/10.1051/0004-6361:20030437}{\color{magenta}\aap},
  \href{https://ui.adsabs.harvard.edu/abs/2003A&A...403..975O}{403, 975}

\bibitem[{{Ostrovskii} \& {Sobolev}(2002)}]{2002IAUS..206..183O}
{Ostrovskii}, A.~B. \& {Sobolev}, A.~M. 2002, in IAU Symposium, Vol. 206,
  Cosmic Masers: From Proto-Stars to Black Holes, ed. V.~{Migenes} \& M.~J.
  {Reid}, \href{https://ui.adsabs.harvard.edu/abs/2002IAUS..206..183O}{183}

\bibitem[{{Parsons} {et~al.}(2018){Parsons}, {Dempsey}, {Thomas}, {Berry},
  {Currie}, {Friberg}, {Wouterloot}, {Chrysostomou}, {Graves}, {Tilanus},
  {Bell}, \& {Rawlings}}]{2018ApJS..234...22P}
{Parsons}, H., {Dempsey}, J.~T., {Thomas}, H.~S., {et~al.} 2018,
  \href{http://dx.doi.org/10.3847/1538-4365/aa989c}{\color{magenta}\apjs},
  \href{https://ui.adsabs.harvard.edu/abs/2018ApJS..234...22P}{234, 22}

\bibitem[{{Pomohaci} {et~al.}(2019){Pomohaci}, {Oudmaijer}, \&
  {Goodwin}}]{2019MNRAS.484..226P}
{Pomohaci}, R., {Oudmaijer}, R.~D., \& {Goodwin}, S.~P. 2019,
  \href{http://dx.doi.org/10.1093/mnras/stz014}{\color{magenta}\mnras},
  \href{https://ui.adsabs.harvard.edu/abs/2019MNRAS.484..226P}{484, 226}

\bibitem[{{Pomohaci} {et~al.}(2017){Pomohaci}, {Oudmaijer}, {Lumsden}, {Hoare},
  \& {Mendigut{\'\i}a}}]{2017MNRAS.472.3624P}
{Pomohaci}, R., {Oudmaijer}, R.~D., {Lumsden}, S.~L., {Hoare}, M.~G., \&
  {Mendigut{\'\i}a}, I. 2017,
  \href{http://dx.doi.org/10.1093/mnras/stx2196}{\color{magenta}\mnras},
  \href{https://ui.adsabs.harvard.edu/abs/2017MNRAS.472.3624P}{472, 3624}

\bibitem[{{Povich} {et~al.}(2011){Povich}, {Smith}, {Majewski}, {Getman},
  {Townsley}, {Babler}, {Broos}, {Indebetouw}, {Meade}, {Robitaille},
  {Stassun}, {Whitney}, {Yonekura}, \& {Fukui}}]{2011ApJS..194...14P}
{Povich}, M.~S., {Smith}, N., {Majewski}, S.~R., {et~al.} 2011,
  \href{http://dx.doi.org/10.1088/0067-0049/194/1/14}{\color{magenta}\apjs},
  \href{https://ui.adsabs.harvard.edu/abs/2011ApJS..194...14P}{194, 14}

\bibitem[{{Pringle}(1981)}]{1981ARA&A..19..137P}
{Pringle}, J.~E. 1981,
  \href{http://dx.doi.org/10.1146/annurev.aa.19.090181.001033}{\color{magenta}\araa},
  \href{https://ui.adsabs.harvard.edu/abs/1981ARA&A..19..137P}{19, 137}

\bibitem[{{Proven-Adzri} {et~al.}(2019){Proven-Adzri}, {MacLeod}, {Heever},
  {Hoare}, {Kuditcher}, \& {Goedhart}}]{2019MNRAS.487.2407P}
{Proven-Adzri}, E., {MacLeod}, G.~C., {Heever}, S.~P. v.~d., {et~al.} 2019,
  \href{http://dx.doi.org/10.1093/mnras/stz1458}{\color{magenta}\mnras},
  \href{https://ui.adsabs.harvard.edu/abs/2019MNRAS.487.2407P}{487, 2407}

\bibitem[{{Rab} {et~al.}(2017){Rab}, {Elbakyan}, {Vorobyov}, {G{\"u}del},
  {Dionatos}, {Audard}, {Kamp}, {Thi}, {Woitke}, \&
  {Postel}}]{2017A&A...604A..15R}
{Rab}, C., {Elbakyan}, V., {Vorobyov}, E., {et~al.} 2017,
  \href{http://dx.doi.org/10.1051/0004-6361/201730812}{\color{magenta}\aap},
  \href{https://ui.adsabs.harvard.edu/abs/2017A&A...604A..15R}{604, A15}

\bibitem[{{Ram{\'\i}rez} {et~al.}(2008){Ram{\'\i}rez}, {Arendt}, {Sellgren},
  {Stolovy}, {Cotera}, {Smith}, \& {Yusef-Zadeh}}]{2008ApJS..175..147R}
{Ram{\'\i}rez}, S.~V., {Arendt}, R.~G., {Sellgren}, K., {et~al.} 2008,
  \href{http://dx.doi.org/10.1086/524015}{\color{magenta}\apjs},
  \href{https://ui.adsabs.harvard.edu/abs/2008ApJS..175..147R}{175, 147}

\bibitem[{{Robitaille}(2011)}]{Robitaille2011}
{Robitaille}, T.~P. 2011,
  \href{http://dx.doi.org/10.1051/0004-6361/201117150}{\color{magenta}\aap},
  \href{https://ui.adsabs.harvard.edu/abs/2011A&A...536A..79R}{536, A79}

\bibitem[{{Robitaille}(2017)}]{2017A&A...600A..11R}
{Robitaille}, T.~P. 2017,
  \href{http://dx.doi.org/10.1051/0004-6361/201425486}{\color{magenta}\aap},
  \href{https://ui.adsabs.harvard.edu/abs/2017A&A...600A..11R}{600, A11}

\bibitem[{{Robitaille} {et~al.}(2007){Robitaille}, {Whitney}, {Indebetouw}, \&
  {Wood}}]{2007ApJS..169..328R}
{Robitaille}, T.~P., {Whitney}, B.~A., {Indebetouw}, R., \& {Wood}, K. 2007,
  \href{http://dx.doi.org/10.1086/512039}{\color{magenta}\apjs},
  \href{https://ui.adsabs.harvard.edu/abs/2007ApJS..169..328R}{169, 328}

\bibitem[{{Robitaille} {et~al.}(2006){Robitaille}, {Whitney}, {Indebetouw},
  {Wood}, \& {Denzmore}}]{2006ApJS..167..256R}
{Robitaille}, T.~P., {Whitney}, B.~A., {Indebetouw}, R., {Wood}, K., \&
  {Denzmore}, P. 2006,
  \href{http://dx.doi.org/10.1086/508424}{\color{magenta}\apjs},
  \href{https://ui.adsabs.harvard.edu/abs/2006ApJS..167..256R}{167, 256}

\bibitem[{{S{\'a}nchez-Portal} {et~al.}(2014){S{\'a}nchez-Portal}, {Marston},
  {Altieri}, {Aussel}, {Feuchtgruber}, {Klaas}, {Linz}, {Lutz}, {Mer{\'\i}n},
  {M{\"u}ller}, {Nielbock}, {Oort}, {Pilbratt}, {Schmidt}, {Stephenson}, \&
  {Tuttlebee}}]{2014ExA....37..453S}
{S{\'a}nchez-Portal}, M., {Marston}, A., {Altieri}, B., {et~al.} 2014,
  \href{http://dx.doi.org/10.1007/s10686-014-9396-z}{\color{magenta}Experimental
  Astronomy}, \href{https://ui.adsabs.harvard.edu/abs/2014ExA....37..453S}{37,
  453}

\bibitem[{{Schlafly} \& {Finkbeiner}(2011)}]{2011ApJ...737..103S}
{Schlafly}, E.~F. \& {Finkbeiner}, D.~P. 2011,
  \href{http://dx.doi.org/10.1088/0004-637X/737/2/103}{\color{magenta}\apj},
  \href{https://ui.adsabs.harvard.edu/abs/2011ApJ...737..103S}{737, 103}

\bibitem[{{Schuller} {et~al.}(2009){Schuller}, {Menten}, {Contreras},
  {Wyrowski}, {Schilke}, {Bronfman}, {Henning}, {Walmsley}, {Beuther},
  {Bontemps}, {Cesaroni}, {Deharveng}, {Garay}, {Herpin}, {Lefloch}, {Linz},
  {Mardones}, {Minier}, {Molinari}, {Motte}, {Nyman}, {Reveret}, {Risacher},
  {Russeil}, {Schneider}, {Testi}, {Troost}, {Vasyunina}, {Wienen}, {Zavagno},
  {Kovacs}, {Kreysa}, {Siringo}, \& {Wei{\ss}}}]{2009AA...504..415S}
{Schuller}, F., {Menten}, K.~M., {Contreras}, Y., {et~al.} 2009,
  \href{http://dx.doi.org/10.1051/0004-6361/200811568}{\color{magenta}\aap},
  \href{https://ui.adsabs.harvard.edu/abs/2009A&A...504..415S}{504, 415}

\bibitem[{{Sobolev} {et~al.}(1997){Sobolev}, {Cragg}, \&
  {Godfrey}}]{1997A&A...324..211S}
{Sobolev}, A.~M., {Cragg}, D.~M., \& {Godfrey}, P.~D. 1997, \aap,
  \href{https://ui.adsabs.harvard.edu/abs/1997A&A...324..211S}{324, 211}

\bibitem[{{Spitzer Science}(2009)}]{2009yCat.2293....0S}
{Spitzer Science}, C. 2009,
  \href{https://ui.adsabs.harvard.edu/abs/2009yCat.2293....0S}{VizieR Online
  Data Catalog, II/293}

\bibitem[{{Stecklum} {et~al.}(2002){Stecklum}, {Brandl}, {Henning}, {Pascucci},
  {Hayward}, \& {Wilson}}]{2002A&A...392.1025S}
{Stecklum}, B., {Brandl}, B., {Henning}, T., {et~al.} 2002,
  \href{http://dx.doi.org/10.1051/0004-6361:20020992}{\color{magenta}\aap},
  \href{https://ui.adsabs.harvard.edu/abs/2002A&A...392.1025S}{392, 1025}

\bibitem[{{Stecklum} {et~al.}(2018{\natexlab{a}}){Stecklum}, {Caratti o
  Garatti}, {Cardenas}, {Cesaroni}, {de Wit}, {Garcia Lopez}, \&
  {Eislöffel}}]{hansfest}
{Stecklum}, B., {Caratti o Garatti}, A., {Cardenas}, M.\, C., {et~al.}
  2018{\natexlab{a}},
  \href{https://www-archive.ph.ed.ac.uk/star-formation/sites/default/files/attachments/stecklum_update.pdf}{in
  The wonders of star formation, online proceedings},
  \url{https://www-archive.ph.ed.ac.uk/star-formation/sites/default/files/attachments/stecklum_update.pdf}

\bibitem[{{Stecklum} {et~al.}(2016){Stecklum}, {Caratti o Garatti}, {Cardenas},
  {Greiner}, {Kruehler}, {Klose}, \& {Eisloeffel}}]{2016ATel.8732....1S}
{Stecklum}, B., {Caratti o Garatti}, A., {Cardenas}, M.~C., {et~al.} 2016, The
  Astronomer's Telegram,
  \href{https://ui.adsabs.harvard.edu/abs/2016ATel.8732....1S}{8732, 1}

\bibitem[{{Stecklum} {et~al.}(2018{\natexlab{b}}){Stecklum}, {Caratti o
  Garatti}, {Hodapp}, {Linz}, {Moscadelli}, \& {Sanna}}]{2018IAUS..336...37S}
{Stecklum}, B., {Caratti o Garatti}, A., {Hodapp}, K., {et~al.}
  2018{\natexlab{b}}, in Astrophysical Masers: Unlocking the Mysteries of the
  Universe, ed. A.~{Tarchi}, M.~J. {Reid}, \& P.~{Castangia}, Vol. 336,
  \href{https://ui.adsabs.harvard.edu/abs/2018IAUS..336...37S}{37--40}

\bibitem[{{Sturm} {et~al.}(2013){Sturm}, {Bouwman}, {Henning}, {Evans},
  {Waters}, {van Dishoeck}, {Green}, {Olofsson}, {Meeus}, {Maaskant},
  {Dominik}, {Augereau}, {Mulders}, {Acke}, {Merin}, \&
  {Herczeg}}]{2013A&A...553A...5S}
{Sturm}, B., {Bouwman}, J., {Henning}, T., {et~al.} 2013,
  \href{http://dx.doi.org/10.1051/0004-6361/201220243}{\color{magenta}\aap},
  \href{https://ui.adsabs.harvard.edu/abs/2013A&A...553A...5S}{553, A5}

\bibitem[{{Sugiyama} {et~al.}(2019){Sugiyama}, {Saito}, {Yonekura}, \&
  {Momose}}]{2019ATel12446....1S}
{Sugiyama}, K., {Saito}, Y., {Yonekura}, Y., \& {Momose}, M. 2019, The
  Astronomer's Telegram,
  \href{https://ui.adsabs.harvard.edu/abs/2019ATel12446....1S}{12446, 1}

\bibitem[{{Szymczak} {et~al.}(2018{\natexlab{a}}){Szymczak}, {Olech},
  {Sarniak}, {Wolak}, \& {Bartkiewicz}}]{2018MNRAS.474..219S}
{Szymczak}, M., {Olech}, M., {Sarniak}, R., {Wolak}, P., \& {Bartkiewicz}, A.
  2018{\natexlab{a}},
  \href{http://dx.doi.org/10.1093/mnras/stx2693}{\color{magenta}\mnras},
  \href{https://ui.adsabs.harvard.edu/abs/2018MNRAS.474..219S}{474, 219}

\bibitem[{{Szymczak} {et~al.}(2018{\natexlab{b}}){Szymczak}, {Olech}, {Wolak},
  {G{\'e}rard}, \& {Bartkiewicz}}]{2018A&A...617A..80S}
{Szymczak}, M., {Olech}, M., {Wolak}, P., {G{\'e}rard}, E., \& {Bartkiewicz},
  A. 2018{\natexlab{b}},
  \href{http://dx.doi.org/10.1051/0004-6361/201833443}{\color{magenta}\aap},
  \href{https://ui.adsabs.harvard.edu/abs/2018A&A...617A..80S}{617, A80}

\bibitem[{{Tabatabaei} {et~al.}(2014){Tabatabaei}, {Braine}, {Xilouris},
  {Kramer}, {Boquien}, {Combes}, {Henkel}, {Relano}, {Verley}, {Gratier},
  {Israel}, {Wiedner}, {R{\"o}llig}, {Schuster}, \& {van der
  Werf}}]{2014A&A...561A..95T}
{Tabatabaei}, F.~S., {Braine}, J., {Xilouris}, E.~M., {et~al.} 2014,
  \href{http://dx.doi.org/10.1051/0004-6361/201321441}{\color{magenta}\aap},
  \href{https://ui.adsabs.harvard.edu/abs/2014A&A...561A..95T}{561, A95}

\bibitem[{{Tapia} {et~al.}(2015){Tapia}, {Roth}, \&
  {Persi}}]{2015MNRAS.446.4088T}
{Tapia}, M., {Roth}, M., \& {Persi}, P. 2015,
  \href{http://dx.doi.org/10.1093/mnras/stu2362}{\color{magenta}\mnras},
  \href{https://ui.adsabs.harvard.edu/abs/2015MNRAS.446.4088T}{446, 4088}

\bibitem[{{Tout} {et~al.}(1996){Tout}, {Pols}, {Eggleton}, \&
  {Han}}]{1996MNRAS.281..257T}
{Tout}, C.~A., {Pols}, O.~R., {Eggleton}, P.~P., \& {Han}, Z. 1996,
  \href{http://dx.doi.org/10.1093/mnras/281.1.257}{\color{magenta}\mnras},
  \href{https://ui.adsabs.harvard.edu/abs/1996MNRAS.281..257T}{281, 257}

\bibitem[{{Ulrich}(1976)}]{1976ApJ...210..377U}
{Ulrich}, R.~K. 1976,
  \href{http://dx.doi.org/10.1086/154840}{\color{magenta}\apj},
  \href{https://ui.adsabs.harvard.edu/abs/1976ApJ...210..377U}{210, 377}

\bibitem[{{Vorobyov} \& {Basu}(2015)}]{2015ApJ...805..115V}
{Vorobyov}, E.~I. \& {Basu}, S. 2015,
  \href{http://dx.doi.org/10.1088/0004-637X/805/2/115}{\color{magenta}\apj},
  \href{https://ui.adsabs.harvard.edu/abs/2015ApJ...805..115V}{805, 115}

\bibitem[{{Weingartner} \& {Draine}(2001)}]{2001ApJ...548..296W}
{Weingartner}, J.~C. \& {Draine}, B.~T. 2001,
  \href{http://dx.doi.org/10.1086/318651}{\color{magenta}\apj},
  \href{https://ui.adsabs.harvard.edu/abs/2001ApJ...548..296W}{548, 296}

\bibitem[{{Wenger} {et~al.}(2000){Wenger}, {Ochsenbein}, {Egret}, {Dubois},
  {Bonnarel}, {Borde}, {Genova}, {Jasniewicz}, {Lalo{\"e}}, {Lesteven}, \&
  {Monier}}]{2000A&AS..143....9W}
{Wenger}, M., {Ochsenbein}, F., {Egret}, D., {et~al.} 2000,
  \href{http://dx.doi.org/10.1051/aas:2000332}{\color{magenta}\aaps},
  \href{https://ui.adsabs.harvard.edu/abs/2000A&AS..143....9W}{143, 9}

\bibitem[{{Whitney} {et~al.}(2013){Whitney}, {Robitaille}, {Bjorkman}, {Dong},
  {Wolff}, {Wood}, \& {Honor}}]{2013ApJS..207...30W}
{Whitney}, B.~A., {Robitaille}, T.~P., {Bjorkman}, J.~E., {et~al.} 2013,
  \href{http://dx.doi.org/10.1088/0067-0049/207/2/30}{\color{magenta}\apjs},
  \href{https://ui.adsabs.harvard.edu/abs/2013ApJS..207...30W}{207, 30}

\bibitem[{{Wiebe} {et~al.}(2019){Wiebe}, {Molyarova}, {Akimkin}, {Vorobyov}, \&
  {Semenov}}]{2019MNRAS.485.1843W}
{Wiebe}, D.~S., {Molyarova}, T.~S., {Akimkin}, V.~V., {Vorobyov}, E.~I., \&
  {Semenov}, D.~A. 2019,
  \href{http://dx.doi.org/10.1093/mnras/stz512}{\color{magenta}\mnras},
  \href{https://ui.adsabs.harvard.edu/abs/2019MNRAS.485.1843W}{485, 1843}

\bibitem[{{Wolf} {et~al.}(2012){Wolf}, {Henning}, \&
  {Stecklum}}]{2012ascl.soft04005W}
{Wolf}, S., {Henning}, T., \& {Stecklum}, B. 2012, {MC3D: Monte-Carlo 3D
  Radiative Transfer Code}

\bibitem[{{Wright} {et~al.}(2010){Wright}, {Eisenhardt}, {Mainzer}, {Ressler},
  {Cutri}, {Jarrett}, {Kirkpatrick}, {Padgett}, {McMillan}, {Skrutskie},
  {Stanford}, {Cohen}, {Walker}, {Mather}, {Leisawitz}, {Gautier}, {McLean},
  {Benford}, {Lonsdale}, {Blain}, {Mendez}, {Irace}, {Duval}, {Liu}, {Royer},
  {Heinrichsen}, {Howard}, {Shannon}, {Kendall}, {Walsh}, {Larsen}, {Cardon},
  {Schick}, {Schwalm}, {Abid}, {Fabinsky}, {Naes}, \&
  {Tsai}}]{2010AJ....140.1868W}
{Wright}, E.~L., {Eisenhardt}, P. R.~M., {Mainzer}, A.~K., {et~al.} 2010,
  \href{http://dx.doi.org/10.1088/0004-6256/140/6/1868}{\color{magenta}\aj},
  \href{https://ui.adsabs.harvard.edu/abs/2010AJ....140.1868W}{140, 1868}

\bibitem[{{Yamamura} {et~al.}(2010){Yamamura}, {Makiuti}, {Ikeda}, {Fukuda},
  {Oyabu}, {Koga}, \& {White}}]{2010yCat.2298....0Y}
{Yamamura}, I., {Makiuti}, S., {Ikeda}, N., {et~al.} 2010,
  \href{https://ui.adsabs.harvard.edu/abs/2010yCat.2298....0Y}{VizieR Online
  Data Catalog, II/298}

\bibitem[{{Young} {et~al.}(2012){Young}, {Becklin}, {Marcum}, {Roellig}, {De
  Buizer}, {Herter}, {G{\"u}sten}, {Dunham}, {Temi}, {Andersson}, {Backman},
  {Burgdorf}, {Caroff}, {Casey}, {Davidson}, {Erickson}, {Gehrz}, {Harper},
  {Harvey}, {Helton}, {Horner}, {Howard}, {Klein}, {Krabbe}, {McLean}, {Meyer},
  {Miles}, {Morris}, {Reach}, {Rho}, {Richter}, {Roeser}, {Sandell}, {Sankrit},
  {Savage}, {Smith}, {Shuping}, {Vacca}, {Vaillancourt}, {Wolf}, \&
  {Zinnecker}}]{2012ApJ...749L..17Y}
{Young}, E.~T., {Becklin}, E.~E., {Marcum}, P.~M., {et~al.} 2012,
  \href{http://dx.doi.org/10.1088/2041-8205/749/2/L17}{\color{magenta}\apjl},
  \href{https://ui.adsabs.harvard.edu/abs/2012ApJ...749L..17Y}{749, L17}

\bibitem[{{Zinnecker} \& {Yorke}(2007)}]{2007ARA&A..45..481Z}
{Zinnecker}, H. \& {Yorke}, H.~W. 2007,
  \href{http://dx.doi.org/10.1146/annurev.astro.44.051905.092549}{\color{magenta}\araa},
  \href{https://ui.adsabs.harvard.edu/abs/2007ARA&A..45..481Z}{45, 481}

\end{thebibliography}


\begin{appendix}
\section{Additional information}\label{app}
\begin{table}
\caption[]{MM3-SED. The fluxes at $\lambda\,{\ge}\,889\,\mu$m, 
are MM3-fluxes, whereas the fluxes at $\lambda\,{\le}\,24\,\mu$m are total fluxes. The contribution of all the other sources, which are less evolved than MM3 at NIR wavelengths, can be neglected (as indicated in Fig. \ref{fig:sed g358 pre}). For the fit of MM3 (see Sec. \ref{drt3}), we use the total fluxes at short wavelengths. Only in the FIR and (sub)mm regimes, where the contribution of the other sources is not negligible, we use the MM3-fluxes. Facilities and instruments are given behind corresponding references.}
\label{tab: MM3 SED}
\begin{tabular*}{8.5cm}{@{\hspace{1.25cm}}ccc}
\hline\hline
\noalign{\smallskip}
Wavelength & Flux & Ref.\\
$[\mu$m] & ${[\rm erg \, cm^{-2} \, s^{-1}}]$& \\
\noalign{\smallskip}
\hline
\noalign{\smallskip}
$ ~~~~1.63 $ & $1.16 \pm 0.07\,{\times}\,10^{-12}$  & 1 \\
$ ~~~~2.13 $ & $1.67 \pm 0.02\,{\times}\,10^{-11}$  & 1 \\
$ ~~~~3.55 $ & $3.64 \pm 0.02\,{\times}\,10^{-10}$  & 2 \\
$ ~~~~4.49 $ & $6.10 \pm 0.03\,{\times}\,10^{-10}$  & 2 \\
$ ~~~~5.73 $ & $9.73 \pm 0.05\,{\times}\,10^{-10}$  & 2 \\
$ ~~~7.0~ $ & $6.90 \pm 0.49\,{\times}\,10^{-10}$  & 3 \\
$ ~~~~7.87 $ & $5.30 \pm 0.04\,{\times}\,10^{-10}$  & 2 \\
$ 11.6 $ & $ 3.54 \pm 0.07\,{\times}\,10^{-10}$  & 4 \\
$ 15.0 $ & $ 6.46 \pm 0.20\,{\times}\,10^{-10}$  & 3 \\
$ 22.1 $ & $ 5.13 \pm 0.13\,{\times}\,10^{-10}$  & 4 \\
$ 23.7 $ & $ 4.49 \pm 0.13\,{\times}\,10^{-10}$  & 5 \\
$ 889~~~~ $ & $ 1.34 \pm 0.04\,{\times}\,10^{-13}$  & 6 \\
$ 1282~~~~~ $ & $ 2.43 \pm 0.10\,{\times}\,10^{-14}$  & 6 \\
$ 1420~~~~~ $ & $ 1.50 \pm 0.19\,{\times}\,10^{-14}$  & 6 \\
$ 1532~~~~~ $ & $ 1.06 \pm 0.08\,{\times}\,10^{-14}$  & 6 \\
\noalign{\smallskip}
\hline
\end{tabular*}
\tablebib{(1)~\cite{2017yCat.2348....0M}  (VISTA/VIRCAM); (2) \cite{2008ApJS..175..147R} ({\em Spitzer}/IRAC)}; (3) \cite{2003AA...403..975O}  (ISO/ISOCAM); (4) \cite{2014yCat.2328....0C} (ALLWISE); (5) \cite{2015AJ....149...64G} ({\em Spitzer}/MIPS); (6) \cite{2019ApJ...881L..39B}  (ALMA, SMA)

\end{table}

\begin{table}
\caption[]{Total- and MM1-pre-burst fluxes. We obtained the MM1 flux densities used for the SED-fit in Sec. \ref{drt1} by removing the contribution from all other sources 
according to Sec. \ref{sedc}. Facilities and instruments are given behind corresponding references.}
\label{tab: preSED}
\begin{tabular}{p{1.25cm} ccc}
\hline\hline
\noalign{\smallskip}
Wavelength & Total flux & Ref. & MM1 flux \\
$[\mu$m] & $[{\rm erg \, cm^{-2} \, s^{-1}}]$ & &$[{\rm erg \, cm^{-2}\, s^{-1}}]$ \\
\noalign{\smallskip}
\hline
\noalign{\smallskip}
~~~~~$2.15$\tablefootmark{*} & $5.86 \pm 0.59\,{\times}\,10^{-13}$& 1 & \\
~~$24$\tablefootmark{*} & $5.25\,{\pm} 0.53\,{\times}\,10^{-11}$  & 1 & \\
~~$65$ & $3.10 \pm 0.06\,{\times}\,10^{-9} $  & 1 & $1.55 \pm 0.03\,{\times}\,10^{-9}$ \\
~~$70$ & $3.13 \pm 0.38\,{\times}\,10^{-9} $  & 1 & $1.57 \pm 0.06\,{\times}\,10^{-9}$ \\
$160$ & $1.90 \pm 0.21\,{\times}\,10^{-9} $  & 1 & $9.50 \pm 0.52\,{\times}\,10^{-10}$ \\
$250$ & $7.82 \pm 1.04\,{\times}\,10^{-10} $  & 1 & $3.91 \pm 0.27\,{\times}\,10^{-10}$ \\
$350$ & $2.34 \pm 0.52\,{\times}\,10^{-10} $  & 1 & $1.17 \pm 0.26\,{\times}\,10^{-10}$ \\
$500$ & $4.98 \pm 2.82\,{\times}\,10^{-11} $  & 1 & $2.49 \pm 0.33\,{\times}\,10^{-11}$ \\
$850$ & $ 4.73\pm 0.15\,{\times}\,10^{-12} $  & 2 & $2.36 \pm 0.08\,{\times}\,10^{-12}$\\ 
$870$ & $4.03 \pm 0.07\,{\times}\,10^{-12} $  & 3 & $2.00 \pm 0.35\,{\times}\,10^{-12}$ \\
\noalign{\smallskip}
\hline
\end{tabular}
\tablebib{(1)~present paper (VISTA/VIRCAM, {\em Spitzer}/MIPS, SOFIA/FIFI-LS); (2) \cite{2018ApJS..234...22P}  (JCMT/SCUBA-2); (3) \citet{2019ApJ...881L..39B} (APEX/LABOCA)
}
\tablefoot{
\tablefoottext{*}{Upper limit}
}

\end{table}

\begin{table}
\caption[]{Total- and MM1-burst fluxes: Similar to Table \ref{tab: preSED} but for the burst epoch. We assume that during the burst only the luminosity of MM1 increased while all other sources remained constant. We note that we do not provide MM1 fluxes for $\lambda>890\mu m$, since the data points at theses wavelengths are discarded for the burst SED fit as discussed in Sec. \ref{misf}.  Facilities and instruments are given behind corresponding references.}
\label{tab: burstSED}
\begin{tabular}{cccc}
\hline\hline
\noalign{\smallskip}
wavelength & total flux & Ref. & MM1 flux \\
$[\mu$m] & $[{\rm erg \, cm^{-2} \, s^{-1}}]$ & & $[{\rm erg \, cm^{-2}\, s^{-1}}]$ \\
\noalign{\smallskip}
\hline
\noalign{\smallskip}
$ ~~~2.15 $\tablefootmark{*}& $ 9.20 \pm 0.92\,{\times}\,10^{-13}$  & 1 &\\
$ ~~3.4~ $\tablefootmark{*}& $ 8.82 \pm 0.89 \,{\times}\,10^{-11}$  & 1 &\\ 
$ ~~4.6~ $\tablefootmark{*}& $ 2.93 \pm 0.30\,{\times}\,10^{-10}$  & 1 &\\
$ 52.0 $& $ 5.51 \pm 0.56\,{\times}\,10^{-9}$  & 1 & $ 4.85 \pm 0.49\,{\times}\,10^{-9}$ \\
$ 54.8 $& $ 6.94 \pm 0.70\,{\times}\,10^{-9}$  & 1 & $ 6.12 \pm 0.62\,{\times}\,10^{-9}$ \\
$ 60.7 $& $ 6.20 \pm 0.62\,{\times}\,10^{-9}$  & 1 & $ 5.04 \pm 0.51\,{\times}\,10^{-9}$ \\ 
$ 87.2 $ & $ 7.48 \pm 0.75\,{\times}\,10^{-9}$  & 1 & $ 5.35 \pm 0.54\,{\times}\,10^{-9}$ \\ 
$ 118.6~ $ & $ 6.45 \pm 0.65\,{\times}\,10^{-9}$  & 1& $ 4.59 \pm 0.46\,{\times}\,10^{-9}$ \\ 
$ 124.2~ $& $ 8.97 \pm 0.90 \,{\times}\,10^{-9}$  & 1  & $ 7.22 \pm 0.72\,{\times}\,10^{-9}$ \\ 
$ 142.2~ $  & $ 5.26 \pm 0.53\,{\times}\,10^{-9}$  & 1 & $ 5.25 \pm 0.53\,{\times}\,10^{-9}$ \\ 
$ 153.3~ $ & $ 5.29 \pm 0.53\,{\times}\,10^{-9}$  & 1 & $ 4.08 \pm 0.41\,{\times}\,10^{-9}$ \\ 
$ 162.8~ $  & $ 5.47 \pm 0.55\,{\times}\,10^{-9}$  & 1& $ 4.41 \pm 0.45\,{\times}\,10^{-9}$ \\ 
$ 186.4~ $& $ 4.58 \pm 0.46 \,{\times}\,10^{-9}$  & 1 & $ 3.82 \pm 0.39\,{\times}\,10^{-9}$ \\ 
$ 889~~~ $ & $ 3.81 \pm 0.11\,{\times}\,10^{-12}$  & 2 & $ 1.72 \pm 0.07\,{\times}\,10^{-12}$ \\ 
$ 1282~~~~ $ & $ 4.21 \pm 0.03\,{\times}\,10^{-13}$  & 2 & \\ 
$ 1420~~~~ $ & $ 2.74\pm 0.28\,{\times}\,10^{-13}$  & 2 & \\ 
$ 1532~~~~ $ & $ 1.88\pm 0.01\,{\times}\,10^{-13}$  & 2 & \\ 
\noalign{\smallskip}
\hline
\end{tabular}
\tablebib{ (1)~present paper  (2.2-m MPG/ESO telescope/GROND, NEOWISE, SOFIA/FIFI-LS); (2) \citet{2019ApJ...881L..39B} (ALMA, SMA)
}
\tablefoot{
\tablefoottext{*}{Upper limit}
}
\end{table}

\begin{table}
\caption[]{Total- and MM1-post-burst fluxes: Similar to Table \ref{tab: preSED} but for the post-burst. We assume that only the luminosity of MM1 has changed while all other sources remained constant. Facilities and instruments are given behind corresponding references.} 
\label{tab: postSED}
\begin{tabular}{cccc}
\hline\hline
\noalign{\smallskip}
wavelength & total flux & Ref.  & MM1 flux  \\
$[\mu$m] & $[{\rm erg \, cm^{-2} \, s^{-1}}]$ & & $[{\rm erg \, cm^{-2}\, s^{-1}}]$ \\
\noalign{\smallskip}
\hline
\noalign{\smallskip}
$ 118.6 $ & $ 5.17 \pm 0.52\,{\times}\,10^{-9}$  & 1& $ 3.55 \pm 0.36\,{\times}\,10^{-9}$ \\ 
$ 124.2 $& $ 7.70 \pm 0.77\,{\times}\,10^{-9}$  & 1  & $ 5.55 \pm 0.56\,{\times}\,10^{-9}$ \\ 
$ 142.2 $  & $ 4.74 \pm 0.48\,{\times}\,10^{-9}$  & 1 & $ 3.28 \pm 0.33\,{\times}\,10^{-9}$ \\ 
$ 153.3 $ & $ 4.35 \pm 0.44\,{\times}\,10^{-9}$  & 1 & $ 3.09 \pm 0.31\,{\times}\,10^{-9}$ \\ 
$ 162.8 $  & $ 3.01 \pm 0.31\,{\times}\,10^{-9}$  & 1& $ 1.90 \pm 0.19\,{\times}\,10^{-9}$ \\ 
$ 186.4 $& $ 2.53 \pm 0.26 \,{\times}\,10^{-9}$  & 1 & $ 1.73 \pm 0.18\,{\times}\,10^{-9}$ \\ 
$ 889~~ $ & $ 3.81 \pm 0.11\,{\times}\,10^{-12}$  & 2 & $ 1.72 \pm 0.07\,{\times}\,10^{-12}$ \\ 
\noalign{\smallskip}
\hline
\end{tabular}
\tablebib{ (1)~present paper (SOFIA/FIFI-LS); (2) \citet{2019ApJ...881L..39B}  (ALMA)
}
\end{table}

\begin{figure*}
    \sidecaption
	\includegraphics[width=12cm]{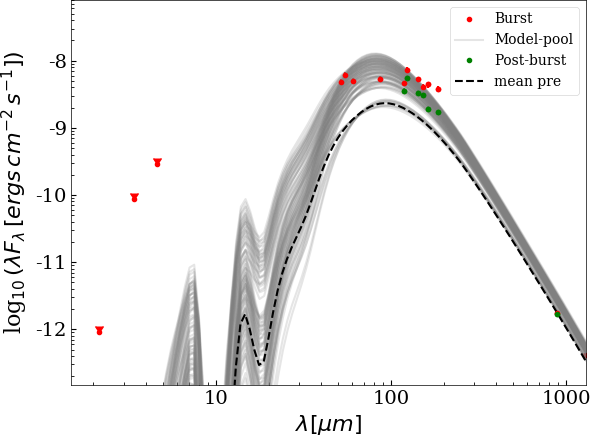}
	\caption{Model pool (in gray) showing all SEDs, used to fit burst and post-burst epochs. The observed fluxes from both epochs are plotted in red (burst) and green (post-burst). They are well covered by the model SEDs (except for the burst data points at $163$ and 186\,$\mu$m, as discussed when introducing the model pool in Sec. \ref{drt1}). The dashed black line indicates the mean model, as obtained from the pre-burst fit alone. The NIR/MIR fluxes of that model have been used to set a lower limit to the post-burst, which was only observed in the red channel of FIFI (at $\lambda\,{\ge}\,118\,\mu$m).} 
 \label{fig:sed g358 modelpool}
\end{figure*}

\begin{figure*}
    \sidecaption
	\includegraphics[width=12cm]{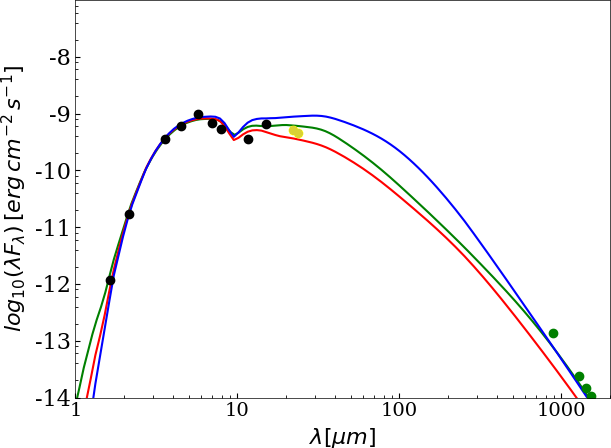}
	\caption{MM3 RT models illustrating constraints on the FIR emission. Filled circles indicate measured fluxes. The model shown in green is the best one (cf. Fig.\,\ref{fig:sed g358 pre}). The best fit with the (sub)mm fluxes (green) omitted is shown in red. When also omitting the WISE W4 and MIPS 24\,$\mu$m  measurements (yellow), the best fit (blue) lies above the nominal one. This confirms that the FIR emission of MM3 and, thus, its contribution to the total FIR fluxes, is well constrained by the WISE, MIPS and ALMA measurements.
	}
 \label{fig:MM3 contri}
\end{figure*}

\clearpage

\begin{sidewaystable}
\vspace*{9cm}
\caption[]{MM3 SED fit results: The table contains the full parameter set 
as well as the derived luminosities (see Sec.\,\ref{rta})
for the ten best models, together wit their respective $\goodchi^2$-values. The weighted mean and standard deviation $\sigma$ is given in the bottom. The confidence interval for log-sampled values extends from $\frac{x}{\sigma}$ to $x \,{\times}\, \sigma$, where for lin-spaced values it is $x \pm \sigma$ as usual. All masses (and densities) are given in unit dust mass, where we use the Weingartner \& Draine (2001) Milky Way grain size distribution A for $R_V=5.5$. 
We note that, some models appear more than ones, but for different inclinations. In our analysis we thread them as different models. The envelope inner radius is set to $r_{disk}^{min}$, its centrifugal radius $r_C$ is set to $r_{disk}^{max}$. All models extends to an outer radius, where density and temperature have dropped to the ambient level. 
}
\label{tab MM3 fit} 
\begin{tabular}{ccccccccccccccccccc} 
\hline\hline
Grid spacing &  & lin & log & log & log & log & log & lin & lin & log & log & lin & lin & log & log & lin & log\\ 
\hline 
Model & $\goodchi^2$ & $\rm A_V$ & d & $\rm R_*$ & $\rm T_*$ & $\rm m_{\rm disk}$  & $\rm r_{\rm disk}^{\rm max}$ & $\beta_{\rm disk}$ & $p_{\rm disk}$ & $h_{100}^{\rm disk}$ & $\rho_0^{\rm env}$ & $p_{\rm cav}$ & $\theta_0^{\rm cav}$ & $\rho_0^{\rm cav}$  & $r_{\rm disk}^{\rm min}$ & inc & $\rm L_*$\\ 
 &  & mag & kpc & ${\rm R}_\odot$ & K & ${\rm M}_\odot$ & au &  &  & au & g/cc &  & $\degr$  & g/cc & $R_{\rm sub}$ & $\degr$ & ${\rm L}_\odot$\\ 
\noalign{\smallskip} 
\hline 
\noalign{\smallskip} 
aer04B9J & $ 5842$ & $ 16.5$ & $ 6.75$ & $ 11.5$ & $ 15720$ & $ 0.0981$ & $ 440$ & $ 1.06$ & $ -0.205$ & $ 9.29$ & $ 7.28\,{\times}\, 10^{-21}$ & $ 1.16$ & $ 40.4$ & $ 1.09\,{\times}\, 10^{-23}$ & $ 3.35$ & $ 60.5$ & $ 7161$ \\ 
aer04B9J & $ 6799$ & $ 17.1$ & $ 6.75$ & $ 11.5$ & $ 15720$ & $ 0.0981$ & $ 440$ & $ 1.06$ & $ -0.205$ & $ 9.29$ & $ 7.28\,{\times}\, 10^{-21}$ & $ 1.16$ & $ 40.4$ & $ 1.09\,{\times}\, 10^{-23}$ & $ 3.35$ & $ 52$ & $ 7161$ \\ 
rlHDFVxN & $ 7251$ & $ 16.6$ & $ 6.07$ & $ 9.32$ & $ 15680$ & $ 0.0321$ & $ 1216$ & $ 1.13$ & $ -1.32$ & $ 11.8$ & $ 9.75\,{\times}\, 10^{-23}$ & $ 1.64$ & $ 32.8$ & $ 8.59\,{\times}\, 10^{-22}$ & $ 3.05$ & $ 59.6$ & $ 4639$ \\ 
aer04B9J & $ 7291$ & $ 17.3$ & $ 7.11$ & $ 11.5$ & $ 15720$ & $ 0.0981$ & $ 440$ & $ 1.06$ & $ -0.205$ & $ 9.29$ & $ 7.28\,{\times}\, 10^{-21}$ & $ 1.16$ & $ 40.4$ & $ 1.09\,{\times}\, 10^{-23}$ & $ 3.35$ & $ 39.5$ & $ 7161$ \\ 
aer04B9J & $ 7694$ & $ 17.5$ & $ 6.75$ & $ 11.5$ & $ 15720$ & $ 0.0981$ & $ 440$ & $ 1.06$ & $ -0.205$ & $ 9.29$ & $ 7.28\,{\times}\, 10^{-21}$ & $ 1.16$ & $ 40.4$ & $ 1.09\,{\times}\, 10^{-23}$ & $ 3.35$ & $ 42.3$ & $ 7161$ \\ 
Ipv9dWMM & $ 9668$ & $ 22.7$ & $ 6.4$ & $ 11.4$ & $ 16700$ & $ 0.0795$ & $ 244$ & $ 1.15$ & $ -0.161$ & $ 13.2$ & $ 2.19\,{\times}\, 10^{-24}$ & $ 1.79$ & $ 11.4$ & $ 1.17\,{\times}\, 10^{-22}$ & $ 1$ & $ 56.7$ & $ 8932$ \\ 
rlHDFVxN & $ 10702$ & $ 16.8$ & $ 6.07$ & $ 9.32$ & $ 15680$ & $ 0.0321$ & $ 1216$ & $ 1.13$ & $ -1.32$ & $ 11.8$ & $ 9.75\,{\times}\, 10^{-23}$ & $ 1.64$ & $ 32.8$ & $ 8.59\,{\times}\, 10^{-22}$ & $ 3.05$ & $ 47.8$ & $ 4639$ \\ 
WFmZWZb5 & $ 10838$ & $ 28.6$ & $ 6.07$ & $ 13.9$ & $ 13460$ & $ 0.0754$ & $ 207$ & $ 1.05$ & $ -0.924$ & $ 13.6$ & $ 2\,{\times}\, 10^{-22}$ & $ 1.23$ & $ 16.9$ & $ 1.53\,{\times}\, 10^{-22}$ & $ 1$ & $ 34.8$ & $ 5571$ \\ 
jZiuuzpK & $ 11033$ & $ 26.3$ & $ 7.11$ & $ 10.9$ & $ 25230$ & $ 0.0653$ & $ 1559$ & $ 1.15$ & $ -1.72$ & $ 4.11$ & $ 7.1\,{\times}\, 10^{-23}$ & $ 1.45$ & $ 14.4$ & $ 2.63\,{\times}\, 10^{-23}$ & $ 1$ & $ 55$ & $ 42775$ \\ 
MR8w9seO & $ 11686$ & $ 28$ & $ 6.07$ & $ 5.01$ & $ 23850$ & $ 0.0321$ & $ 287$ & $ 1.14$ & $ -1.42$ & $ 12.4$ & $ 3.7\,{\times}\, 10^{-22}$ & $ 1.01$ & $ 16.7$ & $ 4.14\,{\times}\, 10^{-21}$ & $ 1$ & $ 51.8$ & $ 7170$ \\ 
\noalign{\smallskip} 
\hline 
mean model \\
\noalign{\smallskip} 
\hline 
\noalign{\smallskip} 
mean & & $ 19.9$ & 6.53 & $ 10.5$ & $ 16672$ & $ 0.0679$ & $ 513$ & $ 1.09$ & $ -0.678$ & $ 9.91$ & $ 6.66\,{\times}\, 10^{-22}$ & $ 1.33$ & $ 30.9$ & $ 6.3\,{\times}\, 10^{-23}$ & $ 2.25$ & $ 50.6$ & $ 7541$ \\ 
sigma & &$ 4.28$ &$ 1.06$ & $ 1.26$ &$ 1.18$ &$ 1.6$ &$ 1.86$ &$ 0.0389$ &$ 0.554$ &$ 1.34$ &$ 13.8$ &$ 0.233$ &$ 10.7$ &$ 7.88$ &$ 1.73$ &$ 7.95$ &$ 1.71$ \\ 
\noalign{\smallskip} 
\hline 
\end{tabular} 
\end{sidewaystable}
\clearpage

\begin{sidewaystable}
\vspace*{9cm}
\caption[]{MM1 SED fit results: The table features the same structure as Table \ref{tab MM3 fit}, but additionally the burst-fit is included. 
The burst names are composed of the original model-name and the adapted luminosity increase. Consequently the weighted mean and $\sigma$ are obtained taking into account both, the pre- and burst-results (see text). Only the parameters, which refers to the source are taken from the pre-burst alone (they change due to the burst). Again, the confidence interval for  log-sampled values extends from $\frac{x}{\sigma}$ to $x \,{\times}\, \sigma$.  
We note that the inner radius of the disk is not a free parameter (unlike for MM3). Instead, it is governed by the sublimation radius (at $T_{\rm sub}\,{=}\,1600$K). An outward shift of the sublimation radius may lower the dust masses for the post-/burst. 
} 
\label{tab MM1 fit} 
\begin{tabular}{ccccccccccccccccc}
\hline\hline
Grid spacing &  & lin & log & log & log & log & log & lin & lin & log & log & lin & lin & log & lin & log\\
\hline 
Model & $\goodchi^2$ & av & d & $\rm R_*$ & $\rm T_*$ & $\rm m_{\rm disk}$ & $\rm r_{\rm disk}^{\rm max}$ & $\beta_{\rm disk}$ & $p_{\rm disk}$ & $h_{\rm 100}^{\rm disk}$ & $\rho_0^{\rm env}$ & $p_{\rm cav}$ & $\theta_0^{\rm cav}$ & $\rho_0^{\rm cav}$ & inc & $\rm L_*$\\ 
 &  & mag & kpc & $R_\odot$ & K & ${\rm M}_\odot$ & au &  &  & au & ${\rm g/cc}$ &  & $\degr$  & ${\rm g/cc}$ & $\degr$ & ${\rm L}_\odot$\\ 
\noalign{\smallskip} 
\hline 
Pre-burst\\
\noalign{\smallskip} 
\hline 
\noalign{\smallskip} 
eGYXcOh8\_02 & $ 42.8$ & $ 67.6$ & $ 7.11$ & $ 3.79$ & $ 25590$ & $ 0.000217$ & $ 3110$ & $ 1.18$ & $ -0.131$ & $ 2.34$ & $ 5.25\,{\times}\,10^{-19}$ & $ 1.56$ & $ 31.5$ & $ 1.81\,{\times}\,10^{-23}$ & $ 19.3$ & $ 5437$ \\ 
90Yt0exl\_03 & $ 46.6$ & $ 65.5$ & $ 6.4$ & $ 5.51$ & $ 19670$ & $ 0.0357$ & $ 1191$ & $ 1.13$ & $ -0.345$ & $ 1.25$ & $ 2.07\,{\times}\,10^{-18}$ & $ 1.39$ & $ 36.4$ & $ 1.1\,{\times}\,10^{-21}$ & $ 24.1$ & $ 4017$ \\ 
PUyhjE8Z\_02 & $ 47.5$ & $ 48.2$ & $ 6.75$ & $ 14.1$ & $ 12130$ & $ 4.01\,{\times}\,10^{-8}$ & $ 278$ & $ 1.28$ & $ -1.46$ & $ 11.4$ & $ 2.07\,{\times}\,10^{-17}$ & $ 1.76$ & $ 23.4$ & $ 2.71\,{\times}\,10^{-22}$ & $ 11.3$ & $ 3797$ \\ 
nTRTrE7X\_02 & $ 51.2$ & $ 59$ & $ 7.11$ & $ 2.71$ & $ 29889$ & $ 8.1\,{\times}\,10^{-6}$ & $ 423$ & $ 1.13$ & $ -0.607$ & $ 1.85$ & $ 9.58\,{\times}\,10^{-18}$ & $ 1.94$ & $ 32.7$ & $ 4.99\,{\times}\,10^{-21}$ & $ 18.1$ & $ 5180$ \\ 
nTRTrE7X\_03 & $ 55$ & $ 47.5$ & $ 6.75$ & $ 2.71$ & $ 29889$ & $ 8.1\,{\times}\,10^{-6}$ & $ 423$ & $ 1.13$ & $ -0.607$ & $ 1.85$ & $ 9.58\,{\times}\,10^{-18}$ & $ 1.94$ & $ 32.7$ & $ 4.99\,{\times}\,10^{-21}$ & $ 23.3$ & $ 5180$ \\ 
TJyUR9bA\_02 & $ 57.8$ & $ 62.6$ & $ 7.11$ & $ 3.91$ & $ 23840$ & $ 3.1\,{\times}\,10^{-7}$ & $ 631$ & $ 1.17$ & $ -0.102$ & $ 12.7$ & $ 7.15\,{\times}\,10^{-18}$ & $ 1.89$ & $ 46$ & $ 9.4\,{\times}\,10^{-23}$ & $ 14.4$ & $ 4366$ \\ 
4kW1TtMH\_04 & $ 59.1$ & $ 41.8$ & $ 6.4$ & $ 47.3$ & $ 7335$ & $ 0.000299$ & $ 140$ & $ 1.24$ & $ -0.752$ & $ 2.49$ & $ 5.14\,{\times}\,10^{-17}$ & $ 1.35$ & $ 32.9$ & $ 8.17\,{\times}\,10^{-22}$ & $ 30.6$ & $ 5715$ \\ 
lQ0YS3aF\_02 & $ 63.2$ & $ 70$ & $ 6.4$ & $ 6.7$ & $ 17980$ & $ 0.00527$ & $ 3798$ & $ 1.06$ & $ -0.775$ & $ 17.3$ & $ 3.24\,{\times}\,10^{-19}$ & $ 1.06$ & $ 12.5$ & $ 8.66\,{\times}\,10^{-22}$ & $ 13.5$ & $ 4150$ \\ 
qWHuujku\_05 & $ 64.4$ & $ 70$ & $ 6.75$ & $ 38$ & $ 8551$ & $ 2.63\,{\times}\,10^{-5}$ & $ 1015$ & $ 1.23$ & $ -1.62$ & $ 1.52$ & $ 2.33\,{\times}\,10^{-18}$ & $ 1.28$ & $ 47.8$ & $ 9.27\,{\times}\,10^{-21}$ & $ 46.8$ & $ 6832$ \\ 
qWHuujku\_06 & $ 65.2$ & $ 36.2$ & $ 6.75$ & $ 38$ & $ 8551$ & $ 2.63\,{\times}\,10^{-5}$ & $ 1015$ & $ 1.23$ & $ -1.62$ & $ 1.52$ & $ 2.33\,{\times}\,10^{-18}$ & $ 1.28$ & $ 47.8$ & $ 9.27\,{\times}\,10^{-21}$ & $ 50$ & $ 6832$ \\ 
\noalign{\smallskip} 
\hline 
Burst\\
\noalign{\smallskip} 
\hline 
\noalign{\smallskip} 
90Yt0exl\_L5.0 & $ 226$ & $ 65.5$ & $ 6.4$ & $ 12.3$ & $ 19670$ & $ 0.0357$ & $ 1191$ & $ 1.13$ & $ -0.345$ & $ 1.25$ & $ 2.07\,{\times}\,10^{-18}$ & $ 1.39$ & $ 36.4$ & $ 1.1\,{\times}\,10^{-21}$ & $ 24.1$ & $ 20078$ \\ 
90Yt0exl\_L5.5 & $ 229$ & $ 65.5$ & $ 6.4$ & $ 12.9$ & $ 19670$ & $ 0.0357$ & $ 1191$ & $ 1.13$ & $ -0.345$ & $ 1.25$ & $ 2.07\,{\times}\,10^{-18}$ & $ 1.39$ & $ 36.4$ & $ 1.1\,{\times}\,10^{-21}$ & $ 24.1$ & $ 22081$ \\ 
eGYXcOh8\_L4.5 & $ 234$ & $ 67.6$ & $ 7.11$ & $ 8.04$ & $ 25590$ & $ 0.000217$ & $ 3110$ & $ 1.18$ & $ -0.131$ & $ 2.34$ & $ 5.25\,{\times}\,10^{-19}$ & $ 1.56$ & $ 31.5$ & $ 1.81\,{\times}\,10^{-23}$ & $ 19.3$ & $ 24470$ \\ 
eGYXcOh8\_L5.0 & $ 234$ & $ 67.6$ & $ 7.11$ & $ 8.47$ & $ 25590$ & $ 0.000217$ & $ 3110$ & $ 1.18$ & $ -0.131$ & $ 2.34$ & $ 5.25\,{\times}\,10^{-19}$ & $ 1.56$ & $ 31.5$ & $ 1.81\,{\times}\,10^{-23}$ & $ 19.3$ & $ 27185$ \\ 
90Yt0exl\_L4.5 & $ 236$ & $ 65.5$ & $ 6.4$ & $ 11.7$ & $ 19670$ & $ 0.0357$ & $ 1191$ & $ 1.13$ & $ -0.345$ & $ 1.25$ & $ 2.07\,{\times}\,10^{-18}$ & $ 1.39$ & $ 36.4$ & $ 1.1\,{\times}\,10^{-21}$ & $ 24.1$ & $ 18077$ \\ 
TJyUR9bA\_L5.0 & $ 236$ & $ 62.6$ & $ 7.11$ & $ 8.74$ & $ 23840$ & $ 3.1\,{\times}\,10^{-7}$ & $ 631$ & $ 1.17$ & $ -0.102$ & $ 12.7$ & $ 7.15\,{\times}\,10^{-18}$ & $ 1.89$ & $ 46$ & $ 9.4\,{\times}\,10^{-23}$ & $ 14.4$ & $ 21828$ \\ 
TJyUR9bA\_L4.5 & $ 243$ & $ 62.6$ & $ 7.11$ & $ 8.3$ & $ 23840$ & $ 3.1\,{\times}\,10^{-7}$ & $ 631$ & $ 1.17$ & $ -0.102$ & $ 12.7$ & $ 7.15\,{\times}\,10^{-18}$ & $ 1.89$ & $ 46$ & $ 9.4\,{\times}\,10^{-23}$ & $ 14.4$ & $ 19644$ \\ 
eGYXcOh8\_L5.5 & $ 244$ & $ 67.6$ & $ 7.11$ & $ 8.88$ & $ 25590$ & $ 0.000217$ & $ 3110$ & $ 1.18$ & $ -0.131$ & $ 2.34$ & $ 5.25\,{\times}\,10^{-19}$ & $ 1.56$ & $ 31.5$ & $ 1.81\,{\times}\,10^{-23}$ & $ 19.3$ & $ 29907$ \\ 
90Yt0exl\_L6.0 & $ 245$ & $ 65.5$ & $ 6.4$ & $ 13.5$ & $ 19670$ & $ 0.0357$ & $ 1191$ & $ 1.13$ & $ -0.345$ & $ 1.25$ & $ 2.07\,{\times}\,10^{-18}$ & $ 1.39$ & $ 36.4$ & $ 1.1\,{\times}\,10^{-21}$ & $ 24.1$ & $ 24108$ \\ 
qWHuujku\_L4.5 & $ 245$ & $ 36.2$ & $ 6.75$ & $ 80.7$ & $ 8551$ & $ 2.63\,{\times}\,10^{-5}$ & $ 1015$ & $ 1.23$ & $ -1.62$ & $ 1.52$ & $ 2.33\,{\times}\,10^{-18}$ & $ 1.28$ & $ 47.8$ & $ 9.27\,{\times}\,10^{-21}$ & $ 50$ & $ 30745$ \\ 
\noalign{\smallskip} 
\hline 
Post-burst\\
\noalign{\smallskip} 
\hline 
\noalign{\smallskip} 
lQ0YS3aF\_L2.5 & $ 92.7$ & $ 70$ & $ 6.4$ & $ 10.6$ & $ 17980$ & $ 0.00527$ & $ 3798$ & $ 1.06$ & $ -0.775$ & $ 17.3$ & $ 3.24\,{\times}\,10^{-19}$ & $ 1.06$ & $ 12.5$ & $ 8.66\,{\times}\,10^{-22}$ & $ 13.5$ & $ 10376$ \\ 
nTRTrE7X\_L2.5 & $ 103$ & $ 59$ & $ 7.11$ & $ 4.29$ & $ 29889$ & $ 8.1\,{\times}\,10^{-6}$ & $ 423$ & $ 1.13$ & $ -0.607$ & $ 1.85$ & $ 9.58\,{\times}\,10^{-18}$ & $ 1.94$ & $ 32.7$ & $ 4.99\,{\times}\,10^{-21}$ & $ 18.1$ & $ 12950$ \\ 
lQ0YS3aF\_L3.0 & $ 105$ & $ 70$ & $ 6.4$ & $ 11.6$ & $ 17980$ & $ 0.00527$ & $ 3798$ & $ 1.06$ & $ -0.775$ & $ 17.3$ & $ 3.24\,{\times}\,10^{-19}$ & $ 1.06$ & $ 12.5$ & $ 8.66\,{\times}\,10^{-22}$ & $ 13.5$ & $ 12448$ \\ 
90Yt0exl\_L3.0 & $ 107$ & $ 65.5$ & $ 6.4$ & $ 9.54$ & $ 19670$ & $ 0.0357$ & $ 1191$ & $ 1.13$ & $ -0.345$ & $ 1.25$ & $ 2.07\,{\times}\,10^{-18}$ & $ 1.39$ & $ 36.4$ & $ 1.1\,{\times}\,10^{-21}$ & $ 24.1$ & $ 12051$ \\ 
TJyUR9bA\_L3.0 & $ 110$ & $ 62.6$ & $ 7.11$ & $ 6.77$ & $ 23840$ & $ 3.1\,{\times}\,10^{-7}$ & $ 631$ & $ 1.17$ & $ -0.102$ & $ 12.7$ & $ 7.15\,{\times}\,10^{-18}$ & $ 1.89$ & $ 46$ & $ 9.4\,{\times}\,10^{-23}$ & $ 14.4$ & $ 13097$ \\ 
90Yt0exl\_L2.5 & $ 110$ & $ 65.5$ & $ 6.4$ & $ 8.71$ & $ 19670$ & $ 0.0357$ & $ 1191$ & $ 1.13$ & $ -0.345$ & $ 1.25$ & $ 2.07\,{\times}\,10^{-18}$ & $ 1.39$ & $ 36.4$ & $ 1.1\,{\times}\,10^{-21}$ & $ 24.1$ & $ 10044$ \\ 
nTRTrE7X\_L3.0 & $ 114$ & $ 59$ & $ 7.11$ & $ 4.69$ & $ 29889$ & $ 8.1\,{\times}\,10^{-6}$ & $ 423$ & $ 1.13$ & $ -0.607$ & $ 1.85$ & $ 9.58\,{\times}\,10^{-18}$ & $ 1.94$ & $ 32.7$ & $ 4.99\,{\times}\,10^{-21}$ & $ 18.1$ & $ 15540$ \\ 
TJyUR9bA\_L3.5 & $ 114$ & $ 62.6$ & $ 7.11$ & $ 7.32$ & $ 23840$ & $ 3.1\,{\times}\,10^{-7}$ & $ 631$ & $ 1.17$ & $ -0.102$ & $ 12.7$ & $ 7.15\,{\times}\,10^{-18}$ & $ 1.89$ & $ 46$ & $ 9.4\,{\times}\,10^{-23}$ & $ 14.4$ & $ 15281$ \\ 
eGYXcOh8\_L2.5 & $ 116$ & $ 67.6$ & $ 7.11$ & $ 5.99$ & $ 25590$ & $ 0.000217$ & $ 3110$ & $ 1.18$ & $ -0.131$ & $ 2.34$ & $ 5.25\,{\times}\,10^{-19}$ & $ 1.56$ & $ 31.5$ & $ 1.81\,{\times}\,10^{-23}$ & $ 19.3$ & $ 13591$ \\ 
nTRTrE7X\_L2.0 & $ 118$ & $ 59$ & $ 7.11$ & $ 3.83$ & $ 29889$ & $ 8.1\,{\times}\,10^{-6}$ & $ 423$ & $ 1.13$ & $ -0.607$ & $ 1.85$ & $ 9.58\,{\times}\,10^{-18}$ & $ 1.94$ & $ 32.7$ & $ 4.99\,{\times}\,10^{-21}$ & $ 18.1$ & $ 10362$ \\ 
\noalign{\smallskip} 
\hline 
mean model \\
\noalign{\smallskip} 
\hline 
\noalign{\smallskip} 
mean & & $ 60.5$ & 6.77 & $ 8.38$* & $ 16834$* & $ 8.42\,{\times}\,10^{-5}$ & $ 952$ & $ 1.16$ & $ -0.592$ & $ 3.37$ & $ 2.98\,{\times}\,10^{-18}$ & $ 1.57$ & $ 33.7$ & $ 6.37\,{\times}\,10^{-22}$ & $ 21.8$ & $ 4984$* \\ 
sigma &&$ 9.7$ &$ 1.05$ & $ 2.86$ &$ 1.66$ &$ 72.2$ &$ 2.51$ &$ 0.06$ &$ 0.480$ &$ 2.72$ &$ 4.14$ &$ 0.31$ &$ 10.2 $ &$ 7.42$ &$ 10.1$ &$ 1.22$ \\ 
\noalign{\smallskip} 
\hline 
\footnotesize
\text{ * Pre-burst-value}
\end{tabular}
\end{sidewaystable}

\clearpage
\begin{table}
\begin{threeparttable}
\caption[]{Optical depth in the $V$-band for the ten best MM1 pre-burst models along the mid-plane and along the line of sight toward the center. The intrinsic extinction dominates the total extinction (including the interstellar extinction) by far, as indicated by the huge values in the $V$-band.
For $\tau^{\rm mid-plane}$ the contribution of the disk is given separately. For some models, $\tau_{\rm disk}$ almost covers the total extinction (along the line of sight), whereas for some models the contribution of the envelope (which is the same as $\tau_{\rm total}^{\rm mid-plane} - \tau_{\rm disk}^{\rm mid-plane}$) is the dominant one. We note that the optical depths are given for the pre-burst, during the burst the optical depths may be slightly lower because of dust sublimation.}
\label{tab: tau_V}
\begin{tabular}{ccccc}
\hline\hline
\noalign{\smallskip}
Model & inc [\degr]& $\tau_{\rm disk,\, V}^{\rm mid-plane}$ & $\tau_{\rm total,\, V}^{\rm mid-plane}$ & $\tau_{\rm total,\, V}^{\rm line\,of\,sight}$ \\
\noalign{\smallskip}
\hline
\noalign{\smallskip}
eGYXcOh8\_{02} & 19.3 & 360 & 3200 & 380 \\ 
90Yt0exl\_{03} &24.1& $1.2\,{\times}\,10^6$ & $1.2\,{\times}\,10^6$ & 390\\
PUyhjE8Z\_{02} &11.3& 30 & 11000 & 360\\ 
nTRTrE7X\_{02} & 18.1& 1700 & 9300 & 420\\ 
nTRTrE7X\_{03} &23.3& 1700 & 9300 & 510\\ 
TJyUR9bA\_{02} & 14.5& 2 & 11000 & 240\\
4kW1TtMH\_{04} &30.6&$3.4\,{\times}\,10^5$ & $3.5\,{\times}\,10^5$ & 830\\ 
lQ0YS3aF\_{02} &13.5&$1.2\,{\times}\,10^5$ &$1.4\,{\times}\,10^6$ & 889\\ 
qWHuujku\_{05} &46.8&$6.2\,{\times}\,10^5$ &$6.6\,{\times}\,10^5$ & 920\\ 
qWHuujku\_{06} &50.0 &$6.2\,{\times}\,10^5$ &$6.6\,{\times}\,10^5$ & 1200\\ 
\noalign{\smallskip}
\hline
mean model & 21.8 & 3300 & 11000 & 430\\
\noalign{\smallskip}
\hline
\end{tabular}
\end{threeparttable}
\end{table}

\end{appendix}

\end{document}